\documentclass[aps,prc,showkeys,nofootinbib]{revtex4}

\usepackage{color} 

\usepackage{graphicx}
\graphicspath{{Figures/}}
\usepackage{dcolumn}
\usepackage{amsthm}
\usepackage{bm}

\usepackage{isotope}
\usepackage{physics}

\newcommand{\xibf}{\mbox{\boldmath $\xi$}}

\newcommand{\varepsilonbf}{\mbox{\boldmath $\varepsilon$}}
\newcommand{\pibf}{\mbox{\boldmath $\pi$}}

\begin{document}


\title{Study of structure of deuteron from analysis of bremsstrahlung emission in proton-deuteron scattering in cluster models
%
}

\author{K.~A.~Shaulskyi$^{1}$}\email{shaulskyi@kinr.kiev.ua}%
\author{S.~P.~Maydanyuk$^{1,2}$}\email{sergei.maydanyuk@wigner.hu}%
\author{V.~S.~Vasilevsky$^{3}$}\email{vsvasilevsky@gmail.com}%

\affiliation{$^{(1)}$Institute for Nuclear Research, National Academy of Sciences of Ukraine, Kyiv, 03680, Ukraine}
\affiliation{$^{(2)}$Wigner Research Centre for Physics, Budapest, 1121, Hungary}
\affiliation{$^{(3)}$Bogolyubov Institute for Theoretical Physics, Metrolohichna str., 14b, Kyiv, 03143, Ukraine}


\date{\small\today}

\begin{abstract}
\textbf{Purpose:}
In this paper we investigated emission of bremsstrahlung photons in the scattering of protons off deuterons within the microscopic cluster models
in a wide region of the beam energy from low energies up to 1.5~GeV.
\textbf{Methods:}
Three-cluster model of bremsstrahlung is constructed for such a reaction.
Formalism of the model includes form factor of deuteron which characterizes dependence of bremsstrahlung cross sections on structure of deuteron.
This gives possibility to investigate the structure of nuclei from analysis of bremsstrahlung cross sections.
\textbf{Results:}
We studied dependence of the bremsstrahlung cross section on the structure of deuteron.
We use three different shapes of the deuteron wave functions. Besides, we also calculate  the cross section by neglecting internal structure of deuteron.
Analysis of dependence of the cross section on such a parameter shows the following.
(1) At beam energies 145 and 195~MeV used in experiments
bremsstrahlung cross section is not sensitive visibly on variations of the shape of the deuteron wave functions. (2) Stable difference between cross sections calculated with and without internal structure of deuteron
is observed at higher energy of beam (larger 500~MeV).
(3) The spectrum is increased as we pass from structureless deuteron (the oscillator length $b$=0) to the deuteron discribed by the shell-model wave function (the realistic oscillator length) 
inside the full energy region of the emitted photons.
\textbf{Conclusion:}
Our cluster model is a suitable tool to study the structure of deuteron with high enough precision from bremsstrahlung analysis.
We propose new experiments for such an investigation.
\end{abstract}



\maketitle

\section{Introduction
\label{sec.introduction}}

The bremsstrahlung emission of photons accompanying nuclear reactions is an important topic of nuclear physics and has been attracted a significant interest of many researchers for a long time
(see reviews~\cite{Amusia.1988.PhysRep,Pluiko.1987.PEPAN,Kamanin.1989.PEPAN,Bertulani.1988.PhysRep}).
This is explained by that the spectra of bremsstrahlung photons are calculated on the basis of nuclear models which include mechanisms of reactions, interactions between nuclei, dynamics, and many other physical issues.
%
%
A lot of aspects of nuclear processes, such as dynamics of nucleons in the nuclear scattering, interactions between nucleons,
mechanisms of reactions, quantum effects, deformations of nuclei,
properties of hypernuclei in reactions, etc.
can be included to the model describing the bremsstrahlung emission
(for example, see
Refs.~\cite{Maydanyuk.2003.PTP,Maydanyuk.2006.EPJA} for general properties of $\alpha$ decay from bremsstrahlung analysis,
Ref.~\cite{Maydanyuk.2009.NPA} for extraction of information about deformation of nuclei in the $\alpha$ decay from experimental bremsstrahlung data,
Ref.~\cite{Maydanyuk.2011.JPG} for bremsstrahlung in the nuclear radioactivity with emission of protons,
Ref.~\cite{Maydanyuk.2010.PRC} for bremsstrahlung in the spontaneous fission of \isotope[252]{Cf},
Ref.~\cite{Maydanyuk.2011.JPCS} for bremsstrahlung in the ternary fission of \isotope[252]{Cf},
Ref.~\cite{Maydanyuk_Zhang_Zou.2018.PRC} for bremsstrahlung in the pion-nucleus scattering
from our research,
there are many investigations of other researchers).
Note perspectives on studying electromagnetic observables of light nuclei based on chiral effective field theory \cite{Pastore.2008.PRC}.
The measurements of photons with analysis provide important information on these phenomena.

Analysis of bremsstrahlung photons accompanying nuclear reactions gives possibility to extract additional information on structure of nuclei.
Study of  the structure of nuclei on the basis of bremsstrahlung analysis is one of the most ambitious aims in nuclear physics.
%
%
Analyzing formalism of models, option to investigate structure of nuclei exists, in principle, and understandable.
Study of structure of nuclei is one of the most promising research directions,
taking into account that photons can be measured in experiments.
However, during long period of investigations of bremsstrahlung photons in nuclear physics, systematic study of structure of nuclei has not been realized yet.
One can explain that by difficulty in development of mathematical formalism of models,
importance to reach stability of numerical calculations that is possible at high precision.
Moreover, it turns out that not all available experimental bremsstrahlung data are well sensitive to structure of nuclei.
%
In this regards, one can remind investigations of bremsstrahlung emission in reactions with light nuclei 
within microscopic two-cluster models~%
%
\cite{
1985NuPhA.443..302B,1990PhRvC..41.1401L,1990PhRvC..42.1895L,%
1991NuPhA.529..467B,1991PhRvC..44.1695L,1992NuPhA.550..250B,Liu.1992.FBS,%
1993nuco.conf..423K,%
Dohet-Eraly.2011.JPCS,%
Dohet-Eraly.2011.PRC,%
Dohet-Eraly.2013.PRC,2013JPhCS.436a2030D,%
Dohet-Eraly.2013.PhD,2014PhRvC..89b4617D,2014PhRvC..90c4611D}.

Summarizing all issues mentioned above, we see perspective problem on realization of such an idea, that is a main aim of this paper.
We would like to understand, which parameters of nuclear structure are more effective to realize such an investigation.
Of course, the best way is to construct this model on the fully quantum basis, with inclusion of realistic nuclear interactions which were well tested experimentally.
A promising way is cluster formalism for description of structure of nuclei and nuclear process.
So, as a basis of this research we will develop the fully cluster model in combination of bremsstrahlung formalism.
The most effective process for such study is proton-deuteron scattering.
%
%
We focus on construction of such a unified cluster formalism,
analysis of available experimental information about bremsstrahlung for proton-deuteron scattering.
This paper  is continuation of our previous research \cite{Maydanyuk_Vasilevsky.2023.PRC},
where we developed cluster model in the folding approximation in study of bremsstrahlung emission
in the scattering of nuclei with the small number of nucleons and
we did not analyze possibility to extract information about structure of nuclei from bremsstrahlung cross sections.

The paper is organized in the following way.
In Sec.~\ref{sec.cluster} 
cluster models of emission of the bremsstrahlung photons in the proton-deuteron scattering is formulated.
Here, we give an explicit form of the operator of the bremsstrahlung emission,
define wave functions of $p+d$ system,
calculate matrix elements of bremsstrahlung emission,
define form factors of deuteron (characterizing its structure),
apply the multiple expansion approach for calculation of matrix elements.
In Sec.~\ref{sec.folding.1} matrix elements of bremsstrahlung emission in folding approximation are reviewed
(following to formalism in Ref.~\cite{Maydanyuk_Vasilevsky.2023.PRC}).
In Sec.~\ref{sec.crosssection} cross section of  the bremsstrahlung emission of photons is determined and resulting formulas are summarized.
In Sec.~\ref{sec.analysis} emission of the bremsstrahlung photons for the proton-deuteron scattering is studied on the basis of the model above.
%
We analyze role of the deuteron wave function and its form factor in calculations of the cross section
at different energies of relative motions between the scattered proton and deuteron.
We also describe the experimental bremsstrahlung data for the proton-deuteron scattering on the basis of the model.
%
%
Conclusions and perspectives are summarized in Sec.~\ref{sec.conclusions}.
Operator of bremsstrahlung emission in three-cluster model is calculated in App.~\ref{sec.app.cluster.operator}.
Useful details of calculation of integrals are presented in App.~\ref{sec.app.integrals}.
Form factor of deuteron in three-cluster approach is derived in App.~\ref{sec.app.clusterformfactor}.


\section{Matrix elements of bremsstrahlung emission in three-cluster model
\label{sec.cluster}}

\subsection{Operator of bremsstrahlung emission in three-cluster model}

Consider the translationally invariant interaction of photon with three-nucleon system
\begin{equation}
  \widehat{H}_{e}\left(  \mathbf{k}_{\gamma},\mathbf{\varepsilon}^{(\alpha)}\right) =
  \frac{1}{2}\frac{e\hbar}{m_{N}c}\sum_{i=1}^{A=3}\frac{1}{2}
  \left( 1+\widehat{\tau}_{iz}\right)  \left[  \widehat{\mathbf{\pi}}_{i}^{\ast} \mathbf{A}^{\ast}
  \left( i\right)  +\mathbf{A}^{\ast}\left(  i\right)
\widehat{\mathbf{\pi}}_{i}^{\ast}\right]
\label{eq:R01}
\end{equation}
where
%
\begin{equation}
\begin{array}{lllllll}
\vspace{1.0mm}
  \mathbf{A}^{\ast}\left(  i\right) =
  \mathbf{\varepsilon}^{(\alpha)}\exp\left\{
  -i\left(  \mathbf{k}_{\gamma}\mathbf{\rho}_{i}\right)  \right\}, & 

  \widehat{\mathbf{\pi}}_{i}^{\ast} = i\nabla_{\mathbf{\rho}_{i}}, & 
  \mathbf{\rho}_{i} = \mathbf{r}_{i}-\mathbf{R}_{\rm cm}, \\ 

  \mathbf{R}_{cm} = \displaystyle\frac{1}{A} \displaystyle\sum_{i=1}^{A}\mathbf{r}_{i}, & 
  \widehat{\mathbf{\pi}}_{i} = \widehat{\mathbf{p}}_{i}-\widehat{\mathbf{P}}_{cm}, & 
  \widehat{\mathbf{P}}_{cm}  = \displaystyle\frac{1}{A} \displaystyle\sum_{i=1}^{A}\widehat{\mathbf{p}}_{i}. 
\end{array}
\label{eq:R02}
\end{equation}
%
Here, $\varepsilonbf^{(\alpha)}$ are unit vectors of \emph{linear} polarization of the photon emitted ($\varepsilonbf^{(\alpha), *} = \varepsilonbf^{(\alpha)}$), $\mathbf{k}_{\gamma}$ is wave vector of the photon and $w_{\gamma} = k_{\gamma} c = \bigl| \mathbf{k}_{\gamma}\bigr|c$. Vectors $\varepsilonbf^{(\alpha)}$ are perpendicular to $\mathbf{k}_{\gamma}$ in Coulomb gauge. We have two independent polarizations $\varepsilonbf^{(1)}$ and $\varepsilonbf^{(2)}$ for the photon with impulse $\mathbf{k}_{\gamma}$ ($\alpha=1,2$).
Also we have properties:
\begin{equation}
\begin{array}{lclc}
  \Bigl[ \mathbf{k}_{\gamma} \times \varepsilonbf^{(1)} \Bigr] = k_{\gamma}\, \varepsilonbf^{(2)}, &
  \Bigl[ \mathbf{k}_{\gamma} \times \varepsilonbf^{(2)} \Bigr] = -\, k_{\gamma}\, \varepsilonbf^{(1)}, &
  \Bigl[ \mathbf{k}_{\gamma} \times \varepsilonbf^{(3)} \Bigr] = 0, &
  \displaystyle\sum\limits_{\alpha=1,2,3}
  \Bigl[ \mathbf{k}_{\gamma} \times \varepsilonbf^{(\alpha)} \Bigr] =
  k_{\gamma}\, (\varepsilonbf^{(2)} - \varepsilonbf^{(1)}).
\end{array}
\label{eq.matrixelement.1.4}
\end{equation}
%
Let us introduce new variables, namely Jacobi vectors $\mathbf{r}$ and $\mathbf{q}$
\begin{equation}
\begin{array}{lllll}
  \mathbf{r} =
  \displaystyle\frac{1}{\sqrt{2}}\left(  \mathbf{\rho}_{1}-\mathbf{\rho}_{2}\right) =
  \displaystyle\frac{1}{\sqrt{2}}\left(  \mathbf{r}_{1}-\mathbf{r}_{2}\right), 
 \quad
  \mathbf{q} =
  \sqrt{\displaystyle\frac{2}{3}}\left(  \mathbf{\rho}_{3}-\frac
  {\mathbf{\rho}_{1}+\mathbf{\rho}_{2}}{2}\right)  =\sqrt{\frac{2}{3}}\left(
  \mathbf{r}_{3} - \displaystyle\frac{\mathbf{r}_{1}+\mathbf{r}_{2}}{2}\right),   \nonumber\\

  \mathbf{q}_{A} =
  \sqrt{\displaystyle\frac{1}{3}}\left(  \mathbf{r}_{1}+\mathbf{r}_{2} + \mathbf{r}_{3}\right) =
  \sqrt{3} \mathbf{R}_{cm}.\nonumber
\end{array}
\label{eq:R02A}
\end{equation}
Inverse relations are
\begin{equation}
\begin{array}{lllll}
  \mathbf{r}_{1} =
    \displaystyle\frac{1}{\sqrt{2}} \mathbf{r} - \displaystyle\frac{1}{\sqrt{6}} \mathbf{q} + \displaystyle\frac{1}{\sqrt{3}}\mathbf{q}_{A}, &
  \mathbf{r}_{2} =
    - \displaystyle\frac{1}{\sqrt{2}} \mathbf{r} - \displaystyle\frac{1}{\sqrt{6}} \mathbf{q} + \displaystyle\frac{1}{\sqrt{3}}\mathbf{q}_{A}, &
  \mathbf{r}_{3} =
    \displaystyle\sqrt{\frac{2}{3}} \mathbf{q} + \displaystyle\frac{1}{\sqrt{3}} \mathbf{q}_{A}.
\end{array}
\label{eq:R02B}
\end{equation}
%
Similar relations can be written for momenta%

\begin{align*}
\mathbf{\pi}_{\mathbf{r}}  &  =\frac{1}{\sqrt{2}}\left(  \mathbf{\pi}%
_{1}-\mathbf{\pi}_{2}\right)  ,\\
\mathbf{\pi}_{\mathbf{q}}  &  =\sqrt{\frac{2}{3}}\left(  \mathbf{\pi}%
_{3}-\frac{\mathbf{\pi}_{1}+\mathbf{\pi}_{2}}{2}\right)  ,\\
\mathbf{\pi}_{A}  &  =\sqrt{\frac{1}{3}}\left(  \mathbf{\pi}_{1}+\mathbf{\pi
}_{2}+\mathbf{\pi}_{3}\right)  .
\end{align*}
Inverse relations
\begin{align*}
\mathbf{\pi}_{1}  &  =\frac{1}{\sqrt{2}}\mathbf{\pi}_{\mathbf{r}}%
\mathbf{-}\frac{1}{\sqrt{6}}\mathbf{\pi}_{\mathbf{q}}\mathbf{+}\frac{1}%
{\sqrt{3}}\mathbf{\pi}_{A},\\
\mathbf{\pi}_{2}  &  =-\frac{1}{\sqrt{2}}\mathbf{\pi}_{\mathbf{r}}%
\mathbf{-}\frac{1}{\sqrt{6}}\mathbf{\pi}_{\mathbf{q}}\mathbf{+}\frac{1}%
{\sqrt{3}}\mathbf{\pi}_{A},\\
\mathbf{\pi}_{3}  &  =\sqrt{\frac{2}{3}}\mathbf{\pi}_{\mathbf{q}}%
\mathbf{+}\frac{1}{\sqrt{3}}\mathbf{\pi}_{A}.
\end{align*}

Now we fix position of nucleons. We assume that vector $\mathbf{r}$ measures
the distance between proton and neutron which form a deuteron. We also assume
that $\mathbf{r}_{1}$ is a coordinate of the first proton and $\mathbf{r}_{2}$
is a coordinate of a neutron. Vector $\mathbf{r}_{3}$ determines the location
of the second proton.
With such definitions, the operator
$\widehat{H}_{e}\left( \mathbf{k}_{\gamma},\mathbf{\varepsilon}^{(\alpha)}\right)$ is
[see App.~\ref{sec.app.cluster.operator} for details, also we take into account that
$\left(  \mathbf{\varepsilon}^{(\alpha)}, \mathbf{k}_{\gamma}\right) = 0$]
%
\begin{equation}
\begin{array}{lllll}
\vspace{1.5mm}
  \widehat{H}_{e}\left(  \mathbf{k}_{\gamma},\mathbf{\varepsilon}^{(\alpha)}\right) & = &
  \displaystyle\frac{1}{2} \displaystyle\frac{e\hbar}{m_{N}c}
  \biggl\{
    \displaystyle\frac{2}{ \sqrt{2}}
    \exp\left\{ -i\displaystyle\frac{1}{\sqrt{2}}\left( \mathbf{k}_{\gamma}\mathbf{r}\right)  \right\}
    \left(  \mathbf{\varepsilon}^{(\alpha)}, \mathbf{\pi}_{\mathbf{r}}^{\ast}\right)
    \exp\left\{ i\frac{1}{\sqrt{6}}\left(  \mathbf{k}_{\gamma}\mathbf{q}\right)  \right\}\, - \\

  & - &
    \sqrt{\displaystyle\frac{2}{3}}\exp\left\{  -i \displaystyle\frac{1}{\sqrt{2}}
    \left( \mathbf{k}_{\gamma}\mathbf{r}\right)  \right\}
    \exp\left\{  i\displaystyle\frac{1}{\sqrt{6}}
    \left( \mathbf{k}_{\gamma}\mathbf{q}\right) \right\}
    \left( \mathbf{\varepsilon}^{(\alpha)},\mathbf{\pi}_{\mathbf{q}}^{\ast}\right)
    + 2\sqrt{\displaystyle\frac{2}{3}}
    \exp\biggl\{  -i\sqrt{\displaystyle\frac{2}{3}} \left( \mathbf{k}_{\gamma}\mathbf{q}\right) \biggr\}
    \left( \mathbf{\varepsilon}^{(\alpha)}, \mathbf{\pi}_{\mathbf{q}}^{\ast}\right)
  \biggr\}.
\end{array}
\label{eq:R19A.add}
\end{equation}

This is the universal and model-independent form of the operator of bremsstrahlung emission for a system comprising from two protons and one neutron. To calculated cross section of bremsstrahlung emission in the process of a proton scattering from a deuteron, we need to formulate model which provides a realistic description of the $p+d$ scattering in economical way, i.e. with minimum of computations but with a reliable output. As the output, we need to determine wave functions of the $p+d$ scattering at selected energies of initial and final states of bremsstrahlung emission. For this aim, we select the resonating group method (RGM), which is the most rigorous and self-consistent realization of  a cluster model. Actually, we will use three different variants of the RGM: two- and three-cluster variants and so-called the folding approximation. These three variants of the RGM are explained in detail in next Section.


\section{Two and three-cluster models of $p+d$ system}

Three-nucleon system $^{3}$He and its decay channel $p+d$ will be studied in
the framework of two-  and three-cluster models. In a two-cluster model, wave
function of the system is 
\begin{equation}
\Psi=\widehat{\mathcal{A}}\left\{  \phi^{(S)}\left(  d,\mathbf{r}\right)
\phi\left(  p\right)  \psi_{E}\left(  \mathbf{q}\right)  \right\}
,\label{eq:C01}%
\end{equation}
where $\phi^{(S)}\left(  d,\mathbf{r}\right)  $ is the deuteron wave function
from the oscillator shell-model, $\phi\left(  p\right)  $ is a wave function
of proton represented by its spin and isospin parts, and $\psi_{E}\left(
\mathbf{q}\right)  $ is a wave function of relative motion of proton and
deuteron. The antisymmetrization operator $\widehat{\mathcal{A}}$ makes wave functions of the $p+d$
system fully antisymmetric. Three-cluster model suggests the following form
for three-nucleon system%
\begin{equation}
\Psi=\widehat{\mathcal{A}}\left\{  \phi\left(  n\right)  \phi\left(
p_{1}\right)  \phi\left(  p_{2}\right)  f\left(  \mathbf{r,q}\right)
\right\}  ,\label{eq:C05}%
\end{equation}
Wave function $\phi\left(  p_{2}\right)  f\left(  \mathbf{r,q}\right)$ of relative motion of nucleons has to be determined by solving
the Schr\"{o}dinger equation or the Faddeev equations.

By assuming that the shape of a deuteron does not change when proton is
approaching, then three-particle wave function can be represented as%
\begin{equation}
\Psi=\widehat{\mathcal{A}}\left\{  \phi\left(  d,\mathbf{r}\right)
\phi\left(  p\right)  \psi_{E}\left(  \mathbf{q}\right)  \right\}
,\label{eq:C07}%
\end{equation}
where  $\phi\left(  d,\mathbf{r}\right)  $ is a wave function of the bound
state of deuteron. The wave function $\phi\left(  d,\mathbf{r}\right)  $ is a
solution of two-body Schr\"{o}dinger equation with selected nucleon-nucleon potential.

Note that in both models, in two-cluster and three-cluster, the wave
function of deuteron is assumed to be antisymmetric, than the
antisymmetrization operator $\widehat{\mathcal{A}}$ in Eqs. (\ref{eq:C01}) and
\ (\ref{eq:C07}) consists of the unit operator and two permutation operators.  As
the results,  the antisymmetrization operator $\widehat{\mathcal{A}}$ creates
in Eqs. (\ref{eq:C01}) and \ (\ref{eq:C07}) three terms which are similar to
the terms in curly brackets.

If one ignores the full antisymmetrization in Eqs. (\ref{eq:C01}) and
 (\ref{eq:C07}) by omitting the operator $\widehat{\mathcal{A}}$, one
obtains a simple version of the two- and three-cluster models which is called
 a folding approximation or folding model. In order to avoid bulky
expressions, we will use this approximation to present matrix elements of the
$\widehat{H}_{e}\left(  \mathbf{k}_{\gamma},\mathbf{\varepsilon}_{\mu}\right)
$ between the initial and final states of the $p+d$ system.

To  construct wave functions of the system $p+d$ in different approximations
(models), we need to solve the appropriate Schr\"{o}dinger equations. For this
aim we employ the algebraic version of the resonating group method (RGM),
formulated in Refs. \cite{kn:Fil_Okhr}, \cite{kn:Fil81}. This version of the
RGM uses the full basis of oscillator functions to expand wave functions of
the relative motion of clusters. As the results, the  Schr\"{o}dinger equation
is reduced to a system of linear algebraic equation for expansion
coefficients. Besides, the algebraic version implements proper boundary
conditions in discrete, oscillator representation.

To study $p+d$ system in three-cluster approximation, we will employ a
three-cluster model developed in Ref.  \cite{2009NuPhA.824...37V}.

\subsection{Wave functions of $p+d$ system in the cluster formalism}

To calculate matrix elements of the operator $\widehat{H}_{e}\left( \mathbf{k}_{\gamma},\mathbf{\varepsilon}^{(\alpha)}\right)$ we need to construct wave
functions of the system $p+d$. If we neglect the Pauli principle and employ an
adiabatic approximation, then wave function of the system can be constructed
in a separable form%
\begin{equation}
\Psi\left(  \mathbf{r},\mathbf{q}\right)  =\phi\left(  \mathbf{r}\right)
\psi\left(  \mathbf{q}\right)  , \label{eq:R21}%
\end{equation}
where wave function of deuteron $\phi\left(  \mathbf{r}\right)  $ is a
solution of the two-body Schr\"{o}dinger equation%
\begin{equation}
\left(  \widehat{H}_{d}-E_{d}\right)  \phi\left(  \mathbf{r}\right)  =0,
\label{eq:R22}%
\end{equation}%
\begin{equation}
\widehat{H}_{d}=-\frac{\hbar^{2}}{2m}\frac{d^{2}}{d\mathbf{r}^{2}}+\widehat
{V}_{NN}\left(  \mathbf{r}\right)  \label{eq:R22A}%
\end{equation}
where $m$ is a mass of nucleon. If the nucleon-nucleon potential ${V}_{NN}\left(  \mathbf{r}\right)$ is used in the form%
\begin{equation}
\widehat{V}_{NN}\left(  \mathbf{r}\right)  =\sum_{S=0,1}\sum_{T=0,1}%
V_{2S+1,2T+1}\left(  \mathbf{r}\right)  \widehat{P}_{S}\widehat{P}%
_{T},\label{eq:R23}%
\end{equation}
where $\widehat{P}_{S}$ ($\widehat{P}_{T}$)\ is the projection operator
projecting onto the spin $S$  (the isospin $T$) of two-nucleon system, then in Eq.
(\ref{eq:R22A}) $\widehat{V}_{NN}\left(  \mathbf{r}\right)  $ should be
replace with the even component $V_{31}\left(  \mathbf{r}\right)  $, as
deuteron has the spin $S$=1 and the isospin $T$=1.

The wave function describing interaction of proton with deuteron obeys the
following equations
\begin{equation}
\left(  \widehat{H}_{p}-E_{p}\right)  \psi\left(  \mathbf{q}\right)  =0,
\label{eq:R24}%
\end{equation}
where%
\[
\widehat{H}_{p}=-\frac{\hbar^{2}}{2m}\frac{d^{2}}{d\mathbf{q}^{2}}+\widehat
{V}_{pd}\left(  \mathbf{q}\right)
\]
and the potential energy $\widehat{V}_{pd}\left(  \mathbf{q}\right)  $ equals%

\[
\widehat{V}_{pd}\left(  \mathbf{q}\right)  =\left\langle \phi\left(
\mathbf{r}\right)  \left\vert \sum_{i=1,2}\widehat{V}_{NN}\left(
\mathbf{r}_{3}-\mathbf{r}_{i}\right)  +\sum_{i=1,2}\widehat{V}_{C}\left(
\mathbf{r}_{3}-\mathbf{r}_{i}\right)  \right\vert \phi\left(  \mathbf{r}%
\right)  \right\rangle.
\]
Here, integration is performed over vector $\mathbf{r}$, and nucleon-nucleon
$\widehat{V}_{NN}$ and Coulomb $\widehat{V}_{C}$ potentials are involved in
definition of $\widehat{V}_{pd}\left(  \mathbf{q}\right) $.
Equation (\ref{eq:R24}) determines both initial $\psi_{E_{i}}\left(
\mathbf{q}\right)  $ and final $\psi_{E_{f}}\left(  \mathbf{q}\right)  $ wave
functions of the $p+d$ system.


\subsection{Matrix elements of bremsstrahlung emission in the cluster formalism}

Based on assumptions made, we have got matrix element of transition from
initial to final states
\[
\left\langle \phi\left(  \mathbf{r}\right)  \psi_{E_{f}}\left(  \mathbf{q}%
\right)  \left\vert \widehat{H}_{e}\left(  \mathbf{k}_{\gamma}%
,\mathbf{\varepsilon}^{(\alpha)}\right)  \right\vert \phi\left(  \mathbf{r}\right)
\psi_{E_{i}}\left(  \mathbf{q}\right)  \right\rangle .
\]
One suggest to calculate this matrix elements on two steps. On the first step,
we calculate the matrix element%
\[
\widehat{\mathcal{H}}_{e}\left(  \mathbf{q}\right)  =\left\langle \phi\left(
\mathbf{r}\right)  \left\vert \widehat{H}_{e}\left(  \mathbf{k}_{\gamma
},\mathbf{\varepsilon}^{(\alpha)}\right)  \right\vert \phi\left(  \mathbf{r}%
\right)  \right\rangle
\]
by integrating over vector $\mathbf{r}$.
By using Eq.~(\ref{eq:R19A.add}), we obtain
\begin{align}
&  \widehat{\mathcal{H}}_{e}\left(  \mathbf{q}\right)  =
\left\langle
\phi\left(  \mathbf{r}\right)  \left\vert \widehat{H}_{e}\left(
\mathbf{k}_{\gamma},\mathbf{\varepsilon}^{(\alpha)}\right)  \right\vert \phi\left(
\mathbf{r}\right)  \right\rangle \label{eq:R30}\\
&  = \frac{1}{2}\frac{e\hbar}{m_{N}c} \biggl\{ 
\frac{2}{\sqrt{2}}\left\langle \phi\left(  \mathbf{r}\right)  \left\vert
\exp\left\{  -i\frac{1}{\sqrt{2}}\left(  \mathbf{k}_{\gamma}\mathbf{r}\right)
\right\}  \left(  \mathbf{\varepsilon}^{(\alpha)},\mathbf{\pi}_{\mathbf{r}}^{\ast
}\right)  \right\vert \phi\left(  \mathbf{r}\right)  \right\rangle
\exp\left\{  i\frac{1}{\sqrt{6}}\left(  \mathbf{k}_{\gamma}\mathbf{q}\right)
\right\}  \nonumber\\
&  \mathbf{-}\sqrt{\frac{2}{3}}\left\langle \phi\left(  \mathbf{r}\right)
\left\vert \exp\left\{  -i\frac{1}{\sqrt{2}}\left(  \mathbf{k}_{\gamma
}\mathbf{r}\right)  \right\}  \right\vert \phi\left(  \mathbf{r}\right)
\right\rangle \exp\left\{  i\frac{1}{\sqrt{6}}\left(  \mathbf{k}_{\gamma
}\mathbf{q}\right)  \right\}  \left(  \mathbf{\varepsilon}^{(\alpha)},\mathbf{\pi
}_{\mathbf{q}}^{\ast}\right)  \nonumber\\
&  +\left.  2\sqrt{\frac{2}{3}}\exp\left\{  -i\sqrt{\frac{2}{3}}\left(
\mathbf{k}_{\gamma}\mathbf{q}\right)  \right\}  \left(  \mathbf{\varepsilon
}^{(\alpha)},\mathbf{\pi}_{\mathbf{q}}^{\ast}\right)  \right\} \nonumber
\end{align}
and then
\begin{align}
&  \left\langle \phi\left(  \mathbf{r}\right)  \psi_{E_{f}}\left(
\mathbf{q}\right)  \left\vert \widehat{H}_{e}\left(  \mathbf{k}_{\gamma
},\mathbf{\varepsilon}^{(\alpha)}\right)  \right\vert \phi\left(  \mathbf{r}%
\right)  \psi_{E_{i}}\left(  \mathbf{q}\right)  \right\rangle =\label{eq:R31} \\
&  = \frac{1}{2}\frac{e\hbar}{m_{N}c} \biggl\{ 
\frac{2}{\sqrt{2}}\left\langle \phi\left(  \mathbf{r}\right)  \left\vert
\exp\left\{  -\frac{i}{\sqrt{2}}\left(  \mathbf{k}_{\gamma}\mathbf{r}\right)
\right\}  \left(  \mathbf{\varepsilon}^{(\alpha)},\mathbf{\pi}_{\mathbf{r}}^{\ast
}\right)  \right\vert \phi\left(  \mathbf{r}\right)  \right\rangle 
%
  \left\langle \psi_{E_{f}}\left(  \mathbf{q}\right)  \left\vert
  \exp\left\{  \frac{i}{\sqrt{6}}\left(  \mathbf{k}_{\gamma}\mathbf{q}\right)
\right\}  \right\vert \psi_{E_{i}}\left(  \mathbf{q}\right)  \right\rangle
\nonumber\\
&  - \sqrt{\frac{2}{3}}\left\langle \phi\left(  \mathbf{r}\right)
\left\vert \exp\left\{  -\frac{i}{\sqrt{2}}\left(  \mathbf{k}_{\gamma
}\mathbf{r}\right)  \right\}  \right\vert \phi\left(  \mathbf{r}\right)
\right\rangle 
%
  \left\langle \psi_{E_{f}}\left(  \mathbf{q}\right)  \left\vert
\exp\left\{  \frac{i}{\sqrt{6}}\left(  \mathbf{k}_{\gamma}\mathbf{q}\right)
\right\}  \left(  \mathbf{\varepsilon}^{(\alpha)},\mathbf{\pi}_{\mathbf{q}}^{\ast
}\right)  \right\vert \psi_{E_{i}}\left(  \mathbf{q}\right)  \right\rangle
\nonumber\\
&  + \left.  2\sqrt{\frac{2}{3}}\left\langle \psi_{E_{f}}\left(  \mathbf{q}%
\right)  \left\vert \exp\left\{  -i\sqrt{\frac{2}{3}}\left(  \mathbf{k}%
_{\gamma}\mathbf{q}\right)  \right\}  \left(  \mathbf{\varepsilon}^{(\alpha)},\mathbf{\pi}_{\mathbf{q}}^{\ast}\right)  \right\vert \psi_{E_{i}}\left(
\mathbf{q}\right)  \right\rangle \right\}. \nonumber
\end{align}
Thus we need to calculate few basic integrals
\begin{align}
&  \left\langle \phi\left(  \mathbf{r}\right)  \left\vert \exp\left\{
-\frac{i}{\sqrt{2}}\left(  \mathbf{k}_{\gamma}\mathbf{r}\right)  \right\}
\right\vert \phi\left(  \mathbf{r}\right)  \right\rangle ,\label{eq:R32}\\
&  \left\langle \phi\left(  \mathbf{r}\right)  \left\vert \exp\left\{
-\frac{i}{\sqrt{2}}\left(  \mathbf{k}_{\gamma}\mathbf{r}\right)  \right\}
\left(  \mathbf{\varepsilon}^{(\alpha)},\mathbf{\pi}_{\mathbf{r}}^{\ast}\right)
\right\vert \phi\left(  \mathbf{r}\right)  \right\rangle ,\nonumber\\
&  \left\langle \psi_{E_{f}}\left(  \mathbf{q}\right)  \left\vert \exp\left\{
\frac{i}{\sqrt{6}}\left(  \mathbf{k}_{\gamma}\mathbf{q}\right)  \right\}
\right\vert \psi_{E_{i}}\left(  \mathbf{q}\right)  \right\rangle ,\nonumber\\
&  \left\langle \psi_{E_{f}}\left(  \mathbf{q}\right)  \left\vert \exp\left\{
\frac{i}{\sqrt{6}}\left(  \mathbf{k}_{\gamma}\mathbf{q}\right)  \right\}
\left(  \mathbf{\varepsilon}^{(\alpha)},\mathbf{\pi}_{\mathbf{q}}^{\ast}\right)
\right\vert \psi_{E_{i}}\left( \mathbf{q} \right)  \right\rangle . \nonumber
\end{align}
Note that with such definition of coordinates (\ref{eq:R02A}),
the wave vectors of initial and final states are defined as
$k_{i} = \sqrt{ 2mE_{i}\, /\, \hbar^{2} }$, 
$k_{f} = \sqrt{ 2mE_{f} \, /\,\hbar^{2}}$, 
%
%
where $E_{i}$ and $E_{f}$ energies are in MeV and in the center of mass motion.


\subsection{Introduction of form factors
\label{sec.matrixelement.1}}

Introduce the following definitions of
\emph{form factors of deuteron} (in formalism of three-cluster model):
\begin{equation}
\begin{array}{lllll}
\vspace{1.0mm}
  F_{1} (\vb{k}_{\gamma}) & = &
    \Bigl\langle \phi (\vb{r}) \Bigl|\,
      \exp\Bigl\{ -\, \displaystyle\frac{i}{\sqrt{2}}\, (\vb{k}_{\gamma} \vb{r}) \Bigr\}
    \Bigr|\,
    \phi (\vb{r})
  \Bigr\rangle, \\

  F_{2,\, \alpha} (\vb{k}_{\gamma}) & = &
    \Bigl\langle \phi (\vb{r}) \Bigl|\,
      \exp\Bigl\{ -\, \displaystyle\frac{i}{\sqrt{2}}\, (\vb{k}_{\gamma} \vb{r}) \Bigr\}\,
      (\varepsilon^{(\alpha)}, \vb{\pi}_{\mathbf{r}}^{*})\,
    \Bigr|\,
    \phi (\vb{r})
  \Bigr\rangle.
\end{array}
\label{eq.matrixelement.1.1}
\end{equation}
Then, the matrix element of emission in Eq.~(\ref{eq:R31}) is rewritten as
\begin{equation}
\begin{array}{lllll}
\vspace{1.0mm}
  & \Bigl\langle \Psi_{E_{f}} (\vb{r}, \vb{q}) \Bigl|\,
    \hat{H}_{\gamma} (\vb{k}_{\gamma}, \varepsilon^{(\alpha)}) \Bigr|\,
    \Psi_{E_{i}} (\vb{r}, \vb{q}) \Bigr\rangle\; = \\

\vspace{1.0mm}
  = &
  \displaystyle\frac{1}{2}\,
  \displaystyle\frac{e\, \hbar}{m_{N}c}
  \Bigl\{

  \displaystyle\frac{2}{\sqrt{2}}\,
    \Bigl\langle \psi_{E_{f}} (\vb{q}) \Bigl|\,
      \exp\Bigl\{ \displaystyle\frac{i}{\sqrt{6}}\, (\vb{k}_{\gamma} \vb{q}) \Bigr\}
    \Bigr|\,
    \phi_{E_{i}} (\vb{q})
  \Bigr\rangle \cdot F_{2,\, \alpha}\; - \\


\vspace{1.0mm}
  - &
    \sqrt{\displaystyle\frac{2}{3}}\,
    \Bigl\langle \psi_{E_{f}} (\vb{q}) \Bigl|\,
      \exp\Bigl\{ \displaystyle\frac{i}{\sqrt{6}}\, (\vb{k}_{\gamma} \vb{q}) \Bigr\}\,
      (\varepsilon^{(\alpha)}, \vb{\pi}_{\mathbf{q}}^{*})\,
    \Bigr|\,
    \psi_{E_{i}} (\vb{q})
  \Bigr\rangle \cdot F_{1}\; + \\

  + &
  2\, \sqrt{\displaystyle\frac{2}{3}}\,
    \Bigl\langle \varphi_{E_{f}} (\vb{q}) \Bigl|\,
      \exp\Bigl\{ -i\, \sqrt{\displaystyle\frac{2}{3}}\, (\vb{k}_{\gamma} \vb{q}) \Bigr\}
      (\varepsilon^{(\alpha)}, \vb{\pi}_{\mathbf{q}}^{*})\,
    \Bigr|
    \varphi_{E_{i}} (\vb{q})
  \Bigr\rangle
  \Bigr\},
\end{array}
\label{eq.matrixelement.1.2}
\end{equation}
where
\begin{equation}
\begin{array}{lllll}
  \Psi_{E_{i}} (\vb{r}, \vb{q}) =
  \phi\left(  \mathbf{r}\right)  \psi_{E_{i}}\left(  \mathbf{q}\right),
\end{array}
\label{eq.matrixelement.1.3.0}
\end{equation}

\begin{equation}
\begin{array}{lllll}
  \Psi_{E_{f}} (\vb{r}, \vb{q}) =
  \phi\left(  \mathbf{r}\right)  \psi_{E_{f}}\left(  \mathbf{q}\right).
\end{array}
\label{eq.matrixelement.1.3}
\end{equation}
%
%
We introduce the following notations for matrix elements as
\begin{equation}
\begin{array}{ll}
  \vb{I}_{1} (\alpha) =
  \biggl< \varphi_{E_{f}} (\vb{q}) \biggl| \,
    e^{-i\alpha\mathbf{k_{\gamma}q}} \displaystyle\frac{\partial}{\partial \mathbf{q}}\,
    \biggr| \, \varphi_{E_{i}} (\vb{q}) \biggr>_\mathbf{q}, &

  I_{2} (\alpha) =
  \Bigl< \varphi_{E_{f}} (\vb{q}) \Bigl| \,  e^{-i\alpha \mathbf{k_{\gamma}q}} \, \Bigr| \, \varphi_{E_{i}} (\vb{q}) \Bigr>_\mathbf{q}.
\end{array}
\label{eq.matrixelement.1.5}
\end{equation}
Then, the full matrix element (\ref{eq.matrixelement.1.2}) can be rewritten as
\begin{equation}
\begin{array}{lllll}
\vspace{1.0mm}
  & \Bigl\langle \Psi_{E_{f}} (\vb{r}, \vb{q}) \Bigl|\,
    \hat{H}_{\gamma} (\vb{k}_{\gamma}, \varepsilon^{(\alpha)}) \Bigr|\,
    \Psi_{E_{i}} (\vb{r}, \vb{q}) \Bigr\rangle\; = \\

\vspace{1.0mm}
  = &
  \displaystyle\frac{1}{2}\,
  \displaystyle\frac{e\, \hbar}{m_{N}c}
  \Bigl\{

    \displaystyle\frac{2}{\sqrt{2}} \cdot
      I_{2} \Bigl( \displaystyle\frac{-1}{\sqrt{6}} \Bigr)
      \cdot F_{2,\, \alpha} - 

    \sqrt{\displaystyle\frac{2}{3}}\,
    \Bigl\langle \psi_{E_{f}} (\vb{q}) \Bigl|\,
      \exp\Bigl\{ \displaystyle\frac{i}{\sqrt{6}}\, (\vb{k}_{\gamma} \vb{q}) \Bigr\}\,
      (\varepsilon^{(\alpha)}, \vb{\pi}_{\mathbf{q}}^{*})\,
    \Bigr|\,
    \psi_{E_{i}} (\vb{q})
  \Bigr\rangle \cdot F_{1}\; + \\

  + &
  2\, \sqrt{\displaystyle\frac{2}{3}}\,
    \Bigl\langle \varphi_{E_{f}} (\vb{q}) \Bigl|\,
      \exp\Bigl\{ -i\, \sqrt{\displaystyle\frac{2}{3}}\, (\vb{k}_{\gamma} \vb{q}) \Bigr\}
      (\varepsilon^{(\alpha)}, \vb{\pi}_{\mathbf{q}}^{*})\,
    \Bigr|
    \varphi_{E_{i}} (\vb{q})
  \Bigr\rangle
  \Bigr\}.
\end{array}
\label{eq.matrixelement.1.6}
\end{equation}

Taking into account
\begin{equation}
\begin{array}{lllll}
  \vb{\pi}_{\mathbf{q}} =
  - i \hbar\, \mathbf{\displaystyle\frac{d}{dq}}, &

  \vb{\pi}_{\mathbf{q}}^{*} =
  i \hbar\, \mathbf{\displaystyle\frac{d}{dq}},
\end{array}
\label{eq.matrixelement.1.7}
\end{equation}
we rewrite
\begin{equation}
\begin{array}{lllll}
  \Bigl\langle \psi_{E_{f}} (\vb{q}) \Bigl|\,
      \exp\Bigl\{ \pm i\, \alpha'\, (\vb{k}_{\gamma} \vb{q}) \Bigr\}\,
      (\varepsilon^{(\alpha)}, \vb{\pi}_{\mathbf{q}}^{*})\,
    \Bigr|\,
    \psi_{E_{i}} (\vb{q})
  \Bigr\rangle =
  i \hbar\, \varepsilon^{(\alpha)}\, \vb{I}_{1} (\mp \alpha').
\end{array}
\label{eq.matrixelement.1.9}
\end{equation}
So, the full matrix element (\ref{eq.matrixelement.1.6}) obtains the following form as
\begin{equation}
\begin{array}{lllll}
\vspace{1.0mm}
  & \Bigl\langle \Psi_{E_{f}} (\vb{r}, \vb{q}) \Bigl|\,
    \hat{H}_{\gamma} (\vb{k}_{\gamma}, \varepsilon^{(\alpha)}) \Bigr|\,
    \Psi_{E_{i}} (\vb{r}, \vb{q}) \Bigr\rangle\; = \\

  = &
  -\, \displaystyle\frac{1}{2}\,
  \displaystyle\frac{e\, \hbar}{m_{N}c}
  \Bigl\{

  \displaystyle\frac{2}{\sqrt{2}}\, F_{2,\, \alpha} \cdot
    I_{2} \Bigl( \displaystyle\frac{-1}{\sqrt{6}} \Bigr) -

  i \hbar\,
    \sqrt{\displaystyle\frac{2}{3}}\, F_{1}\;
    \varepsilon^{(\alpha)}\; \vb{I}_{1} \Bigl( -\displaystyle\frac{1}{\sqrt{6}} \Bigr) +

  2\, i \hbar\, \sqrt{\displaystyle\frac{2}{3}}\,
    \varepsilon^{(\alpha)}\; \vb{I}_{1} \Bigl( \sqrt{\displaystyle\frac{2}{3}} \Bigr)
  \Bigr\}.
\end{array}
\label{eq.matrixelement.1.10}
\end{equation}
%
Note that vectors $\mathbf{\varepsilon}^{(\alpha)}$ are perpendicular to $\mathbf{k}_{\gamma}$ in Coulomb gauge.
Taking this property into account, we obtain
\begin{equation}
\begin{array}{lllll}
  (\varepsilon^{(\alpha)}, \vb{k}_{\gamma}) = 0.
\end{array}
\label{eq.matrixelement.1.11}
\end{equation}
In the case of zero form factor $F_{2,\, \alpha} = 0$,
the matrix element is simplified as
\begin{equation}
\begin{array}{lllll}
  \Bigl\langle \Psi_{E_{f}} (\vb{r}, \vb{q}) \Bigl|\,
    \hat{H}_{\gamma} (\vb{k}_{\gamma}, \varepsilon^{(\alpha)}) \Bigr|\,
    \Psi_{E_{i}} (\vb{r}, \vb{q}) \Bigr\rangle\; = 

  -\, i\,
  \displaystyle\frac{1}{2}\,
  \sqrt{\displaystyle\frac{2}{3}}\,
  \displaystyle\frac{e\, \hbar^{2}}{m_{N}c}
  \Bigl\{
    F_{1}\: \varepsilon_{\mu}\; \vb{I}_{1} \Bigl( -\displaystyle\frac{1}{\sqrt{6}} \Bigr) -
    2\, \varepsilon^{(\alpha)}\; \vb{I}_{1} \Bigl( \sqrt{\displaystyle\frac{2}{3}} \Bigr)
  \Bigr\}.
\end{array}
\label{eq.matrixelement.1.12}
\end{equation}

\subsection{Multipole expansion
\label{sec.multiple}}

In further calculation of Eq.~(\ref{eq.matrixelement.1.12}) it needs to find integrals~(\ref{eq.matrixelement.1.5}).
Applying the multipolar expansion, these integrals obtain form
[see App.~\ref{sec.app.integrals}, Eqs.~(\ref{eq.app.integrals.3}), (\ref{eq.app.integrals.6})]
\begin{equation}
\begin{array}{ll}
  \vspace{1mm}
  \vb{I}_{1} (\alpha) =
  \biggl< \varphi_{E_{f}} (\vb{q}) \biggl| \,
    e^{-i\alpha\mathbf{k_{\gamma}q}} \displaystyle\frac{\partial}{\partial \mathbf{q}}\,
    \biggr| \, \varphi_{E_{i}} (\vb{q}) \biggr>_\mathbf{q} =
  \sqrt{\displaystyle\frac{\pi}{2}}\:
  \displaystyle\sum\limits_{l_{\gamma}=1}\,
    (-i)^{l_{\gamma}}\, \sqrt{2l_{\gamma}+1}\;
  \displaystyle\sum\limits_{\mu = \pm 1}
    \xibf_{\mu}\, \mu\, \times
    \Bigl[ p_{l_{\gamma}\mu}^{M} (\alpha) - i\mu\: p_{l_{\gamma}\mu}^{E} (\alpha) \Bigr], \\

  I_{2} (\alpha) =
  \Bigl< \varphi_{E_{f}} (\vb{q}) \Bigl| \,  e^{-i\alpha \mathbf{k_{\gamma}q}} \, \Bigr| \, \varphi_{E_{i}} (\vb{q}) \Bigr>_\mathbf{q} =
  \sqrt{\displaystyle\frac{\pi}{2}}\:
  \displaystyle\sum\limits_{l_{\gamma}=1}\,
    (-i)^{l_{\gamma}}\, \sqrt{2l_{\gamma}+1}\;
  \displaystyle\sum\limits_{\mu = \pm 1}
    \Bigl[ \mu\,\tilde{p}_{l_{\gamma}\mu}^{M} (\alpha) - i\, \tilde{p}_{l_{\gamma}\mu}^{E} (\alpha) \Bigr],
\end{array}
\label{eq.multiple.1}
\end{equation}
where
[see Eqs.~(\ref{eq.app.integrals.4}), (\ref{eq.app.integrals.7})]
\begin{equation}
\begin{array}{lcl}
\vspace{3mm}
  p_{l_{\gamma}\mu}^{M} (\alpha) 
   & = &
    - I_{M}(0, l_{f}, l_{\gamma}, 1, \mu) \cdot J_{1}(\alpha, 0, l_{f},l_{\gamma}), \\

\vspace{1mm}
  p_{l_{\gamma}\mu}^{E} (\alpha) & = &
    \sqrt{\displaystyle\frac{l_{C}+1}{2l_{\gamma}+1}} \cdot I_{E} (0,l_{f},l_{\gamma}, 1, l_{\gamma}-1, \mu) \cdot J_{1}(\alpha, 0,l_{f},l_{\gamma}-1)\; - \\
  & - &
    \sqrt{\displaystyle\frac{l_{\gamma}}{2l_{\gamma}+1}} \cdot I_{E} (0,l_{f}, l_{\gamma}, 1, l_{\gamma}+1, \mu) \cdot J_{1}(\alpha, 0,l_{f},l_{\gamma}+1),
\end{array}
\label{eq.multiple.2}
\end{equation}
\begin{equation}
\begin{array}{llll}
\vspace{3mm}
  \tilde{p}_{l_{\gamma}\mu}^{M} (\alpha) =
    \tilde{I}\,(0,l_{f},l_{\gamma}, l_{\gamma}, \mu) \cdot \tilde{J}\, (\alpha, 0,l_{f},l_{\gamma}), \\

  \tilde{p}_{l_{\gamma}\mu}^{E} (\alpha) =
    \sqrt{\displaystyle\frac{l_{\gamma}+1}{2l_{\gamma}+1}} \tilde{I}\,(0,l_{f},l_{\gamma},l_{\gamma}-1,\mu) \cdot \tilde{J}\,(\alpha, 0,l_{f},l_{\gamma}-1)\; - 
    \sqrt{\displaystyle\frac{l_{\gamma}}{2l_{\gamma}+1}} \tilde{I}\,(0,l_{f},l_{\gamma},l_{\gamma}+1,\mu) \cdot \tilde{J}\,(\alpha, 0,l_{f},l_{\gamma}+1),
\end{array}
\label{eq.multiple.3}
\end{equation}
and
[see Eqs.~(\ref{eq.app.integrals.5}), (\ref{eq.app.integrals.8})]
\begin{equation}
\begin{array}{lllll}
  J_{1}(\alpha, l_{i},l_{f},n) & = &
  \displaystyle\int\limits^{+\infty}_{0} \displaystyle\frac{dR_{i}(r, l_{i})}{dr}\: R^{*}_{f}(l_{f},r)\, j_{n}(\alpha\, kr)\; r^{2} dr, \\

  \tilde{J}\,(\alpha, l_{i},l_{f},n) & = &
  \displaystyle\int\limits^{+\infty}_{0}
    R_{i}(r)\, R^{*}_{f}(l,r)\, j_{n}(\alpha k_{\gamma}r)\; r^{2} dr.
\end{array}
\label{eq.multiple.4}
\end{equation}
%
%
Here, $\mathbf{\xi}_{\mu}$ are \emph{vectors of circular polarization} with opposite directions of rotation
(see Ref.~\cite{Eisenberg.1973}, (2.39), p.~42).
Also we have the following properties
[see App.~\ref{sec.app.integrals}, Eqs.~(\ref{eq.app.integ.5}), (\ref{eq.app.integ.6})]
\begin{equation}
\begin{array}{llll}
\vspace{0.5mm}
  \displaystyle\sum\limits_{\alpha=1,2} \varepsilonbf^{(\alpha)} \cdot \vb{I}_{1} =
  \sqrt{\displaystyle\frac{\pi}{2}}\:
  \displaystyle\sum\limits_{l_{\gamma}=1}\, (-i)^{l_{\gamma}}\, \sqrt{2l_{\gamma}+1}\;
  \displaystyle\sum\limits_{\mu=\pm 1} \mu\,h_{\mu}\, \bigl(p_{l_{\gamma}, \mu}^{M} + p_{l_{\gamma}, -\mu}^{E} \bigr), \\
%
  (\varepsilonbf_{\rm x} + \varepsilonbf_{\rm z})\,  \displaystyle\sum\limits_{\alpha=1,2} \Bigl[ \vb{I}_{1} \times \varepsilonbf^{(\alpha)} \Bigr] =
  \sqrt{\displaystyle\frac{\pi}{2}}\:
  \displaystyle\sum\limits_{l_{\gamma}=1}\, (-i)^{l_{\gamma}}\, \sqrt{2l_{\gamma}+1}\;
  \displaystyle\sum\limits_{\mu=\pm 1} \mu\,h_{\mu}\, \bigl(p_{l_{\gamma}, \mu}^{M} - p_{l_{\gamma}, -\mu}^{E} \bigr),
\end{array}
\label{eq.multiple.5}
\end{equation}
where
[see App.~\ref{sec.app.integrals}, Eqs.~(\ref{eq.app.polarization.1.2})]
\begin{equation}
\begin{array}{llllll}
  h_{\pm} = \mp \displaystyle\frac{1 \pm i}{\sqrt{2}}, &
  h_{-} + h_{+} = -i\, \sqrt{2}, & 

  \displaystyle\sum\limits_{\mu=\pm 1} \mu\,h_{\mu} =
  - h_{-} + h_{+} =
  - \sqrt{2}, & 

  \displaystyle\sum\limits_{\alpha = 1,2} \varepsilonbf^{(\alpha),*} =
    h_{-1} \mathbf{\xi}_{-1}^{*} + h_{+1} \mathbf{\xi}_{+1}^{*}.
\end{array}
\label{eq.multiple.6}
\end{equation}

\subsection{Case of $l_{i}=0$, $l_{f}=1$, $l_{\gamma}=1$
\label{sec.simplecase}}

In a case of $l_{i}=0$, $l_{f}=1$, $l_{\gamma}=1$ integrals (\ref{eq.multiple.1}) are simplified to
[see App.~\ref{sec.app.integrals}, Eqs.~(\ref{eq.app.simplestcase.1})]
\begin{equation}
\begin{array}{llllll}
\vspace{1mm}
  \vb{I}_{1} =
  -i\, \sqrt{\displaystyle\frac{3\pi}{2}}\:
  \displaystyle\sum\limits_{\mu = \pm 1}
    \xibf_{\mu}\, \mu\, \times
    \Bigl[ p_{l_{\gamma}=1,\, \mu}^{M} - i\mu\: p_{l_{\gamma}=1,\, \mu}^{E} \Bigr], \\

  I_{2} =
  -i\, \sqrt{\displaystyle\frac{3\pi}{2}}\:
  \displaystyle\sum\limits_{\mu = \pm 1}
    \Bigl[ \mu\,\tilde{p}_{l_{\gamma}=1,\, \mu}^{M} - i\, \tilde{p}_{l_{\gamma}=1,\, \mu}^{E} \Bigr],



\end{array}
\label{eq.simplecase.1.1}
\end{equation}
where matrix elements are simplified to
[see details in App.~\ref{sec.app.integrals}, Eqs.~(\ref{eq.app.simplestcase.4})]
%
\begin{equation}
\begin{array}{lllllllll}
\vspace{2mm}
  p_{l_{\gamma}\mu}^{M} = 0, &
  p_{l_{\gamma}\mu}^{E} =
    \displaystyle\frac{1}{6} \sqrt{\displaystyle\frac{1}{\pi}} \cdot J_{1}(0,1,0) -
    \displaystyle\frac{47}{240} \sqrt{\displaystyle\frac{1}{2\pi}} \cdot J_{1}(0,1,2), \\

  \tilde{p}_{1 \mu}^{M} (c) = \displaystyle\frac{\mu}{2\sqrt{2\pi}} \cdot \tilde{J}\, (c, 0,1,1), &
  \tilde{p}_{1 \mu}^{E} (c) = 0.

\end{array}
\label{eq.simplecase.1.2}
\end{equation}
We substitute these solutions to Eq.~(\ref{eq.simplecase.1.1}) and obtain
[see App.~\ref{sec.app.integrals}, Eqs.~(\ref{eq.app.simplestcase.4}), (\ref{eq.app.simplestcase.5})]:
\begin{equation}
\begin{array}{llllll}
  \vb{I}_{1} & = &
  - \displaystyle\frac{1}{6} \cdot
  \sqrt{\displaystyle\frac{3}{2}}\:
  \displaystyle\sum\limits_{\mu = \pm 1}
    \xibf_{\mu} \cdot
    \Bigl(
      J_{1}(0,1,0) -
      \displaystyle\frac{47}{40} \sqrt{\displaystyle\frac{1}{2}} \cdot J_{1}(0,1,2)
    \Bigr).
\end{array}
\label{eq.simplecase.1.3}
\end{equation}
Integrals do not depend on vectors of polarization. So, we simplify further:
\begin{equation}
\begin{array}{llllll}
  \vb{I}_{1} & = &
  - \displaystyle\frac{1}{6} \cdot
  \sqrt{\displaystyle\frac{3}{2}}\:
  \Bigl(
    J_{1}(0,1,0) -
    \displaystyle\frac{47}{40} \sqrt{\displaystyle\frac{1}{2}} \cdot J_{1}(0,1,2)
  \Bigr) \cdot
  \bigl( \xibf_{\mu=+1} + \xibf_{\mu=-1} \bigr).
\end{array}
\label{eq.simplecase.1.4}
\end{equation}
Also from Eqs.~(\ref{eq.simplecase.1.1}) we find
\begin{equation}
\begin{array}{llllll}
  I_{2} (\alpha) & = &



  -i\, \displaystyle\frac{\sqrt{3}}{2}\: \tilde{J}\, (\alpha, 0,1,1).
\end{array}
\label{eq.simplecase.1.5}
\end{equation}


\subsection{Action on vectors of polarization
\label{sec.polarize}}

Now we calculate summation over vectors of polarization.
We use definition of vectors of polarizations
as in Eqs.~(57)--(58) in Ref.~\cite{Maydanyuk_Vasilevsky.2023.PRC}
(see App.~C in that paper for details):
%
\begin{equation}
\begin{array}{cc}
  \varepsilonbf^{(1)} = \displaystyle\frac{1}{\sqrt{2}}\, \bigl(\xibf_{-1} - \xibf_{+1}\bigr), &
  \varepsilonbf^{(2)} = \displaystyle\frac{i}{\sqrt{2}}\, \bigl(\xibf_{-1} + \xibf_{+1}\bigr),
\end{array}
\label{eq.polirize.2}
\end{equation}
and
\begin{equation}
\begin{array}{llll}
  \varepsilonbf^{(1)} \cdot \bigl( \xibf_{\mu=+1} + \xibf_{\mu=-1} \bigr) = 0, &
  \varepsilonbf^{(2)} \cdot \bigl( \xibf_{\mu=+1} + \xibf_{\mu=-1} \bigr) = -\, i\, \sqrt{2}.
\end{array}
\label{eq.polirize.4}
\end{equation}
On such a basis, from Eq.~(\ref{eq.simplecase.1.4}) we find:
\begin{equation}
\begin{array}{llllll}
  \varepsilonbf^{(1)} \cdot \vb{I}_{1} = 0, &

  \varepsilonbf^{(2)} \cdot \vb{I}_{1} (\alpha) =


  i\;
  \displaystyle\frac{\sqrt{3}}{6} \cdot
  \Bigl(
    J_{1}(\alpha, 0,1,0) -
    \displaystyle\frac{47}{40} \sqrt{\displaystyle\frac{1}{2}} \cdot J_{1}(\alpha, 0,1,2)
  \Bigr).
\end{array}
\label{eq.polirize.5}
\end{equation}
Now we can recalculate the matrix element (\ref{eq.matrixelement.1.12}) as
(at $\mu = 1$ it equals to zero)
\begin{equation}
\begin{array}{lllll}
\vspace{1.0mm}
  \Bigl\langle \Psi_{E_{f}} (\vb{r}, \vb{q}) \Bigl|\,
    \hat{H}_{\gamma} (\vb{k}_{\gamma}, \varepsilon^{(\alpha=2)}) \Bigr|\,
    \Psi_{E_{i}} (\vb{r}, \vb{q}) \Bigr\rangle\; & = & 

  \displaystyle\frac{\sqrt{2}\, e\, \hbar^{2}}{12\, m_{N}c}\,
  \Bigl\{
    F_{1}\,
    \Bigl[
      J_{1}\, \Bigl( -\displaystyle\frac{1}{\sqrt{6}}, 0,1,0 \Bigr) -
      \displaystyle\frac{47}{40} \sqrt{\displaystyle\frac{1}{2}} \cdot J_{1}\, \Bigl(-\displaystyle\frac{1}{\sqrt{6}},\, 0,1,2 \bigr)
    \Bigr]\; - \\

  & - &
    2\,
    \Bigl[
      J_{1}\, \Bigl(\sqrt{\displaystyle\frac{2}{3}}, 0,1,0 \Bigr) -
      \displaystyle\frac{47}{40} \sqrt{\displaystyle\frac{1}{2}} \cdot J_{1}\, \Bigl(\sqrt{\displaystyle\frac{2}{3}},\, 0,1,2 \bigr)
    \Bigr]
  \Bigr\},
\end{array}
\label{eq.polirize.7}
\end{equation}
where integrals are defined in Eqs.~(\ref{eq.multiple.4}).

\subsection{Resonating group method}

To study structure of two- and -three-cluster systems, we use the algebraic
version of the resonating group method, which was formulated in Refs.
\cite{kn:Fil_Okhr}, \cite{kn:Fil81}. Two main merits (advantages) of the algebraic version of the RGM are: (i) it
employ a full set of oscillator functions to expand wave functions of relative
motion of clusters and, thus, reduces the many-particle Schr\"{o}dinger
equation a set of linear algebraic equations, and (ii) it implements proper
boundary conditions for bound and continuous-spectrum states in discrete, oscillator space.

\subsection{Wave function of deuteron}

\begin{figure}[htbp]
\centerline{\includegraphics[width=120mm]{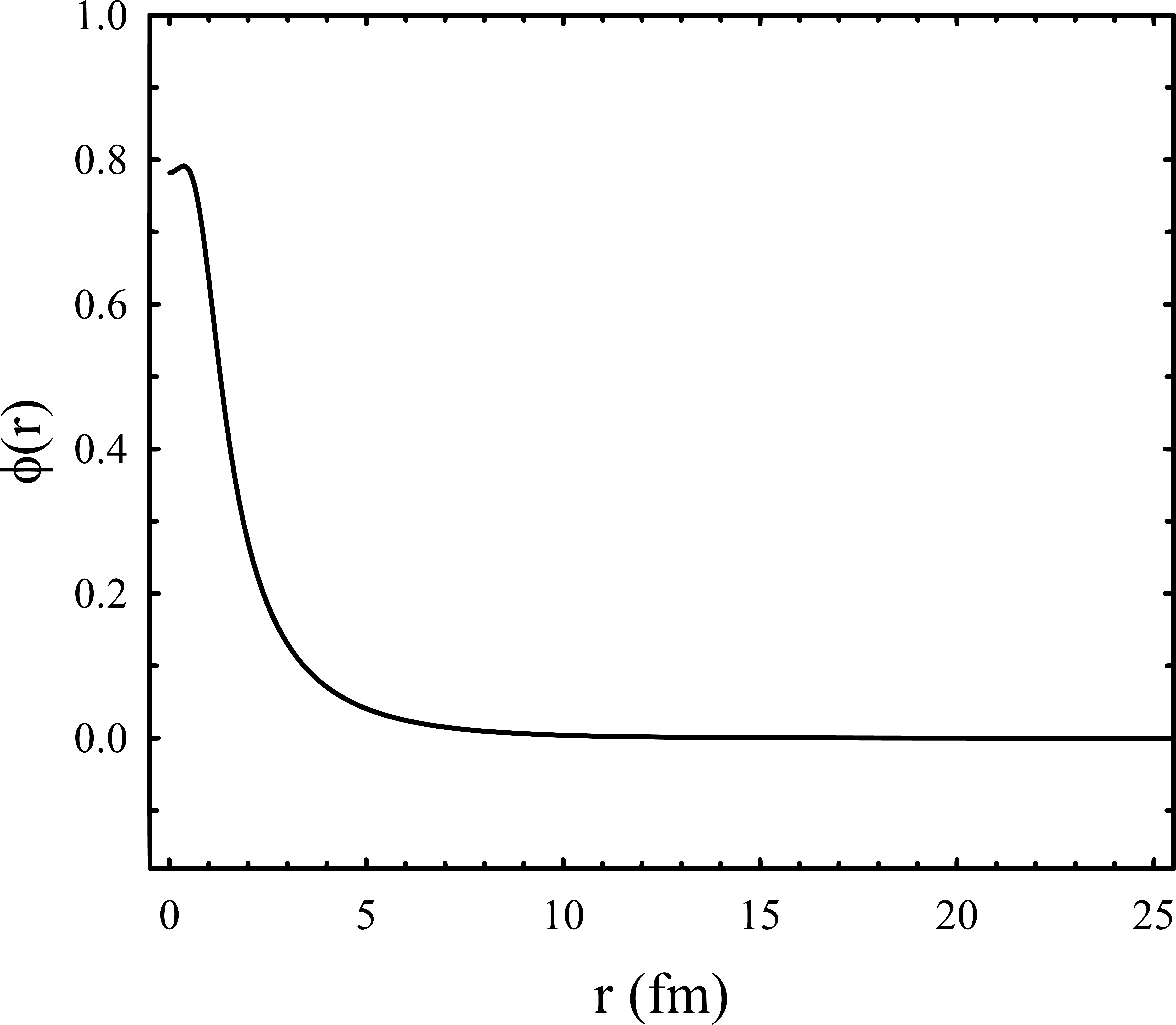}}
\vspace{-4mm}
\caption{\small (Color online)
Wave function of deuteron as a function of coordinate $r$
\label{Fig:WaveFunD}}
\end{figure}

Wave function of the bound state of deuteron was obtained with the Minnesota
NN potential \cite{kn:Minn_pot1}.
This potential creates the bound state at
$E_{d} = - 2.202$~MeV, which has to be compared with experimental value 
$E_{d} = - 2.225$~MeV.
Wave function of deuteron is shown in Fig.~\ref{Fig:WaveFunD}.



It has a long exponential tail
\[
\phi_{E_{d},L=0}\left(  \mathbf{r}\right)  \approx\exp\left\{  -\kappa
r\right\}  /r,
\]
where
\[
\kappa=\sqrt{\frac{2m\left\vert E_{d}\right\vert }{\hbar^{2}}}=\sqrt
{\frac{2\times2.202}{41.47}}=0.325879~\text{fm}^{-1}.
\]
Recall that the vector Jacobi $r$ is measured in fm.

It is interesting to note that the function (\ref{eq:R32}) is an exact
solution of two-body problem with the contact \ interaction%
\[
V(r)=V_{0}\delta\left(  r\right)  .
\]
This interaction is also called as the zero-range interaction \ and is widely
used in atomic and nuclear physics (for more details see Ref.
\cite{kn:DemkovE}). We will use the normalized to unity function
\begin{equation}
\phi\left(  r\right)  =\sqrt{2\kappa}\exp\left\{  -\kappa r\right\}
/r,\label{eq:R53A}%
\end{equation}
to approximate correct wave function of deuteron.

\begin{figure}[htbp]
\centerline{\includegraphics[width=90mm]{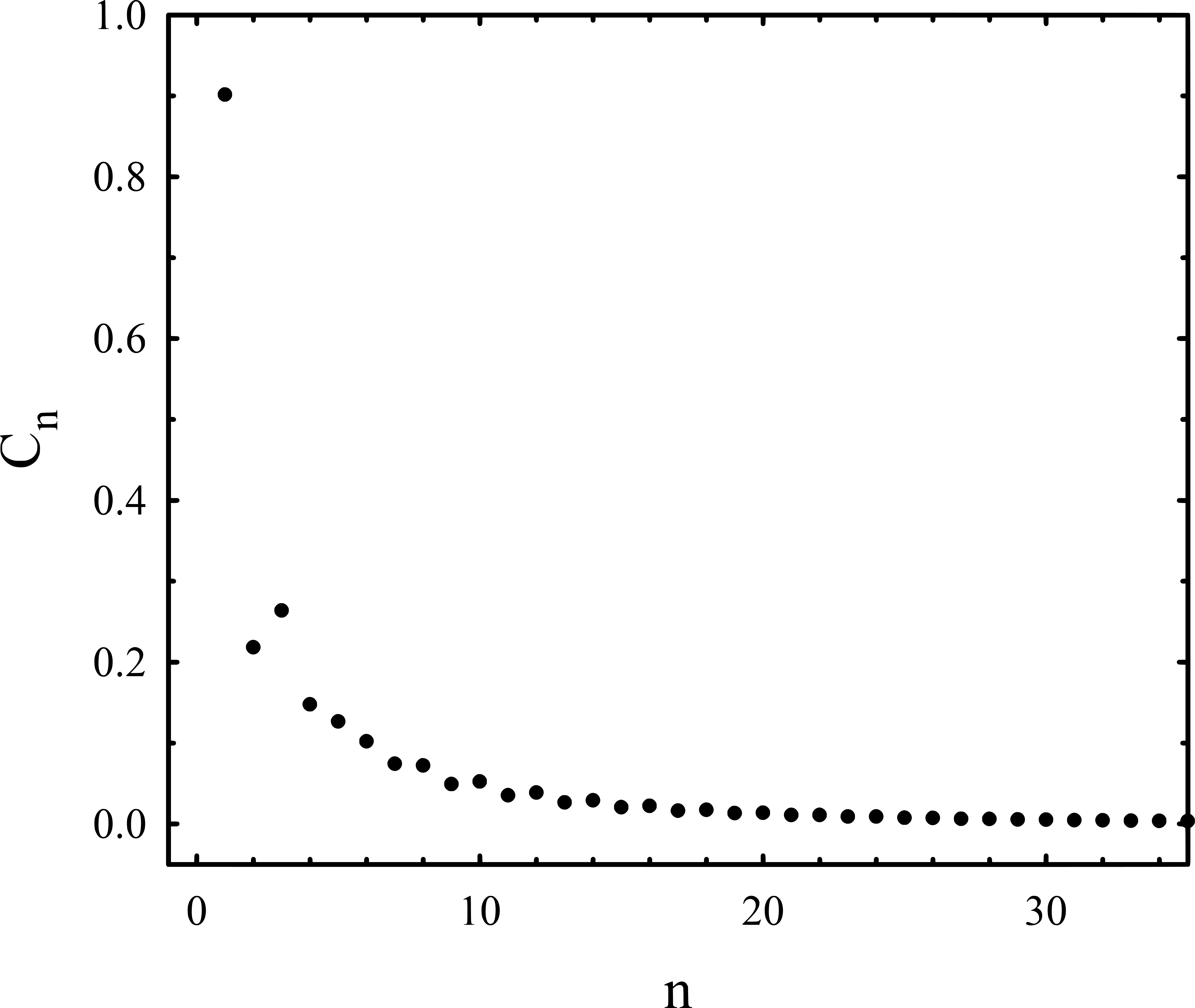}}
\vspace{-4mm}
\caption{\small (Color online)
 Wave function of deuteron in oscillator representation
 \label{Fig:WaveFunDOS}}
\end{figure}

To solve the Schr\"{o}dinger equations (\ref{eq:R22}) and (\ref{eq:R24}) for
deuteron and $p+d$ system, wave functions $\phi_{E_{d},l}\left(  r\right)  $
and $\psi_{E,L}\left(  q\right)  $ are expanded over basis of oscillator
functions%
\begin{eqnarray}
\phi_{E_{d},l}\left(  r\right)   &  =&\sum_{n=0}^{N\max}C_{n}^{\left(
E_{d},l\right)  }\Phi_{nl}\left(  r,b\right)  ,\label{eq:R35A}\\
\psi_{E,L}\left(  q\right)   &  =& \sum_{n=0}^{N\max}C_{n}^{\left(  E,L\right)
}\Phi_{nL}\left(  q,b\right)  , \label{eq:R35}%
\end{eqnarray}
%
where $\Phi_{n}\left(  r,b\right)  $ is an oscillator function%
\begin{equation}
\Phi_{nL}\left(  r,b\right)  =\left(  -1\right)  ^{n}N_{nL}~b^{-3/2}\rho
^{L}e^{-\frac{1}{2}\rho^{2}}L_{n}^{L+1/2}\left(  \rho^{2}\right)  ,\quad
\rho=\frac{r}{b} \label{eq:SP102a}%
\end{equation}
and $b$ is the oscillator length, and%
\[
N_{nL}=\sqrt{\frac{2\Gamma\left(  n+1\right)  }{\Gamma\left(  n+L+3/2\right)
}}.
\]
A set of expansion coefficients $\left\{  C_{n}\right\}  $ can be considered
as deuteron wave function in oscillator representation. 
In Fig.~\ref{Fig:WaveFunDOS} we show deuteron wave function in oscillator
representation.
This wave function was constructed with 200 oscillator
functions ($N\max=199$), however as one can see that only a small number of
basis functions (0$\leq n\leq$25) give a noticeable contribution.


In the shell-model approximation, the wave function of the deuteron bound state is
a Gaussian function
\begin{equation}
\phi\left(  r\right)  =\frac{1}{b^{3/2}}\exp\left\{ - \frac{1}{2}\left(
\frac{r}{b}\right)  ^{2}\right\}.
\label{eq:R51}
\end{equation}

Form factor of deuteron is then
\begin{equation}
  \left\langle \phi\left(  \mathbf{r}\right)  \left\vert
  \exp\left\{ -\frac{i}{\sqrt{2}}\left(  \mathbf{k}_{\gamma}\mathbf{r}\right)  \right\}
  \right\vert
  \phi\left(  \mathbf{r}\right)  \right\rangle =
  \exp\left\{ - \frac{1}{8}\left(  k_{\gamma}b\right)  ^{2}\right\}.
\label{eq:R52}
\end{equation}

If we take the deuteron wave function in the form
\begin{equation}
\phi\left(  r\right)  =\sqrt{2\kappa}\exp\left\{  -\kappa r\right\}  /r,
\label{eq:R53}%
\end{equation}
then we obtain deuteron form factor
\begin{equation}
\left\langle \phi\left(  \mathbf{r}\right)  \left\vert \exp\left\{  -\frac
{i}{\sqrt{2}}\left(  \mathbf{k}_{\gamma}\mathbf{r}\right)  \right\}
\right\vert \phi\left(  \mathbf{r}\right)  \right\rangle =\frac{2\sqrt
{2}\kappa}{k_{\gamma}}\operatorname{arctan}\left(  \frac{k_{\gamma}
}{2\sqrt{2}\kappa}\right)  . \label{eq:R54}%
\end{equation}

Form factor from Eq. (\ref{eq:R54}) as function of $k_{\gamma}$ demonstrates
slower decreasing comparing to the form factor of the shell mode (\ref{eq:R54})



\section{Matrix elements in the folding approximation
\label{sec.folding.1}}

Matrix element of bremsstrahlung emission of photons for two $s$-clusters
(i.e., for clusters with $1 \le A_{\alpha} \le 4$ or
for $n$, $p$, $d$, $^{3}{\rm H}$, $^{3}{\rm He}$, $^{4}{\rm He}$) can be written down as
[see Ref.~\cite{Maydanyuk_Vasilevsky.2023.PRC}, for details]
\begin{equation}
\begin{array}{lllll}
\vspace{1.0mm}
  & \Bigl\langle \Psi_{E_{f} l_{f}} \Bigl| \hat{H}_{\gamma} (\vb{k}_{\gamma}, \varepsilonbf^{(\alpha)}) \Bigr| \Psi_{E_{i} l_{i}} \Bigr\rangle_{\rm fold} = \\

\vspace{1.0mm}
  = &
  \displaystyle\frac{e\hbar}{m_{N}c}
  \biggl\{
  \sqrt{\displaystyle\frac{A_{2}}{A_{1}\, A}}
  \biggl\langle
    R_{E_{f} l_{f}} (r)\, Y_{l_{f} m_{f}} (\widehat{\mathbf{r}}_{i})
    \biggl|
      \exp{-i\, \sqrt{\displaystyle\frac{A_{2}}{A_{1}\, A}}\, (\vb{k}_{\gamma}, \mathbf{r})}\,
      (\varepsilonbf^{(\alpha)}, \hat{\pibf})
    \biggr|
    R_{E_{i} l_{i}} (r)\, Y_{l_{i} m_{i}} (\widehat{\mathbf{r}}_{i})
  \biggr\rangle\, F_{1} - \\

  - &
  \sqrt{\displaystyle\frac{A_{1}}{A_{2}\, A}}
  \biggl\langle
    R_{E_{f} l_{f}} (r)\, Y_{l_{f} m_{f}} (\widehat{\mathbf{r}}_{i})
    \biggl|
      \exp{i\, \sqrt{\displaystyle\frac{A_{1}}{A_{2}\, A}}\, (\vb{k}_{\gamma}, \vb{r})}\,
      (\varepsilonbf^{(\alpha)}, \hat{\pibf})
    \biggr|
    R_{E_{i} l_{i}} (r)\, Y_{l_{i} m_{i}} (\widehat{\mathbf{r}}_{i})
  \biggr\rangle\, F_{2}
  \biggr\}.
\end{array}
\label{eq.folding.1.1}
\end{equation}
In the standard approximation of resonating group method, form factor $F_{n}$ equals ($n=1,2$)
\begin{equation}
\begin{array}{lllll}
  F_{n} = &
  \Bigl\langle \Phi_{n} (A_{n}) \Bigl| F_{0}^{(n)} \Bigr| \Phi_{n}  (A_{n}) \Bigr\rangle =
  Z_{n}\, \exp{ - \displaystyle\frac{1}{4}\, \displaystyle\frac{A_{n} - 1}{A_{n}}\, (k, b)^{2}},
\end{array}
\label{eq.folding.1.2}
\end{equation}
with $b$ is oscillator length.
Using property (\ref{eq.matrixelement.1.9}):
\begin{equation}
\begin{array}{lllll}
  \Bigl\langle \psi_{E_{f}} (\vb{q}) \Bigl|\,
      \exp\Bigl\{ \pm i\, \alpha'\, (\vb{k}_{\gamma} \vb{q}) \Bigr\}\,
      (\varepsilon^{(\alpha)}, \vb{\pi}_{\mathbf{q}}^{*})\,
    \Bigr|\,
    \psi_{E_{i}} (\vb{q})
  \Bigr\rangle =

  i \hbar\, \varepsilon^{(\alpha)}\, \vb{I}_{1} (\mp \alpha'),
\end{array}
\label{eq.folding.1.3}
\end{equation}
matrix element is rewritten as
\begin{equation}
\begin{array}{lllll}
\vspace{1.0mm}
  \Bigl\langle \Psi_{E_{f} l_{f}} \Bigl| \hat{H}_{\gamma} (\vb{k}_{\gamma}, \varepsilon^{(\alpha)}) \Bigr| \Psi_{E_{i} l_{i}} \Bigr\rangle_{\rm fold} & = &

%
%

  i\, \displaystyle\frac{e\hbar^{2}}{m_{N}c}\,
  \varepsilon^{(\alpha)}\,
  \biggl\{
    \sqrt{\displaystyle\frac{A_{2}}{A_{1}\, A}}
    \vb{I}_{1} \Bigl(\sqrt{\displaystyle\frac{A_{2}}{A_{1}\, A}} \Bigr)\, F_{1} -
    \sqrt{\displaystyle\frac{A_{1}}{A_{2}\, A}}
    \vb{I}_{1} \Bigl(- \sqrt{\displaystyle\frac{A_{1}}{A_{2}\, A}} \Bigr)\, F_{2}
  \biggr\}.
\end{array}
\label{eq.folding.1.4}
\end{equation}
Now we take into account property (\ref{eq.polirize.5})
\[ 
\begin{array}{llllll}
  \varepsilonbf^{(1)} \cdot \vb{I}_{1} = 0, &

  \varepsilonbf^{(2)} \cdot \vb{I}_{1} (\alpha) =
  i\: \displaystyle\frac{\sqrt{3}}{6} \cdot
  \Bigl(
    J_{1}(\alpha, 0,1,0) -
    \displaystyle\frac{47}{40} \sqrt{\displaystyle\frac{1}{2}} \cdot J_{1}(\alpha, 0,1,2)
  \Bigr),
\end{array}
\] 
and obtain
\begin{equation}
\begin{array}{lllll}
\vspace{1.5mm}
  \Bigl\langle \Psi_{E_{f} l_{f}} \Bigl| \hat{H}_{\gamma} (\vb{k}_{\gamma}, \varepsilon^{(\alpha)}) \Bigr| \Psi_{E_{i} l_{i}} \Bigr\rangle_{\rm fold} & = &



  \displaystyle\frac{\sqrt{3}\, e\hbar^{2}}{6\, m_{N}c}\,
  \Bigl\{
    \sqrt{\displaystyle\frac{A_{2}}{A_{1}\, A}}
    \Bigl[
      J_{1}\, \Bigl(\sqrt{\displaystyle\frac{A_{2}}{A_{1}\, A}},\, 0,1,0 \Bigr) -
      \displaystyle\frac{47}{40} \sqrt{\displaystyle\frac{1}{2}}\,
        J_{1}\, \Bigl(\sqrt{\displaystyle\frac{A_{2}}{A_{1}\, A}},\, 0,1,2 \Bigr)
    \Bigr]\, F_{1}\; - \\

  & - &
  \sqrt{\displaystyle\frac{A_{1}}{A_{2}\, A}}
    \Bigl[
      J_{1}\, \Bigl(- \sqrt{\displaystyle\frac{A_{1}}{A_{2}\, A}},\, 0,1,0 \Bigr) -
      \displaystyle\frac{47}{40} \sqrt{\displaystyle\frac{1}{2}}\,
        J_{1}\, \Bigl(- \sqrt{\displaystyle\frac{A_{1}}{A_{2}\, A}},\, 0,1,2 \Bigr)
    \Bigr]\, F_{2}
  \Bigr\}.
\end{array}
\label{eq.folding.1.5}
\end{equation}
In particular, for proton-deuteron scattering we have (we choose
the first index --- for proton: $A_{1} = 1$, $F_{1} = F_{\rm p}$,
the second index --- for deuteron: $A_{2} = 2$, $F_{2} = F_{\rm D}$):
\begin{equation}
\begin{array}{lllll}
\vspace{1.5mm}
  \Bigl\langle \Psi_{E_{f} l_{f}} \Bigl| \hat{H}_{\gamma} (\vb{k}_{\gamma}, \varepsilon^{(\alpha)}) \Bigr| \Psi_{E_{i} l_{i}} \Bigr\rangle_{\rm fold} & = &

  \displaystyle\frac{\sqrt{3}\, e\hbar^{2}}{6\, m_{N}c}\,
  \Bigl\{
    \sqrt{\displaystyle\frac{2}{3}}
    \Bigl[
      J_{1}\, \Bigl(\sqrt{\displaystyle\frac{2}{3}},\, 0,1,0 \Bigr) -
      \displaystyle\frac{47}{40} \sqrt{\displaystyle\frac{1}{2}}\, J_{1}\, \Bigl(\sqrt{\displaystyle\frac{2}{3}},\, 0,1,2 \Bigr)
    \Bigr]\, F_{\rm p}\; - \\

  & - &
  \sqrt{\displaystyle\frac{1}{6}}
    \Bigl[
      J_{1}\, \Bigl(- \sqrt{\displaystyle\frac{1}{6}},\, 0,1,0 \Bigr) -
      \displaystyle\frac{47}{40} \sqrt{\displaystyle\frac{1}{2}}\, J_{1}\, \Bigl(- \sqrt{\displaystyle\frac{1}{6}},\, 0,1,2 \Bigr)
    \Bigr]\, F_{\rm D}
  \Bigr\}.
\end{array}
\label{eq.folding.1.6}
\end{equation}
For further analysis it is more convenient to rewrite this solution as
\begin{equation}
\begin{array}{lllll}
\vspace{1.5mm}
  \Bigl\langle \Psi_{E_{f} l_{f}} \Bigl| \hat{H}_{\gamma} (\vb{k}_{\gamma}, \varepsilon^{(\alpha)}) \Bigr| \Psi_{E_{i} l_{i}} \Bigr\rangle_{\rm fold} & = &



  \displaystyle\frac{\sqrt{2}\, e\hbar^{2}}{12\, m_{N}c}\,
  \Bigl\{
    \Bigl[
      J_{1}\, \Bigl(- \sqrt{\displaystyle\frac{1}{6}},\, 0,1,0 \Bigr) -
      \displaystyle\frac{47}{40} \sqrt{\displaystyle\frac{1}{2}}\, J_{1}\, \Bigl(- \sqrt{\displaystyle\frac{1}{6}},\, 0,1,2 \Bigr)
    \Bigr]\, F_{\rm D}\: - \\

  & - &
    2\,
    \Bigl[
      J_{1}\, \Bigl(\sqrt{\displaystyle\frac{2}{3}},\, 0,1,0 \Bigr) -
      \displaystyle\frac{47}{40} \sqrt{\displaystyle\frac{1}{2}}\, J_{1}\, \Bigl(\sqrt{\displaystyle\frac{2}{3}},\, 0,1,2 \Bigr)
    \Bigr]\, F_{\rm p}
  \Bigr\}.
\end{array}
\label{eq.folding.1.7}
\end{equation}

\section{Definition of cross section of bremsstrahlung emission of photons and resulting formulas
\label{sec.crosssection}}

Cross-section of bremsstrahlung emission of photons is~\cite{Maydanyuk_Vasilevsky.2023.PRC} 
\begin{equation}
\begin{array}{lllll}
\vspace{1.0mm}
  \displaystyle\frac{d\, \sigma^{(1)}}{d\Omega_{A_1}\, d\Omega_{A_2}\, d\Omega_{\gamma}} & = &
  \displaystyle\frac{E_{\gamma}}{(2\pi\hbar)^{4}}\,
  \Bigl( \displaystyle\frac{p_{1f}}{\hbar c} \Bigr)\,
    \displaystyle\frac{\sin^{2} \theta_{1}\, \sin^{2} \theta_{2} }{sin^{5}  (\theta_{1} + \theta_{2})}\; \times 

  \displaystyle\frac{1}{2J + 1}
  \displaystyle\sum\limits_{\mu M_{i}}
    \Bigl|
      \Bigl\langle \Psi_{\tilde{E} \tilde{L}} \Bigl| \hat{H}_{\gamma} (\vb{k}_{\gamma}, \varepsilon^{(\alpha)}) \Bigr| \Psi_{EL} \Bigr\rangle
    \Bigr|^{2},
\end{array}
\label{eq.crosssection.1.1}
\end{equation}
%
%
%
where $p_{1}$ is the momentum of the incident nucleus (cluster) $A_{1}$, $\theta_{1}$ and $\theta_{2}$ are scattering angels of the first and second clusters in laboratory frame.

We write down final formulas of matrix elements of bremsstrahlung emission in the proton-deuteron scattering.
It turn out that in first approach
[we will call it as \emph{the three-cluster model},
see Eq.~(\ref{eq.matrixelement.1.12})] and 
in the second approach [we will call it as \emph{the folding model}, see Eq.~(\ref{eq.folding.1.7})]
matrix elements are the same:
\begin{equation}
\begin{array}{lllll}
\vspace{1.0mm}
  \Bigl\langle \Psi_{E_{f}} (\vb{r}, \vb{q}) \Bigl|\,
    \hat{H}_{\gamma} (\vb{k}_{\gamma}, \varepsilon^{(\alpha=2)}) \Bigr|\,
    \Psi_{E_{i}} (\vb{r}, \vb{q}) \Bigr\rangle\; & = & 

  \displaystyle\frac{\sqrt{2}\, e\, \hbar^{2}}{12\, m_{N}c}\,
  \Bigl\{
    \Bigl[
      J_{1}\, \Bigl( -\displaystyle\frac{1}{\sqrt{6}}, 0,1,0 \Bigr) -
      \displaystyle\frac{47}{40} \sqrt{\displaystyle\frac{1}{2}} \cdot J_{1}\, \Bigl(-\displaystyle\frac{1}{\sqrt{6}},\, 0,1,2 \bigr)
    \Bigr]\; F_{1}\, - \\

  & - &
    2\,
    \Bigl[
      J_{1}\, \Bigl(\sqrt{\displaystyle\frac{2}{3}}, 0,1,0 \Bigr) -
      \displaystyle\frac{47}{40} \sqrt{\displaystyle\frac{1}{2}} \cdot J_{1}\, \Bigl(\sqrt{\displaystyle\frac{2}{3}},\, 0,1,2 \bigr)
    \Bigr]
  \Bigr\}.
\end{array}
\label{eq.crosssection.1.3}
\end{equation}
%
%
%
%
%
Integrals are [see Eqs.~(\ref{eq.multiple.4})]
\begin{equation}
\begin{array}{lllll}
  J_{1}(\alpha, l_{i},l_{f},n) & = &
  \displaystyle\int\limits^{+\infty}_{0} \displaystyle\frac{dR_{i}(r, l_{i})}{dr}\: R^{*}_{f}(l_{f},r)\, j_{n}(\alpha\, k_{\gamma} r)\; r^{2} dr.

\end{array}
\label{eq.crosssection.1.5}
\end{equation}

\section{Analysis, numerical calculations
\label{sec.analysis}}

To understand the role of the deuteron structure, we are going to perform three
(four) types of calculations. They are distinguished by the wave function of
the deuteron and are labeled by indexes $0$, ($C$), $S$, and $R$. The index
$0$ means that the deuteron is considered as a structureless particle and,
thus, its internal structure is ignored. (If the wave function of deuteron is
approximated by the case of the contact interaction (\ref{eq:R53A}), we will use the
index C). The index $S$ stands for the shell-model approximation (\ref{eq:R51}) of the
wave function of deuteron,  and the last case $R$  means that the realistic
wave function (\ref{eq:R35}) of deuteron is involved in calculations.

\subsection{Deuteron wave functions and form factor}

\begin{figure}
[ptbh]
\begin{center}
\includegraphics[
height=11.7959cm,
width=13.4543cm
]%
{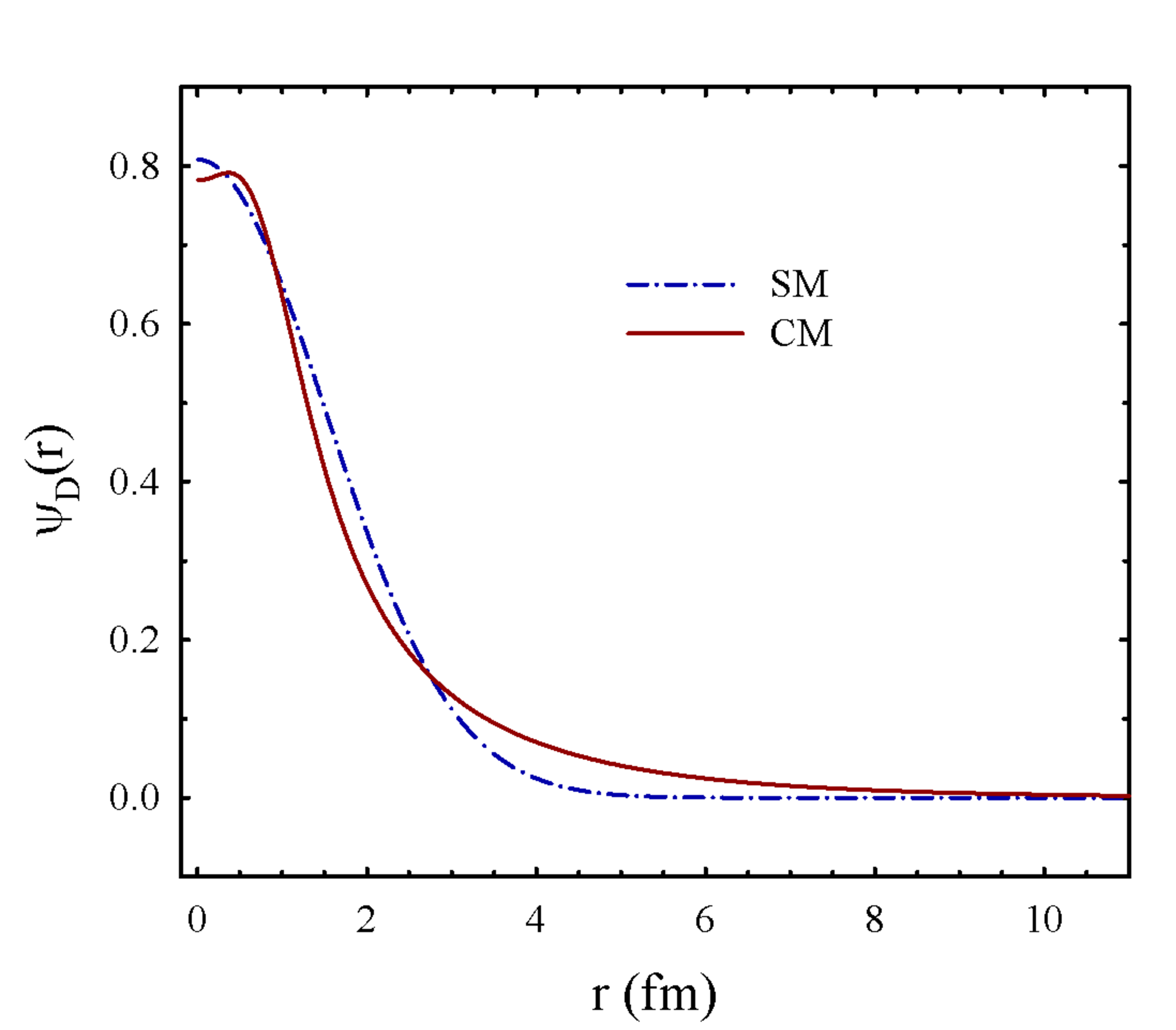}%
\caption{Wave functions of deuteron obtained in the shell model (SM) and
cluster model (CM).}%
\label{Fig:WaveFunsD2}%
\end{center}
\end{figure}

\begin{figure}
[ptbh]
\begin{center}
\includegraphics[
height=11.9584cm,
width=13.3577cm
]%
{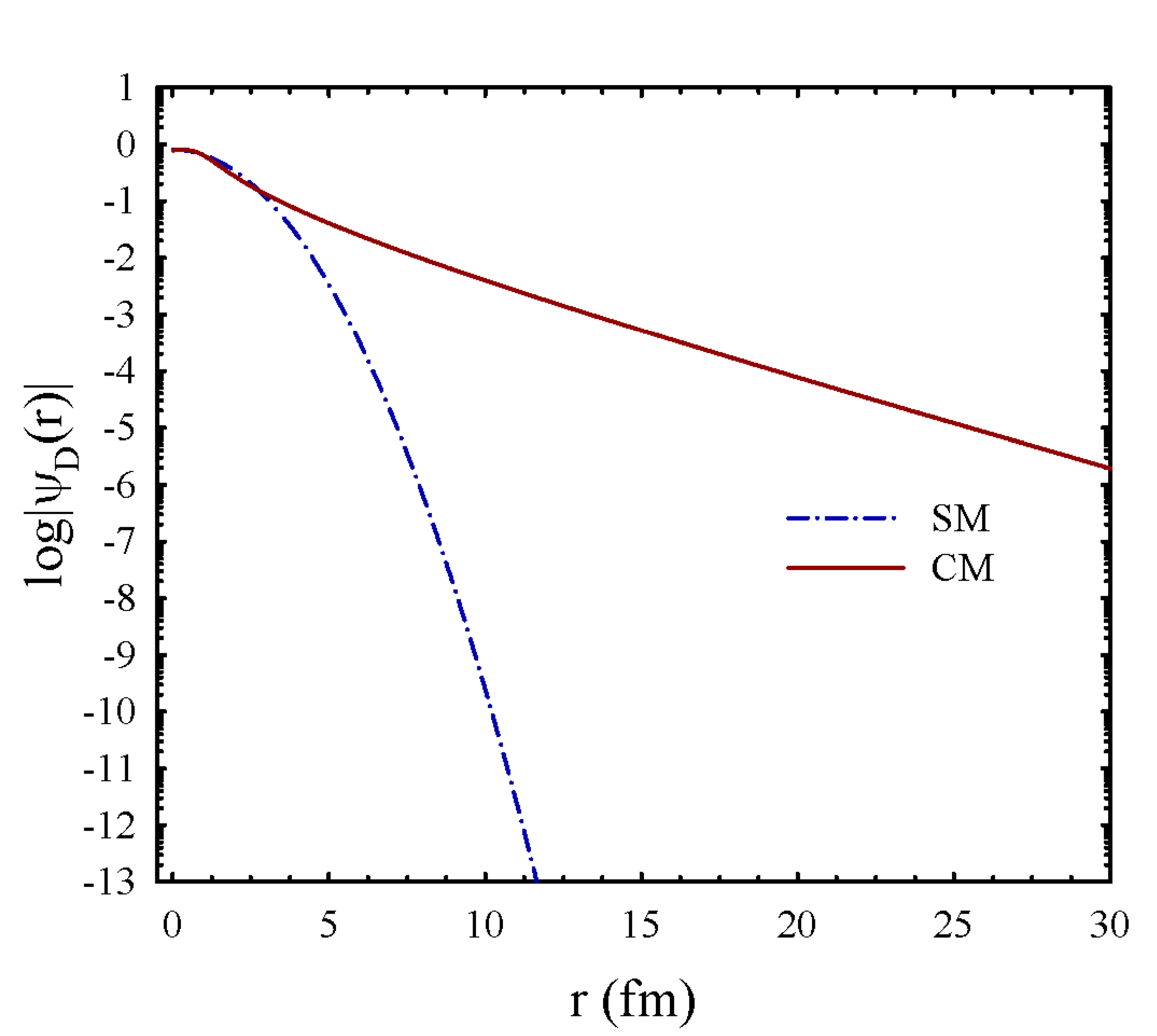}%
\caption{Asymptotic form of the deuteron wave functions constructed in the
shell model (SM) and cluster model (CM).}%
\label{Fig:WaveFunsD2L}%
\end{center}
\end{figure}

The key element of our model is a wave function of the deuteron bound state.
This function will determine the interaction of proton and deuteron. Thus we
start our analysis from deuteron wave functions in different model and
approximations. In Figs. \ref{Fig:WaveFunsD2}\ and \ref{Fig:WaveFunsD2L} we
display wave functions of deuteron obtained with the Minnesota potential. The
shell model (SM) and cluster model (CM) visually are very similar. However,
displaying wave functions in a logarithmic scale, we see that they have quite
different asymptotic behavior.%
%

\begin{figure}
[ptb]
\begin{center}
\includegraphics[
height=12.1342cm,
width=13.395cm
]%
{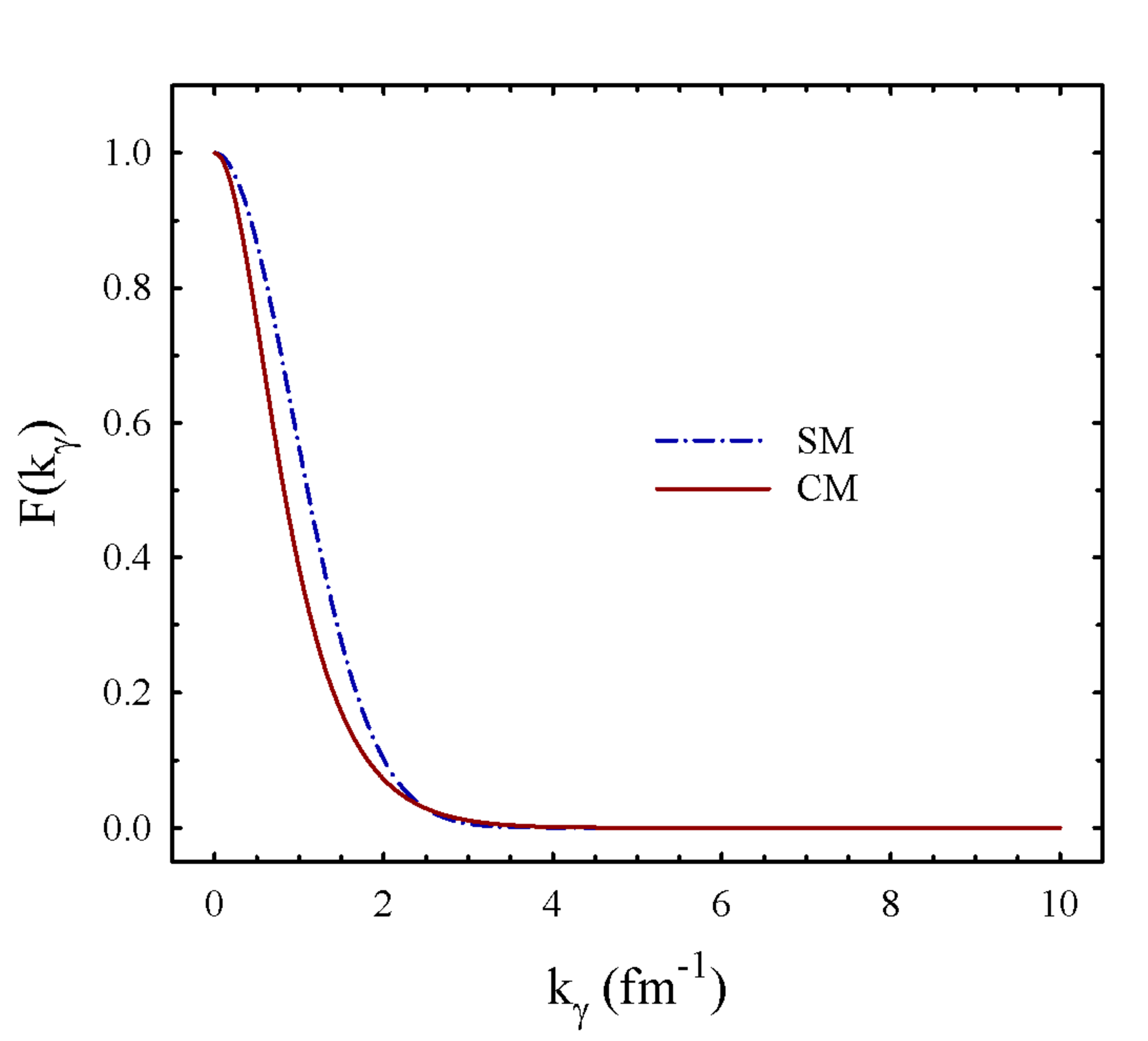}%
\caption{Form factors of deuteron determined in the shell and cluster
models}%
\label{Fig:FormFactD2}%
\end{center}
\end{figure}

\begin{figure}
[ptb]
\begin{center}
\includegraphics[
height=11.9584cm,
width=13.3577cm
]%
{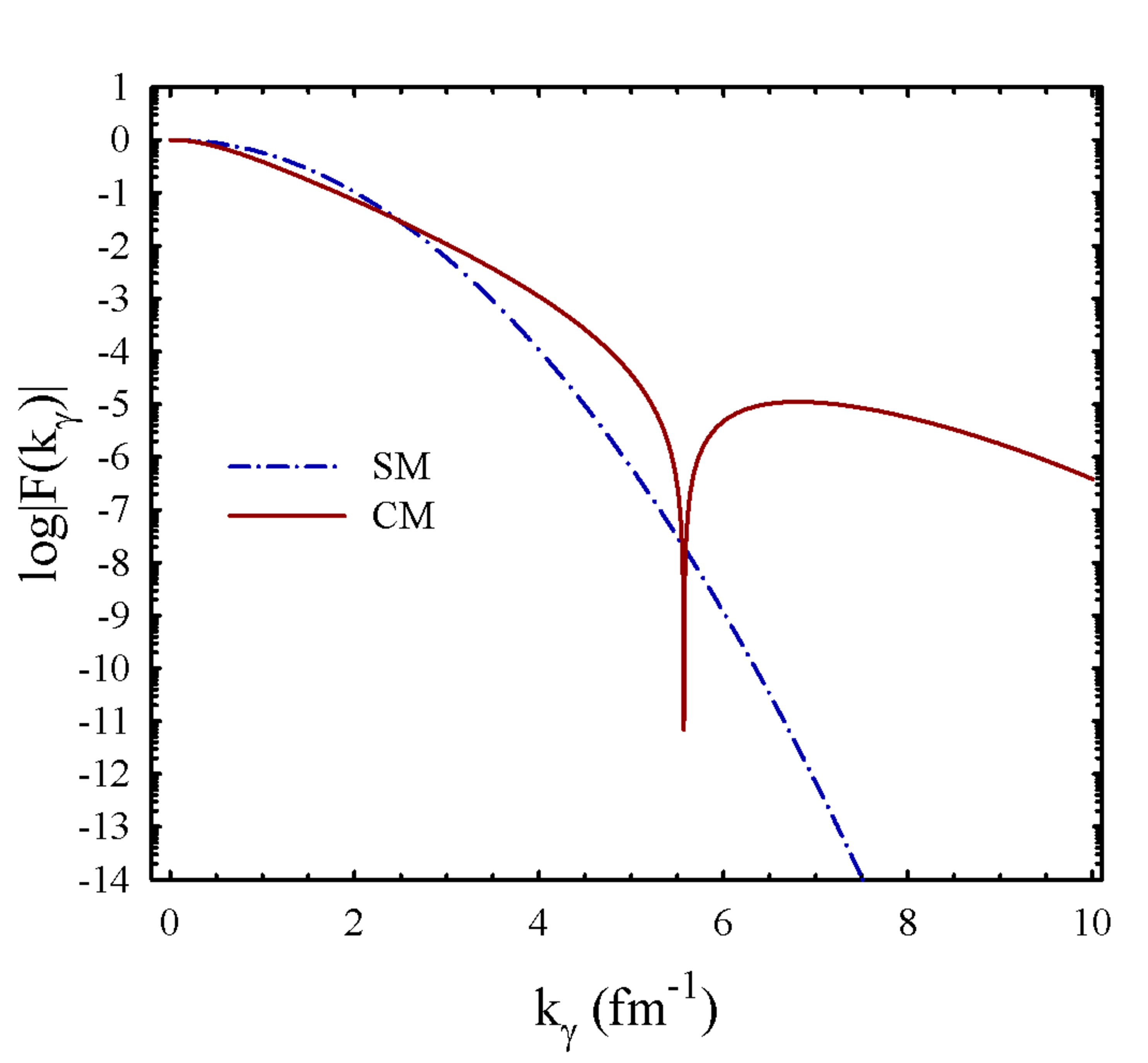}%
\caption{Form factors of deuteron in logarithmic scale calculated in the shell
and cluster models}%
\label{Fig:FormFactD2L}%
\end{center}
\end{figure}

Deuteron form factors calculated within the shell model  and cluster model
are shown in Figs. \ref{Fig:FormFactD2} and \ref{Fig:FormFactD2L}.

%

\begin{figure}[htbp]
\centerline{\includegraphics[width=82mm]{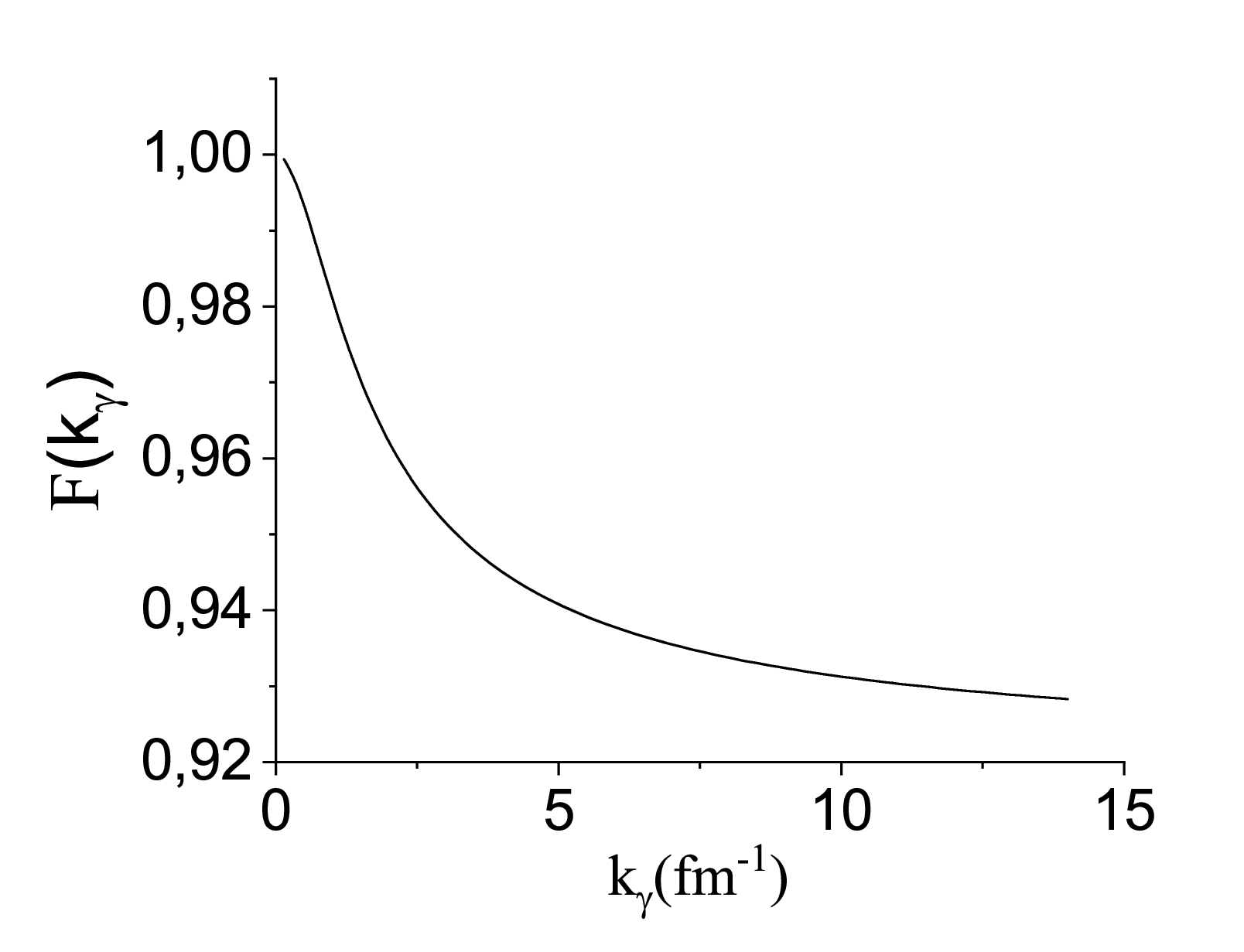}
\hspace{-1mm}\includegraphics[width=82mm]{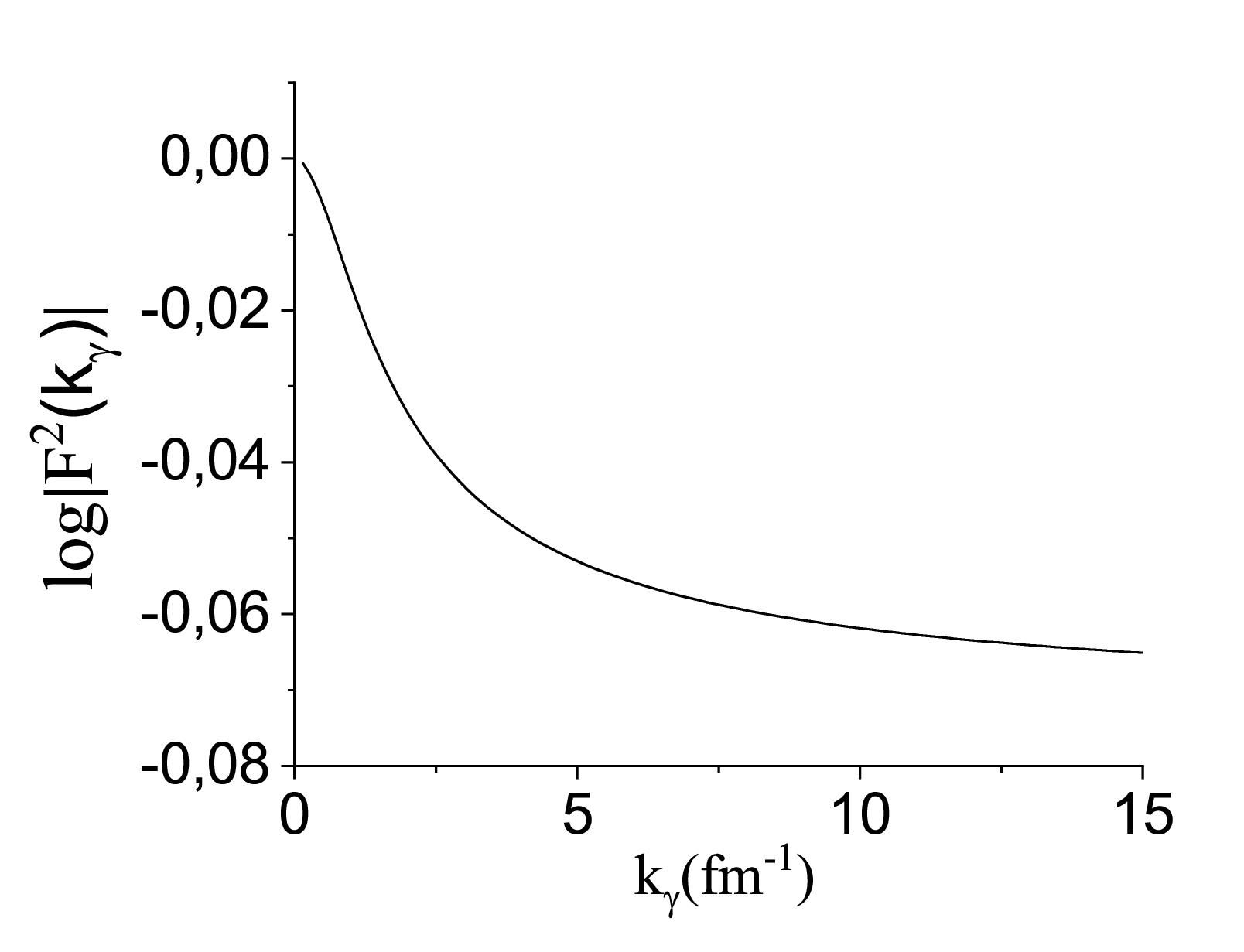}}
\vspace{-2mm}
\caption{\small
Form factor $F_1$ (in left panel) and square of this form factor $F_1^2$ (in right panel) at $\kappa = 0.325879$ fm\textsuperscript{-1}.}
\label{fig.1}
\end{figure}

If the zero-range interaction is used to determine wave function of deuteron (\ref{eq:R53A}), then the deuteron form factor is
(see App.~\ref{sec.app.clusterformfactor})
\begin{equation}
  F_1 (\pmb{k}_\gamma) =  
  \dfrac{2\sqrt{2}\kappa}{ k_\gamma}
  \arctan \biggl(\dfrac{k_\gamma}{2\sqrt{2}\kappa}\biggr).
\label{formfactor_new}
\end{equation}
This form factor is shown in Figs.~\ref{fig.1}.

\subsection{Wave functions of $p+d$ system}

\begin{figure}
[ptb]
\begin{center}
\includegraphics[
height=11.6355cm,
width=13.3752cm
]%
{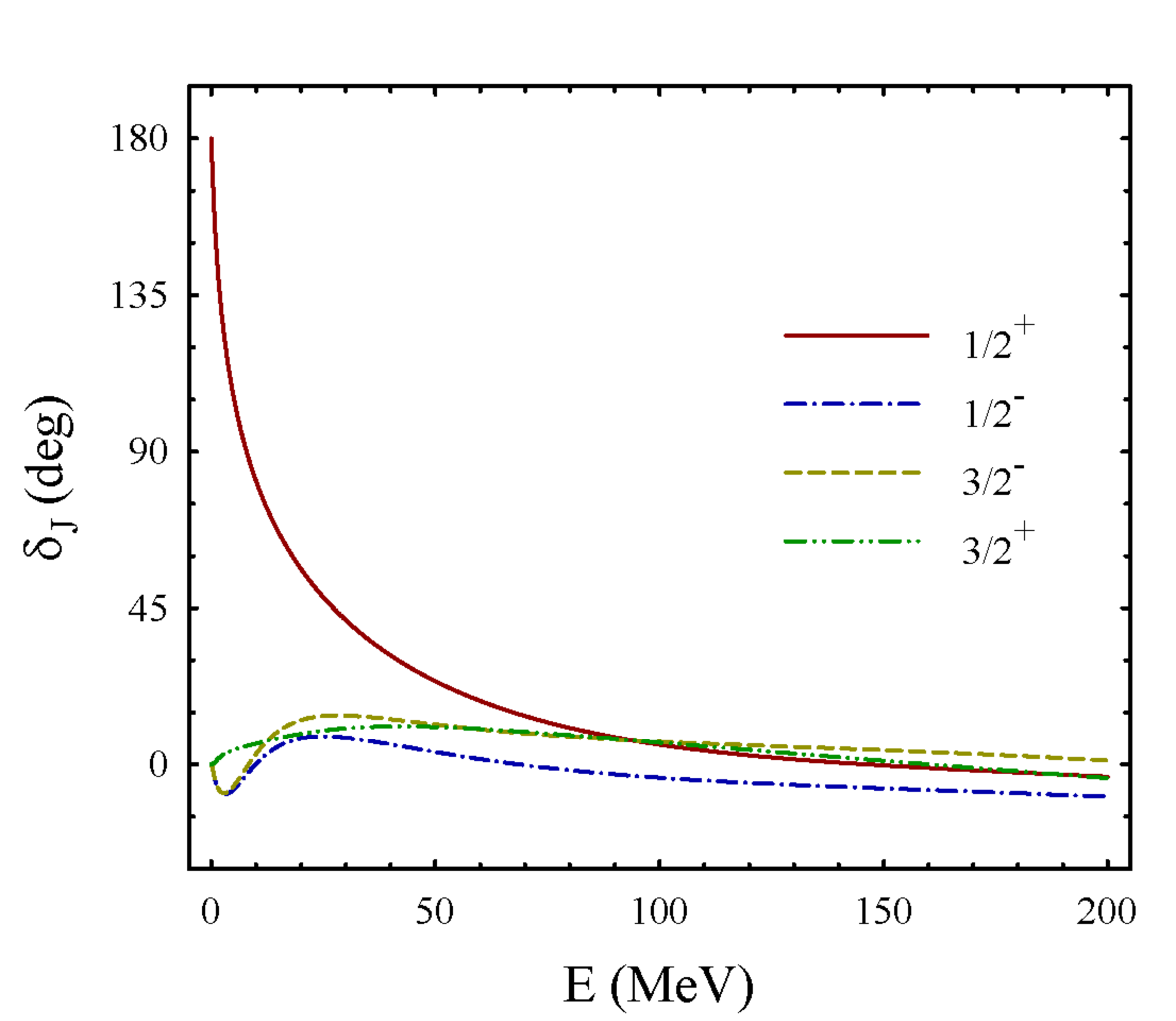}%
\caption{Phase shifts of the elastic $p+d$ scattering, obtained with the MP for different $J^{\pi}$ states  }%
\label{Fig:PhasesPD}%
\end{center}
\end{figure}

In Fig. \ref{Fig:PhasesPD} we display phase shifts of the elastic $p+d$
scattering. One can see that the strongest interaction is observed in the
1/2$^{+}$ state, where the nucleus $^{3}$He has a bound state. For energy $E>$%
100 MeV, all displayed phase shifts are very close to zero. This is an
additional indication that potential of the $p+d$ interaction is weak and that
the Born approximation can be used for this energy range.%

\begin{figure}
[ptb]
\begin{center}
\includegraphics[
height=11.7959cm,
width=13.395cm
]%
{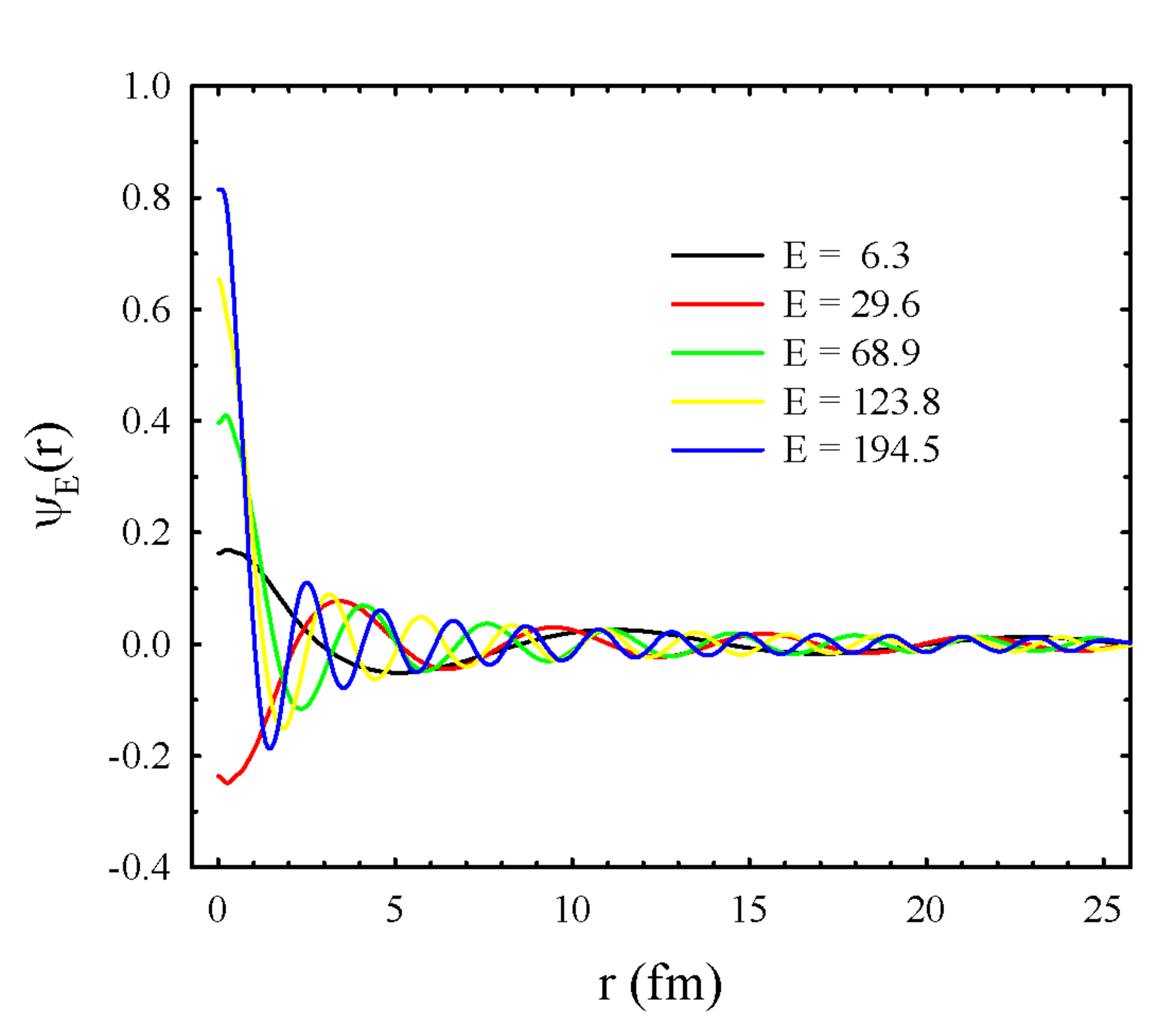}%
\caption{Wave functions of continuous spectrum states \ in the 1/2$^{+}$ state
of $p+d$ system}%
\label{Fig:WaveFuns3He12P}%
\end{center}
\end{figure}

\begin{figure}
[ptb]
\begin{center}
\includegraphics[
height=11.7959cm,
width=13.395cm
]%
{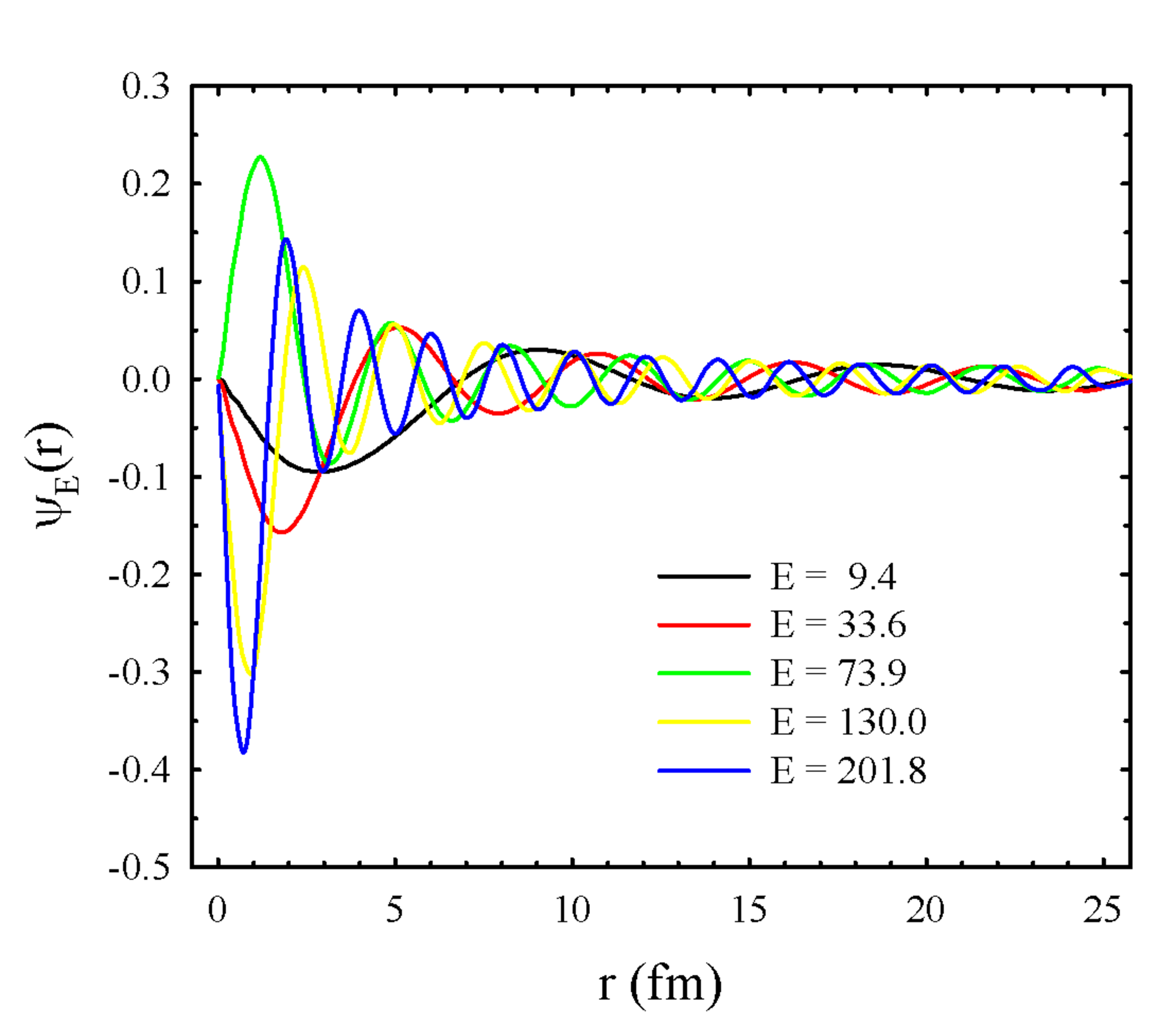}%
\caption{Wave functions of relative motion of proton with respect to deuteron
in the 1/2$^{-}$ state}%
\label{Fig:WaveFuns3He12M}%
\end{center}
\end{figure}

We constructed wave functions \ of the continuous spectrum states using
diagonalization procedure of the 100$\times$100 matrix of Hamiltonian. Details
and justification of this procedure can be found, for example, in Refs.
\cite{2015NuPhA.941..121L}, \cite{2023UkrJPh..68..3K}. Fig.
\ref{Fig:WaveFuns3He12P} shows wave functions of 1/2$^{+}$ states as a
function of distance between proton and deuteron. In Fig.
\ref{Fig:WaveFuns3He12M} we display wave functions for the 1/2$^{-}$ state.
Note that the states 1/2$^{+} $ and 1/2$^{-}$ can be connected by dipole
transition operator. General features of the displayed wave functions are that
they have large amplitude at relatively small distances between clusters
($r<$5 fm) and that they slowly decreasing as $1/r$.%

%

\subsection{Different NN potentials}

\begin{figure}
[ptb]
\begin{center}
\includegraphics[
height=11.7959cm,
width=13.395cm
]%
{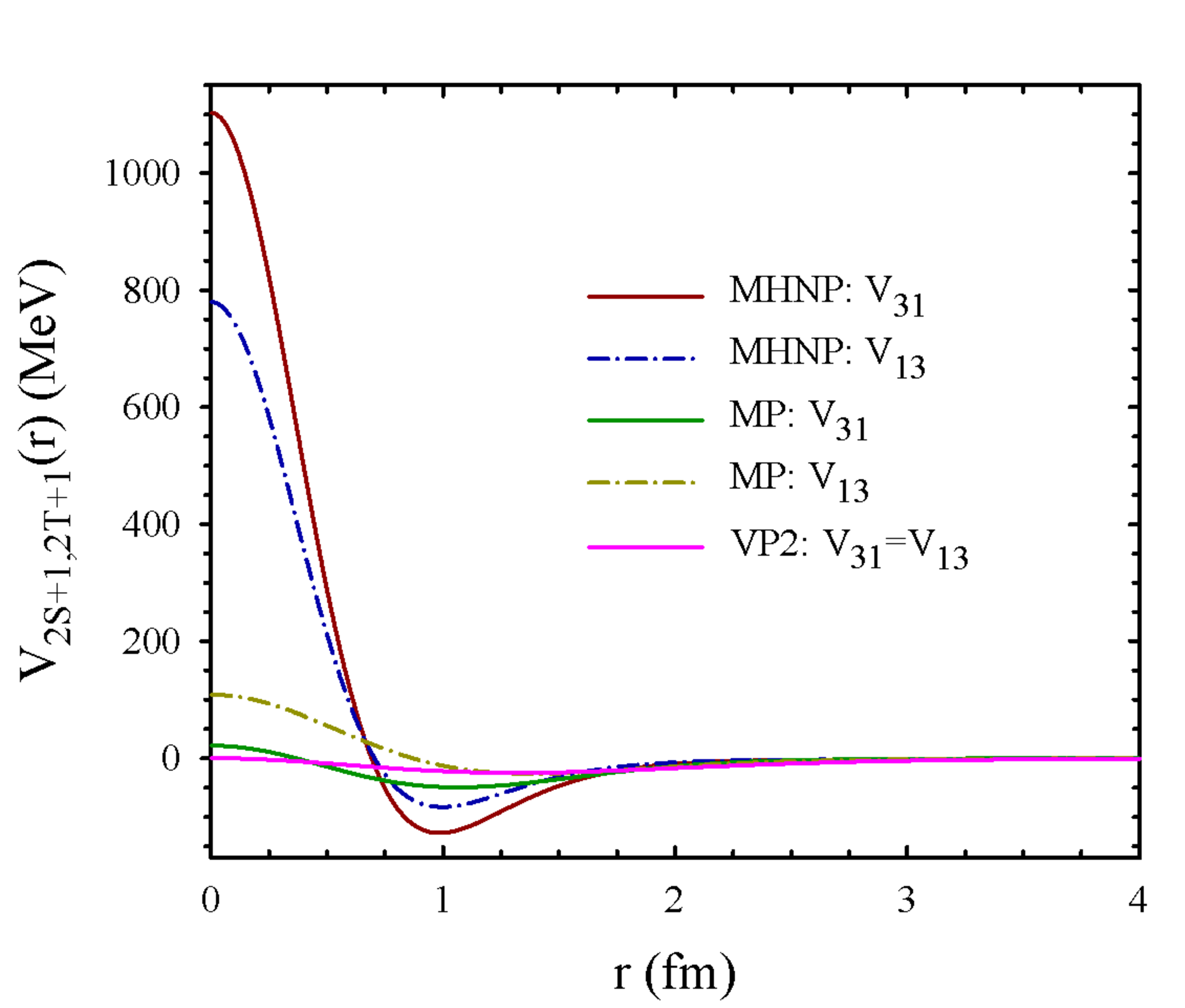}%
\caption{The even components  $V_{31}$ and
$V_{13}$ of the MHNP, MP and VP nucleon-nucleon potentials as a function of distance between nucleons}%
\label{Fig:NNPoten}%
\end{center}
\end{figure}

In this section we consider how the shape of nucleon-nucleon potential affects
phase shifts of the elastic $p+d$ scattering. For this aim we involved in our
calculations two new NN potentials. They are the Volkov potential (VP)
\cite{kn:Volk65} and modified Hasegawa-Nagata potential (MHNP) \cite{potMHN1,
potMHN2}. These potentials alongside with the Minnesota potential are often
used in different cluster models.
It is demonstrated in Fig. \ref{Fig:NNPoten}, where the even components $V_{31}$ and
$V_{13}$ of three nucleon-nucleon potentials are displayed, that the MHNP has
the largest repulsive core at small distance between nucleons, the VP has
smallest repulsive core and the MP represents intermediate case among three
selected potentials.

\begin{figure}
[ptb]
\begin{center}
\includegraphics[
height=11.7959cm,
width=13.395cm
]%
{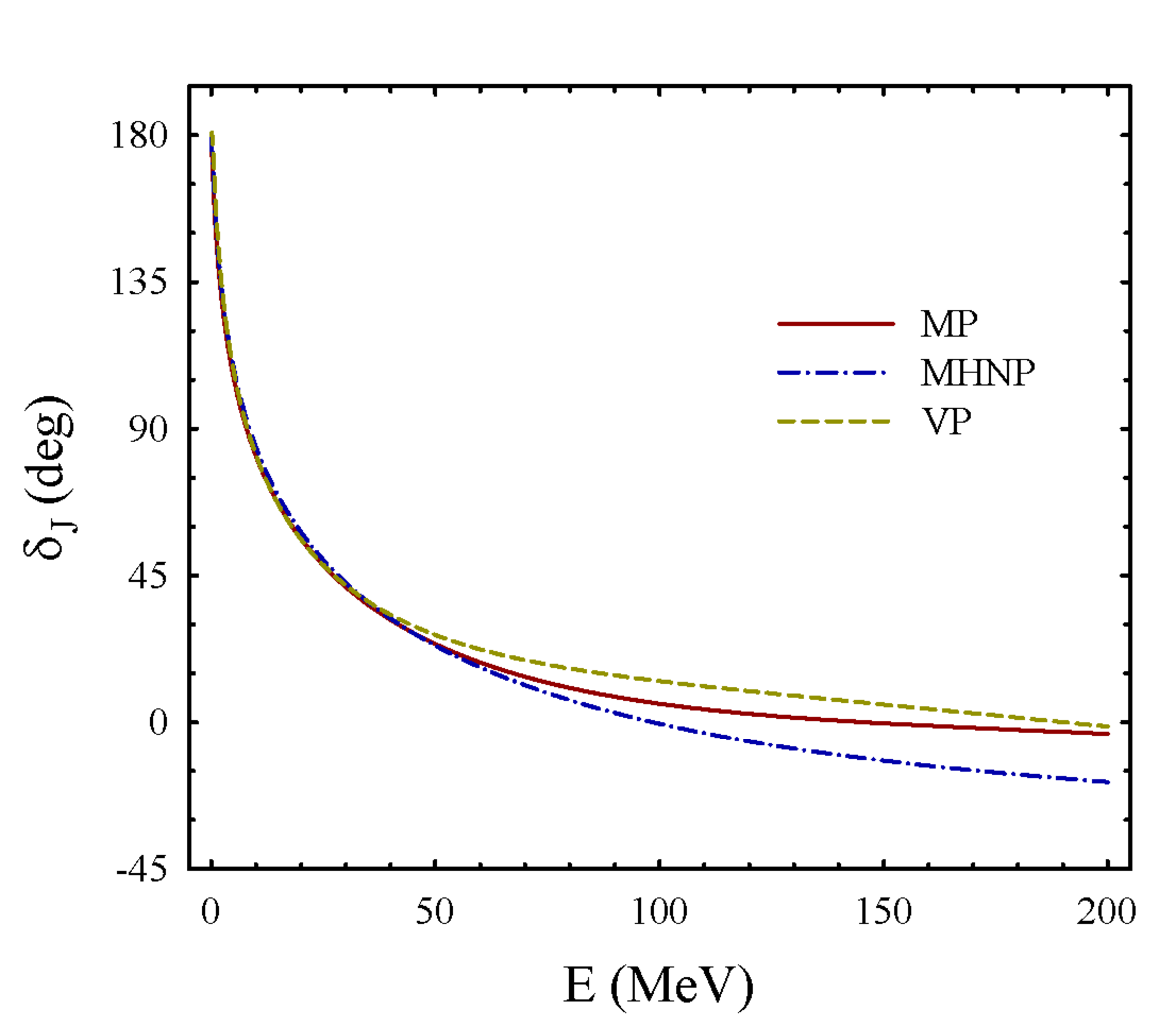}%
\caption{Phase shifts of the $p+d$ scattering in the 1/2$^{+}$ state
calculated with three nucleon-nucleon potentials}%
\label{Fig:PhasesPD3P}%
\end{center}
\end{figure}

In Fig. \ref{Fig:PhasesPD3P} we display phase shifts of the elastic $p+d$
scattering in the 1/2$^{+}$ state. We can see that the phase shifts slightly
depend on the shape of the nucleon-nucleon potentials especially at the energy
region $E<$50 MeV. At the energy range 150$<E<$200 differences of phase
shifts for different potentials are less than 20 degrees.%

\begin{figure}
[ptb]
\begin{center}
\includegraphics[
height=11.7959cm,
width=13.395cm
]%
{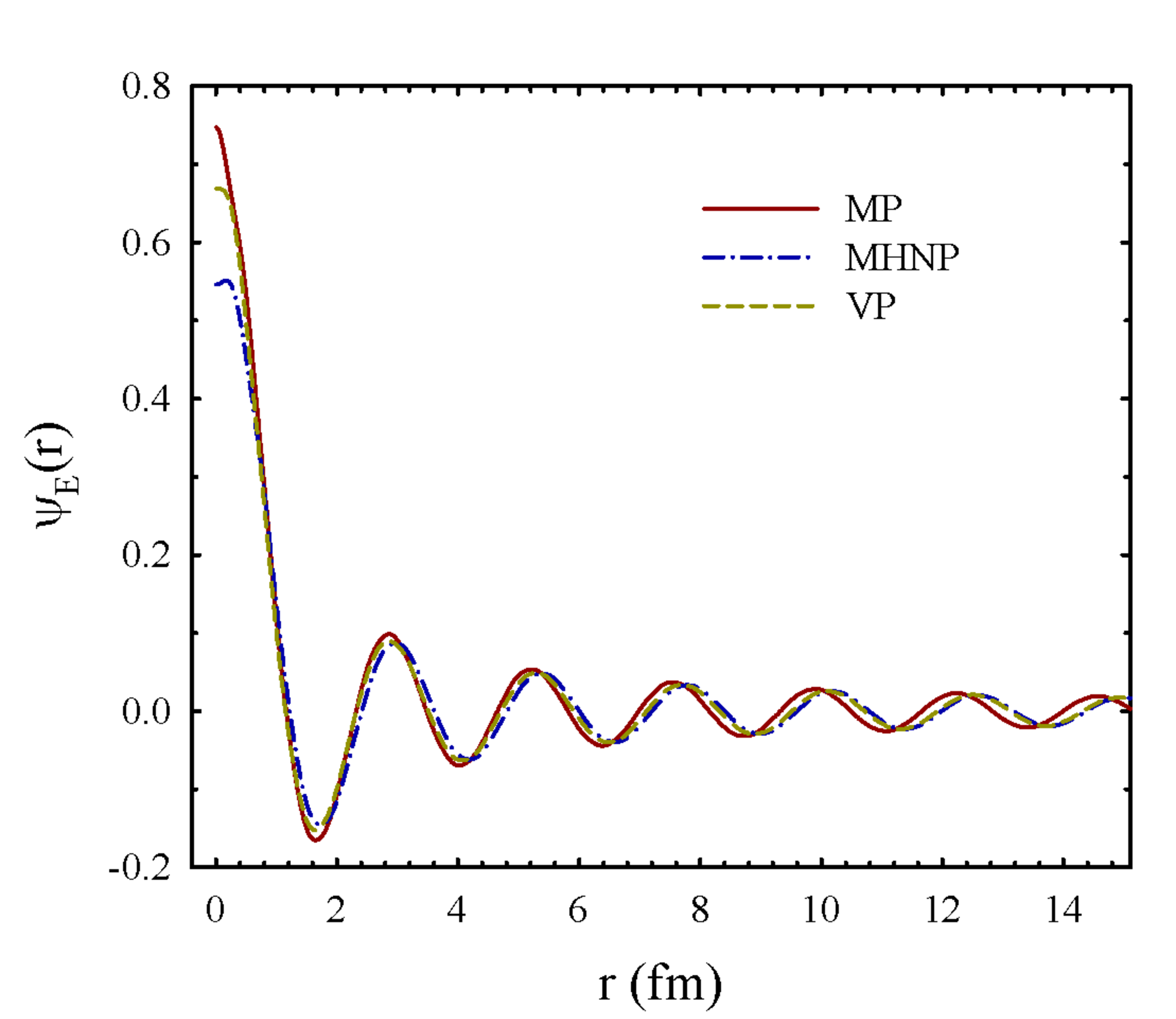}%
\caption{Wave functions of the $p+d$ system at energy 147 MeV obtained with
three different nucleon-nucleon potentials}%
\label{Fig:WaveFuns12P3NNP}%
\end{center}
\end{figure}

As a results, wave functions of the elastic  $p+d$ scattering obtained with
different NN potentials are very close to one other. In Fig.
\ref{Fig:WaveFuns12P3NNP} we display wave functions of $p+d$ system for the
energy 147 MeV. The noticeable difference of wave functions is observed at
small distances $r<$ 1 fm.%

\begin{figure}
[ptb]
\begin{center}
\includegraphics[
height=11.7959cm,
width=13.4851cm
]%
{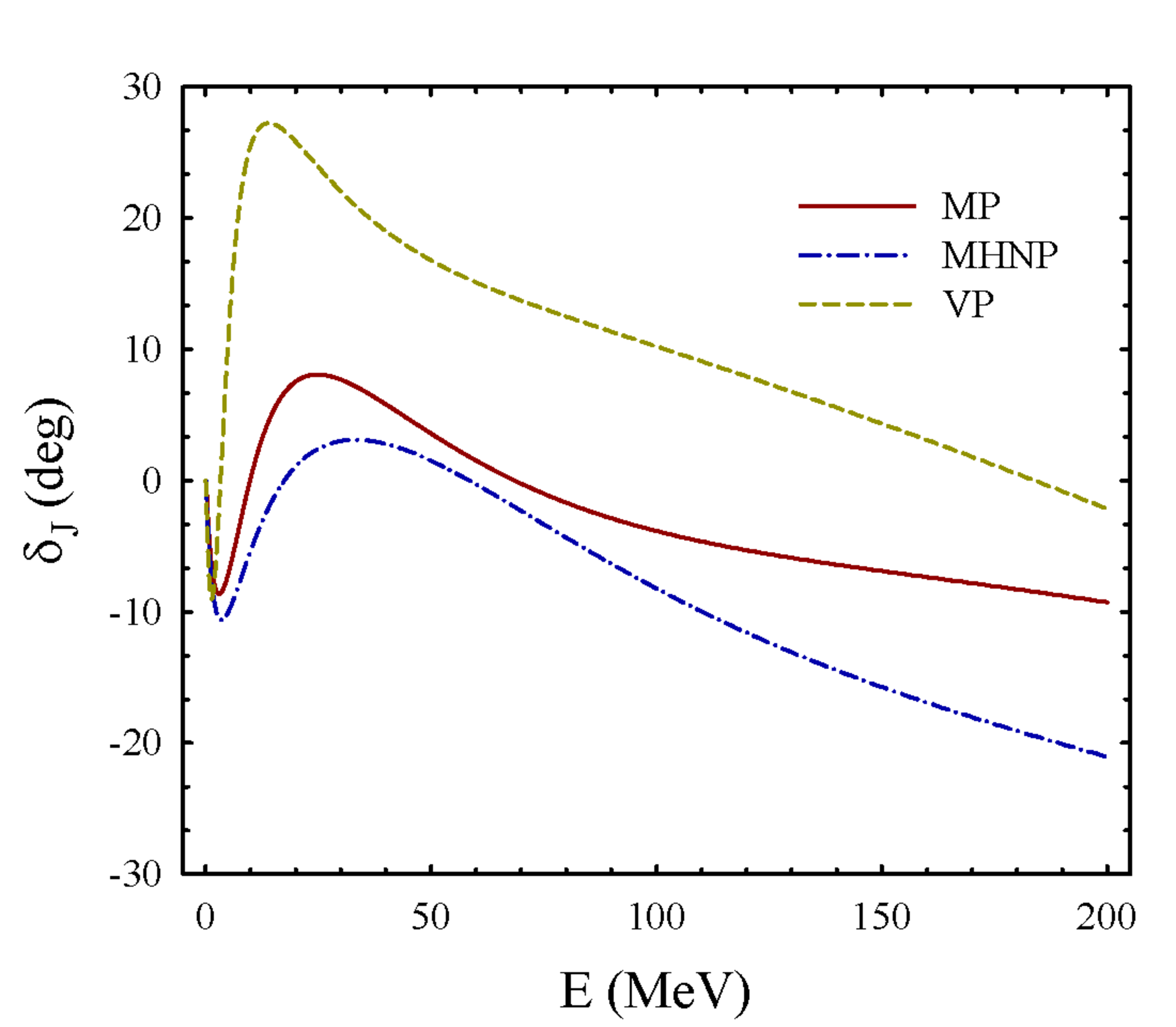}%
\caption{Phase shifts of the elastic $p+d$ scattering in the 1/2$^{-}$ state
obtained with the MP, MHNP and VP}%
\label{Fig:PhasesPD3P12M}%
\end{center}
\end{figure}

Let us consider the 1/2$^{-}$ states in $^{3}$He and in the $p+d$ elastic
scattering. The 1/2$^{-}$ states can be connected to the 1/2$^{+}$ states by
the dipole transition operator. In Fig. \ref{Fig:PhasesPD3P12M} we display
phase shifts of $p+d$ elastic scattering in the 1/2$^{-}$ states, calculated
with three nucleon-nucleon potentials. At low energy range, phase shifts
exhibit resonance-like behavior, when phase shifts rapidly growing with
increasing of energy $E$. However, amplitudes of growing are small, besides we
estimate that the widths of such resonance states are larger than 20 MeV and
their energies are less then 8 MeV. Thus, such states cannot be considered as
resonance states.%

\begin{figure}
[ptb]
\begin{center}
\includegraphics[
height=11.306cm,
width=13.3928cm
]%
{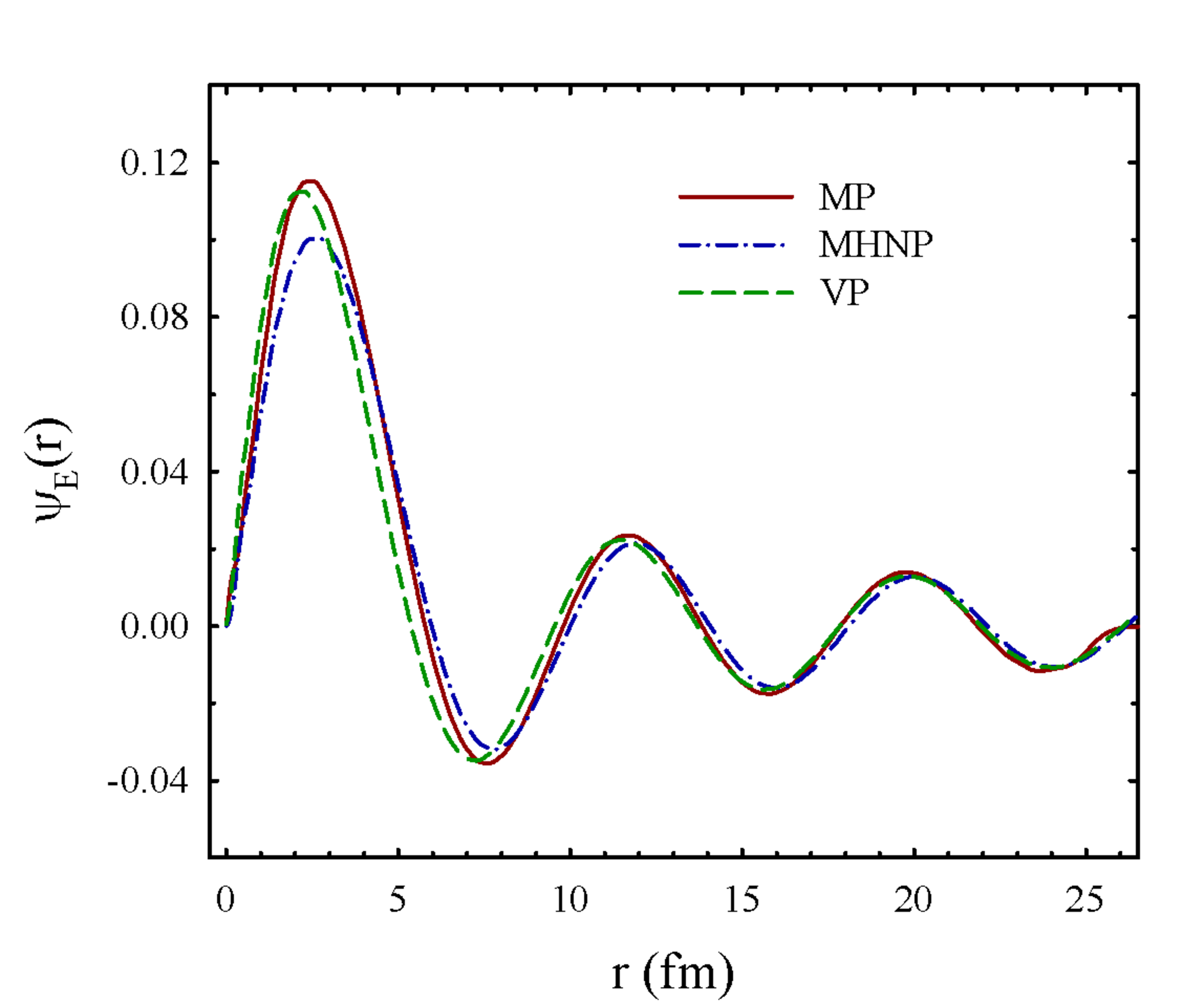}%
\caption{Wave functions of relative motion of proton and deuteron with energy
$E$=12.7 MeV and the total angular momentum $J^{\pi}=$1/2$^{-}$}%
\label{Fig:WaveFuns3He12ME23NNP}%
\end{center}
\end{figure}

\begin{figure}
[ptb]
\begin{center}
\centerline{\includegraphics[width=120mm]{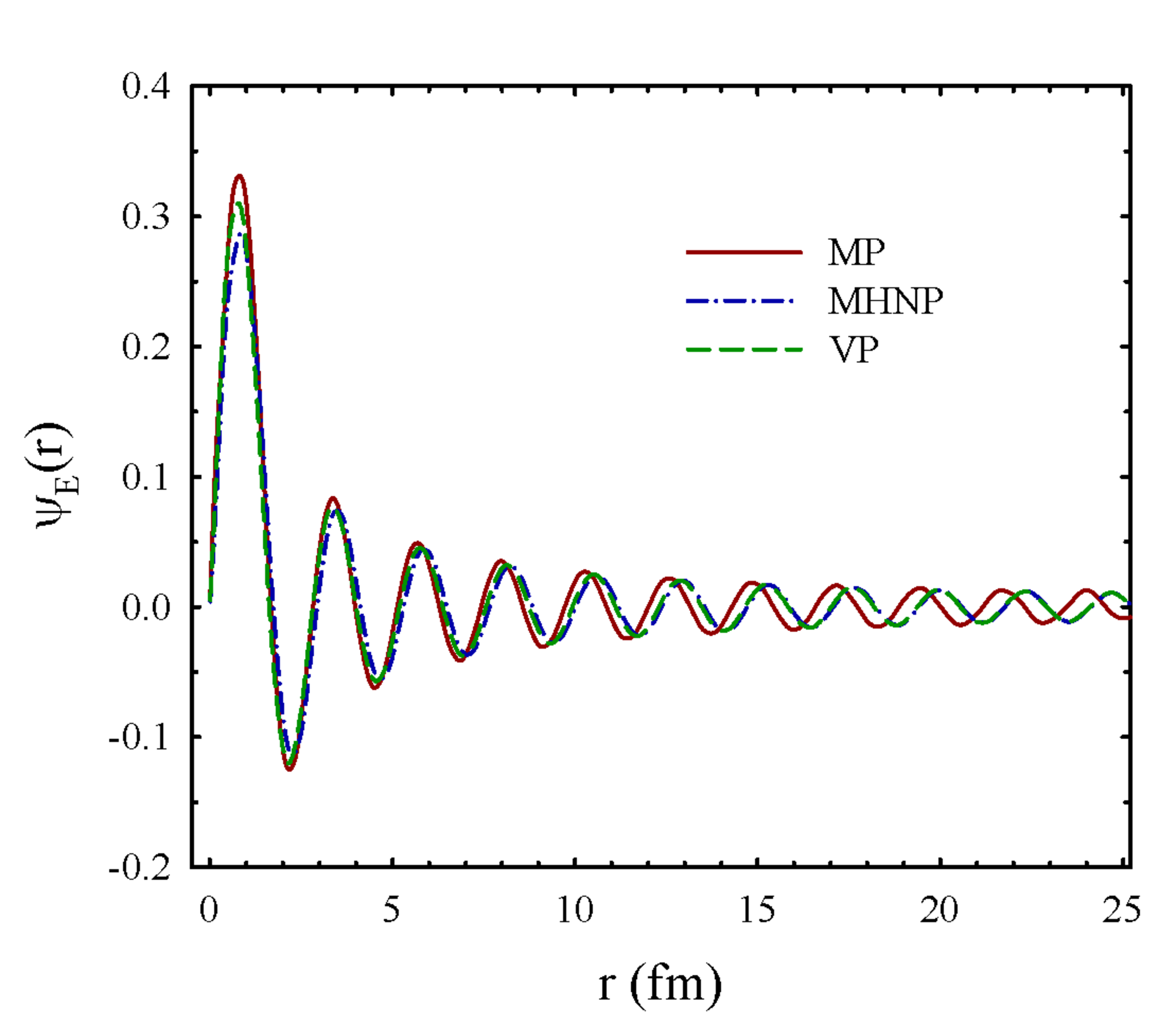}}
\caption{Wave functions of relative motion of proton and deuteron with energy
$E$=147 MeV and the total angular momentum $J^{\pi}=$1/2$^{-}$}%
\label{Fig:WaveFuns3He12ME13NNP}%
\end{center}
\end{figure}

This conclusion can be partially confirmed by behavior of wave functions in
the 1/2$^{-}$ states. In Fig. \ref{Fig:WaveFuns3He12ME23NNP} we display wave
functions obtained with three nucleon-nucleon potentials at the energy
$E$=12.7 MeV and in Fig. \ref{Fig:WaveFuns3He12ME13NNP}  are shown wave
functions for larger value of the energy $E$=147 MeV. As we can see, the wave
functions at relatively small and large energies have noticeable maxima at
relatively small distances (0.8$<r<$2.5 fm) between proton and deuteron. We
may conclude that maxima of wave functions in the 1/2$^{-}$ state at small
distances are due to interplay between effects of nucleon-nucleon and Coulomb
interactions from one side and effects of the Pauli principle from another side.%

%

\subsection{Dependence of the bremsstrahlung cross section on the structure of deuteron
\label{sec.analysis.osc.1}}

We are interesting in question if the bremsstrahlung cross section is dependent on the structure of deuteron. 
In the previous section we have discussed several forms of the deuteron wave
functions. Two of them are presented in analytic form, and one of them is
obtained numerically by solving two-body Schr\"{o}dinger equation with the
Minnesota potential. The deuteron wave function of the oscillator shell model,
displayed in Eq. (\ref{eq:R51}), allows us in a simple way to study effects of
the deuteron structure on the bremsstrahlung cross section.  Indeed, this wave
function depends on the oscillator length $b$. Recall, that the oscillator
length is selected to minimize the ground state energy of deuteron with
selected potential. If in Eq. (\ref{eq:R51})  put $b$=0, then we obtain
structureless deuteron or, in other words, we disregard of the internal
structure of deuteron.

If to suppose existence of such a dependence, it can be small or even not visible for analysis.
If this is correct, then it is unclear, if this is not so at other energies.
In particular, we would like to find energies, where the spectra are already dependent visibly on the internal
structure of deuteron.

Some information about the internal structure of deuteron is included in its form factor, which is presented in the folding and cluster approaches.
The matrix element of  bremsstrahlung emission in both approaches is defined by
Eq.~(\ref{eq.crosssection.1.3}) as ($F_{\rm p} = 1$),
where we will consider form factor of deuteron in folding approach given by Eq.~(\ref{eq.folding.1.2})
[
$Z_{D}=1$, $A_{D}=2$ for deuteron]
%
%
\begin{equation}
\begin{array}{lllll}
  F_{D} = &
  Z_{D}\, \exp{ - \displaystyle\frac{1}{4}\, \displaystyle\frac{A_{D} - 1}{A_{D}}\, (k, b)^{2}} =
  \exp{ - \displaystyle\frac{1}{8}\, (k, b)^{2}}
\end{array}
\label{eq.analysis.osc.2}
\end{equation}
with the oscillator length $b$.

\begin{figure}[htbp]
\centerline{\includegraphics[width=120mm]{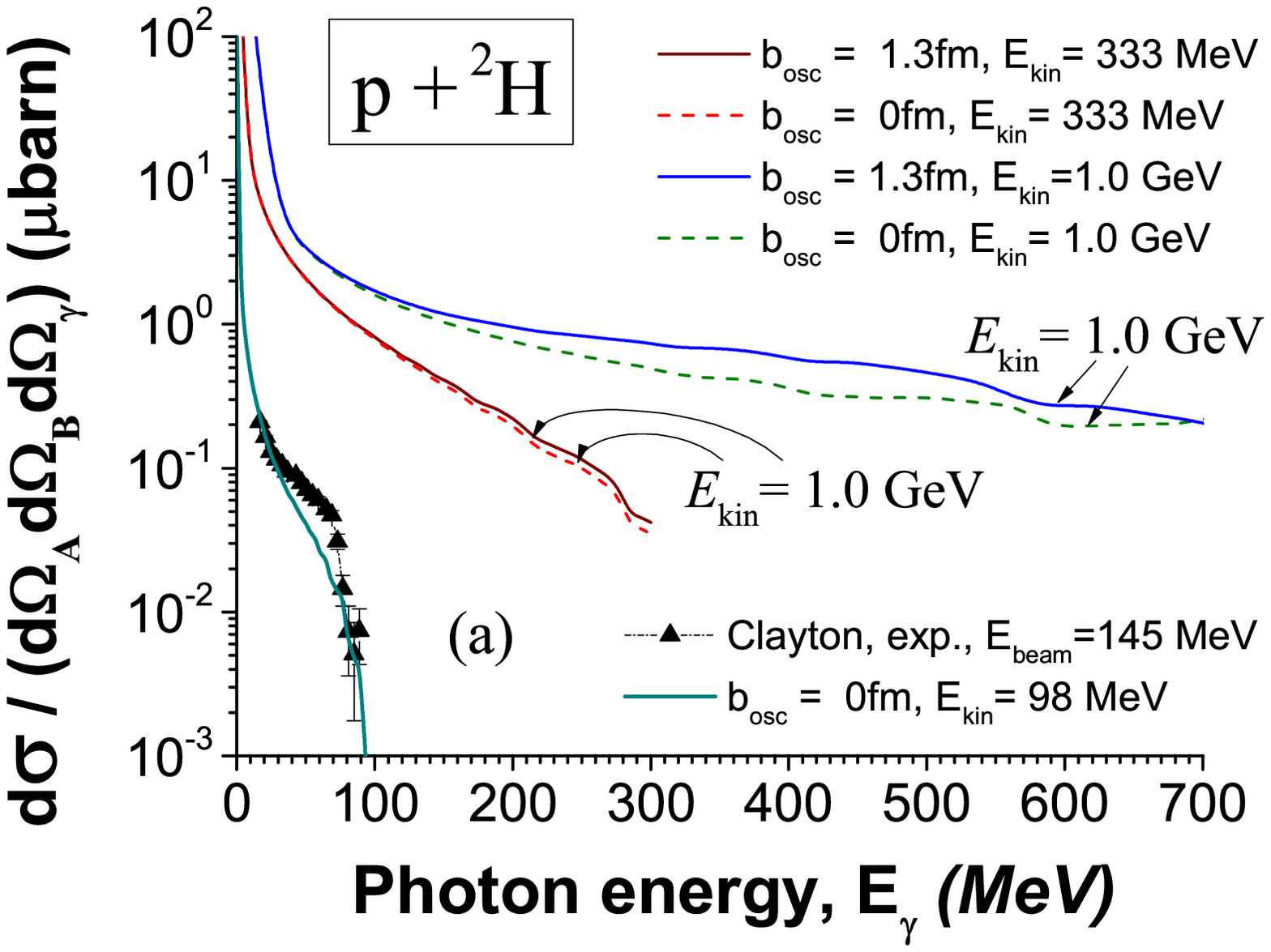}
}
\vspace{-4mm}
\caption{\small (Color online)
Cross sections of bremsstrahlung emission for $p + \isotope[2]{H}$
with included structure of deuteron and without it
calculated at energies of beam of
145~MeV ($E_{\rm kin} = 98~MeV$),
500~MeV ($E_{\rm kin} = 333~MeV$) and
1.5~GeV ($E_{\rm kin} = 1.0~GeV$)
[Parameters of calculations:
cross section is defined in Eq.~(\ref{eq.crosssection.1.1}) and
then averaged over all angles with exception of photon emission angle $\theta$ used in experiments ($\theta=60^{\circ}$,
see also Ref.~\cite{Nakayama.1992.PRC.v45}, Fig.~5),
as non-zero oscillator length for deuteron we choose $b=1.3$~fm,
$R_{\rm max}= 20000\, {\rm fm}$ and 2500000~intervals are used in the numerical integration;
time of computer calculations is 2--4 min for 40 points of each calculated spectrum,
kinetic energy $E_{\rm kin}$ of relative motion of proton and deuteron is used in calculations of bremsstrahlung matrix elements,
which is $E_{\rm kin} = 2/3 \cdot E_{\rm beam}$].
%
Here,
experimental data at 145~MeV of beam energy given by black triangles are extracted from Ref.~\cite{Clayton.1992.PRC.p1810}.
\label{fig.2}}
\end{figure}

Results of such calculations at different energies of beam are presented in Fig.~\ref{fig.2}.
%
%
From such results we conclude the following.
\begin{enumerate}
\item
Difference between cross sections calculated with included structure of deuteron and without it 
at the same beam energy becomes visible and stable at higher energy of beam
(larger 500~MeV).

\item
Calculation with realistic wave functions (with included structure of deuteron) gives larger cross section of bremsstrahlung
than cross section without inclusion of structure of deuteron.

\item
Experiment \cite{Clayton.1992.PRC.p1810} at used beam energy of 145~MeV is not effective for such a study
(it is demonstrated by cross sections at 1.5~GeV in comparison with cross section at 145~MeV in this figure).
At the same time,
possible new measurement of bremsstrahlung cross sections but at higher energies (about 0.5--1.5 GeV of beam energy)
will allow to extract information about structure of deuteron (realistic oscillator length, and wave function).

\item
More precise information about structure of deuteron can be obtained if to organize unified experiments in measurement of bremsstrahlung cross section at two different energies of beam (for example, at 145~MeV and 500~MeV).
Then, our model will estimate ratio between two bremsstrahlung cross sections at such energies and provide value about realistic oscillator length with high precision.

\end{enumerate}


\subsection{Dependence of bremsstrahlung cross section on new form factor of deuteron
\label{sec.analysis.formfactor}}

\begin{figure}[htbp]
\centerline{\includegraphics[width=120mm]{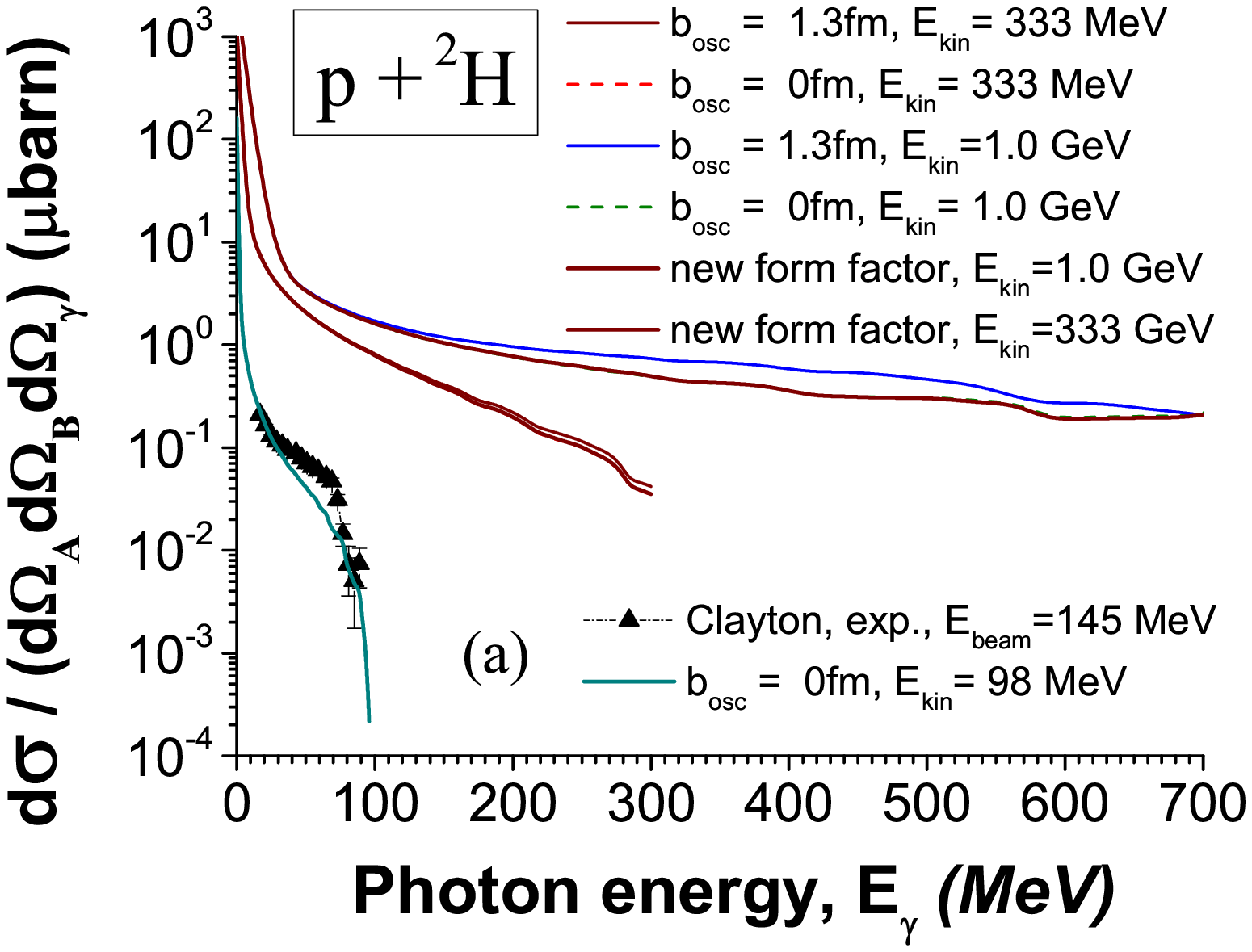}
}
\vspace{-4mm}
\caption{\small (Color online)
Cross sections of bremsstrahlung emission for $p + \isotope[2]{H}$
with included new form factor
calculated at energies of beam of
145~MeV ($E_{\rm kin} = 98~MeV$),
500~MeV ($E_{\rm kin} = 333~MeV$) and
1.5~GeV ($E_{\rm kin} = 1.0~GeV$)
[Parameters of calculations:
cross section is defined in Eq.~(\ref{eq.crosssection.1.1}),
matrix element is defined in Eq.~(\ref{eq.crosssection.1.3}),
as non-zero oscillator length for deuteron we choose $b=1.3$~fm,
$R_{\rm max}= 20000\, {\rm fm}$ and 2500000~intervals are used in the numerical integration;
time of computer calculations is 2--4 min for 40 points of each calculated spectrum,
kinetic energy $E_{\rm kin}$ of relative motion of proton and deuteron is used in calculations of bremsstrahlung matrix elements,
which is $E_{\rm kin} = 2/3 \cdot E_{\rm beam}$].
%
Here,
experimental data at 145~MeV of beam energy given by black triangles are extracted from Ref.~\cite{Clayton.1992.PRC.p1810}.
\label{fig.3}}
\end{figure}

Now we will analyze how much the spectrum is changed if to use new form factor of deuteron instead of previous calculations.
So, we have the matrix element in form~(\ref{eq.crosssection.1.3}) as
with form factor of deuteron given by Eq.~(\ref{formfactor_new}).
Note that this form factor of deuteron does not include the oscillator length.
Results of such calculations with new form factor are presented in Fig.~\ref{fig.3}.
%
%
From these calculations one can see that inclusion of new form factor reduces full cross section a little.
But, general dependence of the cross section on this form of form factor is observed at higher energies.


\section{Conclusions and perspectives
\label{sec.conclusions}}

In this paper we investigated emission of bremsstrahlung photons in the scattering of protons off deuterons on the fully cluster basis
in a wide region of the beam energy from low energies till 1.5~GeV.
To realize this investigations, 
we developed a new model.
%
On the basis of such a model we obtain the following results:

\begin{itemize}
\item
It is demonstrated that the matrix elements of bremsstrahlung emission in the deuteron-proton scattering in the three-cluster formalism  coincides with the corresponding matrix elements in the folding approximation given in Ref.~\cite{Maydanyuk_Vasilevsky.2023.PRC}.


\item
Formalism of the model includes form factor of a deuteron which affects behavior  
of bremsstrahlung cross sections and reflects  the structure of deuteron and influence of parameters of nucleon-nucleon interactions.
This gives possibility to investigate structure of nuclei and properties of interactions from analysis of bremsstrahlung cross sections.

\item
We studied dependence of the bremsstrahlung cross section on structure of deuteron.
We find that the oscillator length $b$, related to the shell-model description of the deuteron, is convenient parameter for such a study.
Analysis of dependence of the cross section on such a parameter shows the following.
At beam energies used at experiment~\cite{Clayton.1992.PRC.p1810} the cross section is not sensitive visibly on variations of oscillator length, i.e. on the internal structure of deuteron.
However, 
stable difference between cross sections calculated at zero and non-zero oscillator lengths
at the same beam energy is observed at higher energy of beam
(larger 500~MeV).
The spectrum is increased at increasing of the oscillator length inside the full energy region of the emitted photons.
However, the usage of new the deuteron form factor in the cluster formalism [see Eq.~(\ref{formfactor_new})] reduces the bremsstrahlung cross section a little.

\item
More precise information about structure of deuteron can be obtained if to organize unified new experiments in measurement of bremsstrahlung cross section at two different energies of beam (for example, at 145~MeV and 500~MeV or above).
Then, our model will estimate ratio between two bremsstrahlung cross sections at such energies and provide information about  realistic  value of the oscillator length with high precision.

\end{itemize}

\noindent
As a perspective, we see that
formalism of our model provides strict basis for description of the deuteron-proton scattering and emission of virtual photons in study of dilepton productions in the deuteron-proton scattering
(see, for example,  Refs.~\cite{Ernst.1998.PRC,Wilson.1998.PRC}).
This can be interesting for further investigations and applications.


\section*{Acknowledgements
\label{sec.acknowledgements}}

Authors are highly appreciated to
Prof. Gyorgy~Wolf for fruitful discussions concerning to different aspects of nuclear collisions at high energies,
Dr.~S.~A.~Omelchenko for fruitful discussions concerning to aspects of different nuclear models for scattering and decays.
%
This work was supported in part by the Program of Fundamental Research of the
Physics and Astronomy Department of the National Academy of Sciences of
Ukraine (Project No. 0117U000239 and Project No. 0122U000889).
S.~P.~M. thanks the support of OTKA grant K138277.
V.V. is grateful to the Simons foundation for financial support (Award No. 1030284).
\appendix
\section{Calculation of operator of bremsstrahlung emission in three-cluster model
\label{sec.app.cluster.operator}}

In this Appendix we will calculate operator of emission of bremsstrahlung photons in three-cluster formalism.
We fix position of nucleons. We assume that vector $\mathbf{r}$ measures
the distance between proton and neutron which form a deuteron. We also assume
that $\mathbf{r}_{1}$ is a coordinate of the first proton and $\mathbf{r}_{2}$
is a coordinate of a neutron. Vector $\mathbf{r}_{3}$ determines the location
of the second proton.
Starting from Eq.~(\ref{eq:R01}), this operator is obtained the following form:
%
\begin{align*}
&  \widehat{H}_{e}\left(  \mathbf{k}_{\gamma},\mathbf{\varepsilon}^{(\alpha)}\right)  =
\frac{1}{2}\frac{e\hbar}{m_{N}c}\left[  \widehat{\mathbf{\pi}}%
_{1}^{\ast}\mathbf{A}^{\ast}\left(  1\right)  +\mathbf{A}^{\ast}\left(
1\right)  \widehat{\mathbf{\pi}}_{1}^{\ast}+\widehat{\mathbf{\pi}}_{3}^{\ast
}\mathbf{A}^{\ast}\left(  3\right)  +\mathbf{A}^{\ast}\left(  3\right)
\widehat{\mathbf{\pi}}_{3}^{\ast}\right]  \\
%
%
&  =\frac{1}{2}\frac{e\hbar}{m_{N}c}\left\{  \widehat{\mathbf{\pi}}_{1}^{\ast
}\mathbf{\varepsilon}^{(\alpha)}\exp\left\{  -i\left(  \mathbf{k}_{\gamma
}\mathbf{\rho}_{1}\right)  \right\}  +\mathbf{\varepsilon}^{(\alpha)}\exp\left\{
-i\left(  \mathbf{k}_{\gamma}\mathbf{\rho}_{1}\right)  \right\}
\widehat{\mathbf{\pi}}_{1}^{\ast}\right.  \\
&  +\left.  \widehat{\mathbf{\pi}}_{3}^{\ast}\mathbf{\varepsilon}^{(\alpha)}%
\exp\left\{  -i\left(  \mathbf{k}_{\gamma}\mathbf{\rho}_{3}\right)  \right\}
+\mathbf{\varepsilon}^{(\alpha)}\exp\left\{  -i\left(  \mathbf{k}_{\gamma
}\mathbf{\rho}_{3}\right)  \right\}  \widehat{\mathbf{\pi}}_{3}^{\ast
}\right\}  \\
&  =\frac{1}{2}\frac{e\hbar}{m_{N}c}\left\{  \left(  \mathbf{\varepsilon}%
^{(\alpha)}\widehat{\mathbf{\pi}}_{1}^{\ast}\right)  \exp\left\{  -i\frac{1}%
{\sqrt{2}}\left(  \mathbf{k}_{\gamma}\mathbf{r}\right)  \mathbf{+}i\frac
{1}{\sqrt{6}}\left(  \mathbf{k}_{\gamma}\mathbf{q}\right)  \right\}  \right.
\\
&  +\exp\left\{  -i\frac{1}{\sqrt{2}}\left(  \mathbf{k}_{\gamma}%
\mathbf{r}\right)  \mathbf{+}i\frac{1}{\sqrt{6}}\left(  \mathbf{k}_{\gamma
}\mathbf{q}\right)  \right\}  \left(  \mathbf{\varepsilon}^{(\alpha)}%
\widehat{\mathbf{\pi}}_{1}^{\ast}\right)  \\
&  +\left.  \left(  \mathbf{\varepsilon}^{(\alpha)}\widehat{\mathbf{\pi}}_{3}%
^{\ast}\right)  \exp\left\{  -i\sqrt{\frac{2}{3}}\left(  \mathbf{k}_{\gamma
}\mathbf{q}\right)  \right\}  +\exp\left\{  -i\sqrt{\frac{2}{3}}\left(
\mathbf{k}_{\gamma}\mathbf{q}\right)  \right\}  \left(  \mathbf{\varepsilon
}^{(\alpha)}\widehat{\mathbf{\pi}}_{3}^{\ast}\right)  \right\}
\end{align*}
%
and then
\begin{align*}
&  \widehat{H}_{e}\left(  \mathbf{k}_{\gamma},\mathbf{\varepsilon}^{(\alpha)}\right)  \\
&  =\frac{1}{2}\frac{e\hbar}{m_{N}c}\left\{  \left( \mathbf{\varepsilon}^{(\alpha)}, 
\frac{1}{\sqrt{2}}\mathbf{\pi}_{\mathbf{r}}^{\ast}\mathbf{-}\frac
{1}{\sqrt{6}}\mathbf{\pi}_{\mathbf{q}}^{\ast}\right)  \exp\left\{  -i\frac
{1}{\sqrt{2}}\left(  \mathbf{k}_{\gamma}\mathbf{r}\right)  \mathbf{+}i\frac
{1}{\sqrt{6}}\left(  \mathbf{k}_{\gamma}\mathbf{q}\right)  \right\}  \right.
\\
&  +\exp\left\{  -i\frac{1}{\sqrt{2}}\left(  \mathbf{k}_{\gamma}%
\mathbf{r}\right)  \mathbf{+}i\frac{1}{\sqrt{6}}\left(  \mathbf{k}_{\gamma
}\mathbf{q}\right)  \right\}  \left(  \mathbf{\varepsilon}^{(\alpha)},\frac
{1}{\sqrt{2}}\mathbf{\pi}_{\mathbf{r}}^{\ast}\mathbf{-}\frac{1}{\sqrt{6}%
}\mathbf{\pi}_{\mathbf{q}}^{\ast}\right)  \\
&  +\left.  \sqrt{\frac{2}{3}}\left(  \mathbf{\varepsilon}^{(\alpha)} \mathbf{\pi
}_{\mathbf{q}}^{\ast}\right)  \exp\left\{  -i\sqrt{\frac{2}{3}}\left(
\mathbf{k}_{\gamma}\mathbf{q}\right)  \right\}  +\sqrt{\frac{2}{3}}%
\exp\left\{  -i\sqrt{\frac{2}{3}}\left(  \mathbf{k}_{\gamma}\mathbf{q}\right)
\right\}  \left(  \mathbf{\varepsilon}^{(\alpha)}\mathbf{\pi}_{\mathbf{q}}^{\ast
}\right)  \right\}
\end{align*}
Collecting similar terms we obtain%
\begin{align*}
  & \widehat{H}_{e}\left(  \mathbf{k}_{\gamma},\mathbf{\varepsilon}^{(\alpha)} \right) 
  =
  \frac{1}{2}\frac{e\hbar}{m_{N}c}\left\{
  \left(  \mathbf{\varepsilon}^{(\alpha)}, \frac{1}{2}\mathbf{k}_{\gamma}\mathbf{+}\frac{1}{6}\mathbf{k}_{\gamma}\right)
  \exp\left\{  -i\frac{1}{\sqrt{2}}\left(  \mathbf{k}_{\gamma}\mathbf{r}\right)  +
  i\frac{1}{\sqrt{6}}\left(  \mathbf{k}_{\gamma}\mathbf{q}\right)  \right\}  \right.  \\
  & +
  2\exp\left\{  -i\frac{1}{\sqrt{2}}\left(  \mathbf{k}_{\gamma}%
\mathbf{r}\right)  \mathbf{+}i\frac{1}{\sqrt{6}}\left(  \mathbf{k}_{\gamma
}\mathbf{q}\right)  \right\}  \left(  \mathbf{\varepsilon}^{(\alpha)}, \frac
{1}{\sqrt{2}}\mathbf{\pi}_{\mathbf{r}}^{\ast}\mathbf{-}\frac{1}{\sqrt{6}%
}\mathbf{\pi}_{\mathbf{q}}^{\ast}\right)  \\
  & +
  \sqrt{\frac{2}{3}}\left(  \mathbf{\varepsilon}^{(\alpha)} i\left[  -i\sqrt
{\frac{2}{3}}\mathbf{k}_{\gamma}\right]  \right)  \exp\left\{  -i\sqrt
{\frac{2}{3}}\left(  \mathbf{k}_{\gamma}\mathbf{q}\right)  \right\}  \\
  & +
  \left.  2\sqrt{\frac{2}{3}}
  \exp\left\{  -i\sqrt{\frac{2}{3}}\left(\mathbf{k}_{\gamma}\mathbf{q}\right)  \right\}
  \left( \mathbf{\varepsilon}^{(\alpha)} \mathbf{\pi}_{\mathbf{q}}^{\ast}\right)  \right\}
\end{align*}
or
\begin{align*}
&  \widehat{H}_{e}\left(  \mathbf{k}_{\gamma},\mathbf{\varepsilon}^{(\alpha)}\right)  =\frac{1}{2}\frac{e\hbar}{m_{N}c}\left\{  \frac{2}{3}\left(
\mathbf{\varepsilon}^{(\alpha)},\mathbf{k}_{\gamma}\right)  \exp\left\{  -i\frac
{1}{\sqrt{2}}\left(  \mathbf{k}_{\gamma}\mathbf{r}\right)  \mathbf{+}i\frac
{1}{\sqrt{6}}\left(  \mathbf{k}_{\gamma}\mathbf{q}\right)  \right\}  \right.
\\
&  +2\exp\left\{  -i\frac{1}{\sqrt{2}}\left(  \mathbf{k}_{\gamma}%
\mathbf{r}\right)  \mathbf{+}i\frac{1}{\sqrt{6}}\left(  \mathbf{k}_{\gamma
}\mathbf{q}\right)  \right\}  \left(  \mathbf{\varepsilon}^{(\alpha)}, \frac
{1}{\sqrt{2}}\mathbf{\pi}_{\mathbf{r}}^{\ast}\mathbf{-}\frac{1}{\sqrt{6}%
}\mathbf{\pi}_{\mathbf{q}}^{\ast}\right)  \\
&  +\frac{2}{3}\left(  \mathbf{\varepsilon}^{(\alpha)} \mathbf{k}_{\gamma}\right)
\exp\left\{  -i\sqrt{\frac{2}{3}}\left(  \mathbf{k}_{\gamma}\mathbf{q}\right)
\right\}  \\
&  +\left.  2\sqrt{\frac{2}{3}}\exp\left\{  -i\sqrt{\frac{2}{3}}\left(
\mathbf{k}_{\gamma}\mathbf{q}\right)  \right\}  \left(  \mathbf{\varepsilon
}^{(\alpha)} \mathbf{\pi}_{\mathbf{q}}^{\ast}\right)  \right\}
\end{align*}
or
\begin{align*}
&  \widehat{H}_{e}\left(  \mathbf{k}_{\gamma},\mathbf{\varepsilon}^{(\alpha)}\right)  =\frac{1}{2}\frac{e\hbar}{m_{N}c}\left\{  \frac{2}{3}\left(
\mathbf{\varepsilon}^{(\alpha)}, \mathbf{k}_{\gamma}\right)  \exp\left\{  -i\frac
{1}{\sqrt{2}}\left(  \mathbf{k}_{\gamma}\mathbf{r}\right)  \mathbf{+}i\frac
{1}{\sqrt{6}}\left(  \mathbf{k}_{\gamma}\mathbf{q}\right)  \right\}  \right.
\\
&  +\frac{2}{3}\left(  \mathbf{\varepsilon}^{(\alpha)} \mathbf{k}_{\gamma}\right)
\exp\left\{  -i\sqrt{\frac{2}{3}}\left(  \mathbf{k}_{\gamma}\mathbf{q}\right)
\right\}  \\
&  +2\exp\left\{  -i\frac{1}{\sqrt{2}}\left(  \mathbf{k}_{\gamma}%
\mathbf{r}\right)  \mathbf{+}i\frac{1}{\sqrt{6}}\left(  \mathbf{k}_{\gamma
}\mathbf{q}\right)  \right\}  \left(  \mathbf{\varepsilon}^{(\alpha)}, \frac
{1}{\sqrt{2}}\mathbf{\pi}_{\mathbf{r}}^{\ast}\mathbf{-}\frac{1}{\sqrt{6}%
}\mathbf{\pi}_{\mathbf{q}}^{\ast}\right)  \\
&  +\left.  2\sqrt{\frac{2}{3}}\exp\left\{  -i\sqrt{\frac{2}{3}}\left(
\mathbf{k}_{\gamma}\mathbf{q}\right)  \right\}  \left(  \mathbf{\varepsilon
}^{(\alpha)},\mathbf{\pi}_{\mathbf{q}}^{\ast}\right)  \right\}
\end{align*}
Final form of the operator
\begin{align}
&  \widehat{H}_{e}\left(  \mathbf{k}_{\gamma},\mathbf{\varepsilon}^{(\alpha)}\right)  =\frac{1}{2}\frac{e\hbar}{m_{N}c}\left\{  \frac{2}{3}\left(
\mathbf{\varepsilon}^{(\alpha)}, \mathbf{k}_{\gamma}\right)  \exp\left\{  -i\frac
{1}{\sqrt{2}}\left(  \mathbf{k}_{\gamma}\mathbf{r}\right)  \mathbf{+}i\frac
{1}{\sqrt{6}}\left(  \mathbf{k}_{\gamma}\mathbf{q}\right)  \right\}  \right.
\label{eq:R19}\\
&  +\frac{2}{\sqrt{2}}\exp\left\{  -i\frac{1}{\sqrt{2}}\left(  \mathbf{k}%
_{\gamma}\mathbf{r}\right)  \mathbf{+}\right\}  \exp\left\{  \mathbf{+}%
i\frac{1}{\sqrt{6}}\left(  \mathbf{k}_{\gamma}\mathbf{q}\right)  \right\}
\left(  \mathbf{\varepsilon}^{(\alpha)}, \mathbf{\pi}_{\mathbf{r}}^{\ast}\right)
\nonumber\\
&  +\frac{2}{3}\left(  \mathbf{\varepsilon}^{(\alpha)} \mathbf{k}_{\gamma}\right)
\exp\left\{  -i\sqrt{\frac{2}{3}}\left(  \mathbf{k}_{\gamma}\mathbf{q}\right)
\right\}  \nonumber\\
&  \mathbf{-}\sqrt{\frac{2}{3}}\exp\left\{  -i\frac{1}{\sqrt{2}}\left(
\mathbf{k}_{\gamma}\mathbf{r}\right)  \right\}  \exp\left\{  \mathbf{+}%
i\frac{1}{\sqrt{6}}\left(  \mathbf{k}_{\gamma}\mathbf{q}\right)  \right\}
\left(  \mathbf{\varepsilon}^{(\alpha)},\mathbf{\pi}_{\mathbf{q}}^{\ast}\right)
\nonumber\\
&  +\left.  2\sqrt{\frac{2}{3}}\exp\left\{  -i\sqrt{\frac{2}{3}}\left(
\mathbf{k}_{\gamma}\mathbf{q}\right)  \right\}  \left(  \mathbf{\varepsilon
}^{(\alpha)},\mathbf{\pi}_{\mathbf{q}}^{\ast}\right)  \right\}  \nonumber
\end{align}
or by taking into account that
$\left( \mathbf{\varepsilon}^{(\alpha)},\mathbf{k}_{\gamma}\right) =0$,
we obtain
\begin{align}
&  \widehat{H}_{e}\left(  \mathbf{k}_{\gamma},\mathbf{\varepsilon}^{(\alpha)}\right)  =\frac{1}{2}\frac{e\hbar}{m_{N}c}\left\{  {}\right.  \label{eq:R19A}%
\\
&  +\frac{2}{\sqrt{2}}\exp\left\{  -i\frac{1}{\sqrt{2}}\left(  \mathbf{k}%
_{\gamma}\mathbf{r}\right)  \right\}  \left(  \mathbf{\varepsilon}^{(\alpha)},\mathbf{\pi}_{\mathbf{r}}^{\ast}\right)  \exp\left\{  i\frac{1}{\sqrt{6}%
}\left(  \mathbf{k}_{\gamma}\mathbf{q}\right)  \right\}  \nonumber\\
&  \mathbf{-}\sqrt{\frac{2}{3}}\exp\left\{  -i\frac{1}{\sqrt{2}}\left(
\mathbf{k}_{\gamma}\mathbf{r}\right)  \right\}  \exp\left\{  i\frac{1}%
{\sqrt{6}}\left(  \mathbf{k}_{\gamma}\mathbf{q}\right)  \right\}  \left(
\mathbf{\varepsilon}^{(\alpha)}, \mathbf{\pi}_{\mathbf{q}}^{\ast}\right)
\nonumber\\
&  +\left.  2\sqrt{\frac{2}{3}}\exp\left\{  -i\sqrt{\frac{2}{3}}\left(
\mathbf{k}_{\gamma}\mathbf{q}\right)  \right\}  \left(  \mathbf{\varepsilon
}^{(\alpha)}, \mathbf{\pi}_{\mathbf{q}}^{\ast}\right)  \right\}  .\nonumber
\end{align}


\section{Calculations of integrals
\label{sec.app.integrals}}

\subsection{A general case
\label{sec.app.integrals.1}}

In this Appendix we calculate integrals~(\ref{eq.matrixelement.1.5}):
\begin{equation}
\begin{array}{lllll}
\vspace{0.5mm}
  \vb{I}_{1} = \biggl\langle\: \Phi_{f} (\vb{r})\; \biggl|\, e^{-i\, \vb{k}_{\gamma} \vb{r}}\; \vb{\displaystyle\frac{d}{dr}} \biggr|\: \Phi_{i} (\vb{r})\: \biggr\rangle_\mathbf{r}, \\

\vspace{0.5mm}
  I_{2} = \Bigl\langle \Phi_{f} (\vb{r})\; \Bigl|\, e^{i\, c_{p}\, \vb{k_{\gamma}} \vb{r}}\, \Bigr|\, \Phi_{i} (\vb{r})\: \Bigr\rangle_\mathbf{r}, \\

\vspace{0.5mm}
  I_{3} = \Bigl\langle \Phi_{f} (\vb{r})\; \Bigl|\, e^{-i\, c_{A}\, \vb{k_{\gamma}} \vb{r}}\, \Bigr|\, \Phi_{i} (\vb{r})\: \Bigr\rangle_\mathbf{r}, \\
  I_{4} = \Bigl\langle \Phi_{f} (\vb{r})\; \Bigl|\, e^{- i\, c_{A}\, \vb{k_{\gamma}} \vb{r}}\, V(\vb{r})\, \Bigr|\, \Phi_{i} (\vb{r})\: \Bigr\rangle_\mathbf{r}.
\end{array}
\label{eq.app.integrals.1}
\end{equation}
Here,
to two integrals in Eqs.~(\ref{eq.matrixelement.1.5})
we have added two new types of integrals else,
which are used in calculations in other problems of bremsstrahlung emission.
$V(\vb{r})$ is arbitrary potential function.

We apply multipole expansion of wave function of photons, following to formalism in Sect.~D in Ref.~\cite{Maydanyuk.2012.PRC}
[see Eqs.~(29)--(31) and (24)--(28) in that paper].
Here, we obtain the following formulas for matrix elements:
%
\begin{equation}
\begin{array}{ll}
  \vspace{1mm}
  \Bigl< k_{f} \Bigl| \,  e^{-i\mathbf{k_{\gamma}r}} \, \Bigr| \,k_{i} \Bigr>_\mathbf{r} =
  \sqrt{\displaystyle\frac{\pi}{2}}\:
  \displaystyle\sum\limits_{l_{\gamma}=1}\,
    (-i)^{l_{\gamma}}\, \sqrt{2l_{\gamma}+1}\;
  \displaystyle\sum\limits_{\mu = \pm 1}
    \Bigl[ \mu\,\tilde{p}_{l_{\gamma}\mu}^{M} - i\, \tilde{p}_{l_{\gamma}\mu}^{E} \Bigr], \\

  \biggl< k_{f} \biggl| \,  e^{-i\mathbf{k_{\gamma}r}} \displaystyle\frac{\partial}{\partial \mathbf{r}}\,
  \biggr| \,k_{i} \biggr>_\mathbf{r} =
  \sqrt{\displaystyle\frac{\pi}{2}}\:
  \displaystyle\sum\limits_{l_{\gamma}=1}\,
    (-i)^{l_{\gamma}}\, \sqrt{2l_{\gamma}+1}\;
  \displaystyle\sum\limits_{\mu = \pm 1}
    \xibf_{\mu}\, \mu\, \times
    \Bigl[ p_{l_{\gamma}\mu}^{M} - i\mu\: p_{l_{\gamma}\mu}^{E} \Bigr].
\end{array}
\label{eq.app.integrals.2}
\end{equation}
On the basis of these formulas, we write solutions for integrals (for simplicity, we study case of $l_{i}=0$).

\vspace{1.5mm}
According to the second formula in Eqs.~(\ref{eq.app.integrals.2}), the first integral is
\begin{equation}
\begin{array}{ll}
  \vb{I}_{1} =
  \biggl< \Phi_{f} \biggl| \,  e^{-i\mathbf{k_{\gamma}r}} \displaystyle\frac{\partial}{\partial \vb{r}}\, \biggr| \,\Phi_{i} \biggr>_\mathbf{r} =
  \sqrt{\displaystyle\frac{\pi}{2}}\:
  \displaystyle\sum\limits_{l_{\gamma}=1}\,
    (-i)^{l_{\gamma}}\, \sqrt{2l_{\gamma}+1}\;
  \displaystyle\sum\limits_{\mu = \pm 1}
    \xibf_{\mu}\, \mu\, \times
    \Bigl[ p_{l_{\gamma}\mu}^{M} - i\mu\: p_{l_{\gamma}\mu}^{E} \Bigr],
\end{array}
\label{eq.app.integrals.3}
\end{equation}
where [see Eqs.~(38) at $l_{i}=0$ in Ref.~\cite{Maydanyuk.2012.PRC}]
%
\begin{equation}
\begin{array}{lcl}
\vspace{3mm}
  p_{l_{\gamma}\mu}^{M} 
   & = &
    - I_{M}(0, l_{f}, l_{\gamma}, 1, \mu) \cdot J_{1}(0, l_{f},l_{\gamma}), \\

\vspace{1mm}
  p_{l_{\gamma}\mu}^{E} & = &
    \sqrt{\displaystyle\frac{l_{\gamma}+1}{2l_{\gamma}+1}} \cdot I_{E} (0,l_{f},l_{\gamma}, 1, l_{\gamma}-1, \mu) \cdot J_{1}(0,l_{f},l_{\gamma}-1)\; - \\
  & - &
    \sqrt{\displaystyle\frac{l_{\gamma}}{2l_{\gamma}+1}} \cdot I_{E} (0,l_{f}, l_{\gamma}, 1, l_{\gamma}+1, \mu) \cdot J_{1}(0,l_{f},l_{\gamma}+1)
\end{array}
\label{eq.app.integrals.4}
\end{equation}
and [see Eqs.~(39) in Ref.~\cite{Maydanyuk.2012.PRC}]
\begin{equation}
\begin{array}{ccl}
  J_{1}(l_{i},l_{f},n) & = & \displaystyle\int\limits^{+\infty}_{0} \displaystyle\frac{dR_{i}(r, l_{i})}{dr}\: R^{*}_{f}(l_{f},r)\, j_{n}(k_{\gamma}r)\; r^{2} dr.
\end{array}
\label{eq.app.integrals.5}
\end{equation}

For the other integrals from Eqs.~(\ref{eq.app.integrals.1}) one can get similar solutions
[those are directly derived from the first expansion in Eqs.~(\ref{eq.app.integrals.2}), where the other corresponding radial integrals should be used]:
%
%
\begin{equation}
\begin{array}{lllll}
\vspace{0.5mm}
  I_{2} =
  \Bigl\langle \Phi_{f}\; \Bigl|\, e^{i\, c_{p}\, \vb{k_{\gamma}} \vb{r}}\, \Bigr|\, \Phi_{i}\: \Bigr\rangle_\mathbf{r} =

  \sqrt{\displaystyle\frac{\pi}{2}}\:
  \displaystyle\sum\limits_{l_{\gamma}=1}\,
    (-i)^{l_{\gamma}}\, \sqrt{2l_{\gamma}+1}\;
  \displaystyle\sum\limits_{\mu = \pm 1}
    \Bigl[ \mu\,\tilde{p}_{l_{\gamma}\mu}^{M} (-c_{p}) - i\, \tilde{p}_{l_{\gamma}\mu}^{E} (-c_{p}) \Bigr], \\

\vspace{0.5mm}
  I_{3} =
  \Bigl\langle \Phi_{f}\; \Bigl|\, e^{-i\, c_{A}\, \vb{k_{\gamma}} \vb{r}}\, \Bigr|\, \Phi_{i}\: \Bigr\rangle_\mathbf{r} =
  \sqrt{\displaystyle\frac{\pi}{2}}\:
  \displaystyle\sum\limits_{l_{\gamma}=1}\,
    (-i)^{l_{\gamma}}\, \sqrt{2l_{\gamma}+1}\;
  \displaystyle\sum\limits_{\mu = \pm 1}
    \Bigl[ \mu\,\tilde{p}_{l_{\gamma}\mu}^{M} (c_{A}) - i\, \tilde{p}_{l_{\gamma}\mu}^{E} (c_{A}) \Bigr], \\

  I_{4} =
  \Bigl\langle \Phi_{f}\; \Bigl|\, e^{- i\, c_{A}\, \vb{k_{\gamma}} \vb{r}}\, V(\vb{r})\, \Bigr|\, \Phi_{i}\: \Bigr\rangle_\mathbf{r} =

  \sqrt{\displaystyle\frac{\pi}{2}}\:
  \displaystyle\sum\limits_{l_{\gamma}=1}\,
    (-i)^{l_{\gamma}}\, \sqrt{2l_{\gamma}+1}\;
  \displaystyle\sum\limits_{\mu = \pm 1}
    \Bigl[ \mu\, \breve{p}_{l_{\gamma}\mu}^{M} (c_{A}) - i\, \breve{p}_{l_{\gamma}\mu}^{E} (c_{A}) \Bigr],
\end{array}
\label{eq.app.integrals.6}
\end{equation}
where
[see solutions (40) at $l_{i}=0$ in Ref.~\cite{Maydanyuk.2012.PRC}
for $\tilde{p}_{l_{\gamma}\mu}^{M}$, $\tilde{p}_{l_{\gamma}\mu}^{E}$,
Eqs.~(F14) and (F26) in Ref.~\cite{Maydanyuk_Zhang_Zou.2019.PRC.microscopy}
for all matrix elements]
\begin{equation}
\begin{array}{llll}
\vspace{3mm}
  \tilde{p}_{l_{\gamma}\mu}^{M} (c) =
    \tilde{I}\,(0,l_{f},l_{\gamma}, l_{\gamma}, \mu) \cdot \tilde{J}\, (c, 0,l_{f},l_{\gamma}), \\

\vspace{3mm}
  \tilde{p}_{l_{\gamma}\mu}^{E} (c) =
    \sqrt{\displaystyle\frac{l_{\gamma}+1}{2l_{\gamma}+1}} \tilde{I}\,(0,l_{f},l_{\gamma},l_{\gamma}-1,\mu) \cdot \tilde{J}\,(c, 0,l_{f},l_{\gamma}-1)\; - 
    \sqrt{\displaystyle\frac{l_{\gamma}}{2l_{\gamma}+1}} \tilde{I}\,(0,l_{f},l_{\gamma},l_{\gamma}+1,\mu) \cdot \tilde{J}\,(c, 0,l_{f},l_{\gamma}+1), \\

\vspace{3mm}
  \breve{p}_{l_{\gamma}\mu}^{M} (c_{A}) =
    \tilde{I}\,(0,l_{f},l_{\gamma}, l_{\gamma}, \mu) \cdot \breve{J}\, (c_{A}, 0,l_{f},l_{\gamma}), \\

  \breve{p}_{l_{\gamma}\mu}^{E} (c_{A}) =
    \sqrt{\displaystyle\frac{l_{\gamma}+1}{2l_{\gamma}+1}} \tilde{I}\,(0,l_{f},l_{\gamma},l_{\gamma}-1,\mu) \cdot \breve{J}\,(c_{A}, 0,l_{f},l_{\gamma}-1)\; - 
    \sqrt{\displaystyle\frac{l_{\gamma}}{2l_{\gamma}+1}} \tilde{I}\,(0,l_{f},l_{\gamma},l_{\gamma}+1,\mu) \cdot \breve{J}\,(c_{A}, 0,l_{f},l_{\gamma}+1)
\end{array}
\label{eq.app.integrals.7}
\end{equation}
and
[see solutions (41) in Ref.~\cite{Maydanyuk.2012.PRC} for $\tilde{J}$ and corresponding angular integral,
Eqs.~(F13) and (F21) in Ref.~\cite{Maydanyuk_Zhang_Zou.2019.PRC.microscopy} for all matrix elements]
\begin{equation}
\begin{array}{ccl}
  \tilde{J}\,(c, l_{i},l_{f},n) & = & \displaystyle\int\limits^{+\infty}_{0} R_{i}(r)\, R^{*}_{f}(l,r)\, j_{n}(c\, k_{\gamma}r)\; r^{2} dr, \\
  \breve{J}\,(c_{A}, l_{i}, l_{f},n) & = & \displaystyle\int\limits^{+\infty}_{0} R_{i}(r)\, R^{*}_{l,f}(r)\, V(\mathbf{r})\, j_{n}(c_{A}\,kr)\; r^{2} dr.
\end{array}
\label{eq.app.integrals.8}
\end{equation}


\subsection{Linear and circular polarizations of the photon emitted
\label{sec.app.polarization}}

Rewrite vectors of \emph{linear} polarization $\varepsilonbf^{(\alpha)}$ through \emph{vectors of circular polarization} $\mathbf{\xi}_{\mu}$ with opposite directions of rotation (see Ref.~\cite{Eisenberg.1973}, (2.39), p.~42):
\begin{equation}
\begin{array}{ccc}
  \mathbf{\xi}_{-1} = \displaystyle\frac{1}{\sqrt{2}}\,
                      \bigl(\varepsilonbf^{(1)} - i\varepsilonbf^{(2)}\bigr), &
  \mathbf{\xi}_{+1} = -\displaystyle\frac{1}{\sqrt{2}}\,
                      \bigl(\varepsilonbf^{(1)} + i\varepsilonbf^{(2)}\bigr), &
  \mathbf{\xi}_{0} = \varepsilonbf^{(3)}.
\end{array}
\label{eq.app.polarization.1.1}
\end{equation}
%
where
\begin{equation}
\begin{array}{ccc}
  h_{\pm} = \mp \displaystyle\frac{1 \pm i}{\sqrt{2}}, &
  h_{-1} + h_{+1} = -i\sqrt{2}, &
  \sum\limits_{\alpha = 1,2} \varepsilonbf^{(\alpha),*} =
    h_{-1} \mathbf{\xi}_{-1}^{*} + h_{+1} \mathbf{\xi}_{+1}^{*}.
\end{array}
\label{eq.app.polarization.1.2}
\end{equation}
We have (in Coulomb gauge at $\varepsilonbf^{(3)}=0$)
%
\begin{equation}
\begin{array}{cc}
  \varepsilonbf^{(1)} = \displaystyle\frac{1}{\sqrt{2}}\, \bigl(\xibf_{-1} - \xibf_{+1}\bigr), &
  \varepsilonbf^{(2)} = \displaystyle\frac{i}{\sqrt{2}}\, \bigl(\xibf_{-1} + \xibf_{+1}\bigr),
\end{array}
\label{eq.app.polarization.1.3}
\end{equation}
\begin{equation}
\begin{array}{ccc}
  \displaystyle\sum\limits_{\mu = \pm 1} \xi_{\mu}^{*} \cdot \xi_{\mu} =
  \displaystyle\frac{1}{2}\,
    \bigl(\varepsilonbf^{(1)} - i\varepsilonbf^{(2)}\bigr)\, \bigl(\varepsilonbf^{(1)} - i\varepsilonbf^{(2)}\bigr)^{*} +
  \displaystyle\frac{1}{2}\,
    \bigl(\varepsilonbf^{(1)} + i\varepsilonbf^{(2)}\bigr)\, \bigl(\varepsilonbf^{(1)} + i\varepsilonbf^{(2)}\bigr)^{*} = 2.
\end{array}
\label{eq.app.polarization.1.4}
\end{equation}

Also we will find vectorial products of vectors $\xibf_{\pm 1}$. From Eqs.~(\ref{eq.app.polarization.1.1}) we obtain
%
\begin{equation}
\begin{array}{cc}
  \xibf_{-1}^{*} = - \xibf_{+1}, &
  \xibf_{+1}^{*} = - \xibf_{-1}.
\end{array}
\label{eq.app.polarization.1.5} 
\end{equation}
From here we obtain vector multiplications as
%
\begin{equation}
\begin{array}{lcl}
  \Bigl[\xibf_{-1} \times \xibf_{+1}\Bigr] & = &
  - i\, \bigl[\varepsilonbf^{(1)} \times \varepsilonbf^{(2)}\bigr] =
  - i\, \varepsilonbf_{\rm z},
\end{array}
\label{eq.app.polarization.1.6} 
\end{equation}

\vspace{-2mm}
\begin{equation}
\begin{array}{lcllcl}
  \vspace{2mm}
  \Bigl[\xibf_{-1}^{*} \times \xibf_{+1}\Bigr] & = & -\, \Bigl[\xibf_{+1} \times \xibf_{+1}\Bigr] = 0, &

  \hspace{10mm}
  \Bigl[\xibf_{-1}^{*} \times \xibf_{-1}\Bigr] & = & -\, \Bigl[\xibf_{+1} \times \xibf_{-1}\Bigr] =
  i\, \varepsilonbf_{\rm z}, \\

  \Bigl[\xibf_{+1}^{*} \times \xibf_{-1}\Bigr] & = & -\, \Bigl[\xibf_{-1} \times \xibf_{-1}\Bigr] = 0, &

  \hspace{10mm}
  \Bigl[\xibf_{+1}^{*} \times \xibf_{+1}\Bigr] & = & -\, \Bigl[\xibf_{-1} \times \xibf_{+1}\Bigr] =
  -i\, \varepsilonbf_{\rm z}.
\end{array}
\label{eq.app.polarization.1.7} 
\end{equation}



\subsection{Summation over vectors of polarizations
\label{sec.app.integ}}

In this section we will calculate multiplications of integrals on vectors of polarizations.
Let' consider the first integral $\vb{I}_{1}$ which has form [see Eqs.~(\ref{eq.app.integrals.3})]
\begin{equation}
\begin{array}{ll}
  \vb{I}_{1} =
  \biggl< \Phi_{f} \biggl| \,  e^{-i\mathbf{k_{\gamma}r}} \displaystyle\frac{\partial}{\partial \vb{r}}\, \biggr| \,\Phi_{i} \biggr>_\mathbf{r} =
  \sqrt{\displaystyle\frac{\pi}{2}}\:
  \displaystyle\sum\limits_{l_{\gamma}=1}\,
    (-i)^{l_{\gamma}}\, \sqrt{2l_{\gamma}+1}\;
  \displaystyle\sum\limits_{\mu = \pm 1}
    \xibf_{\mu}\, \mu\, \times
    \Bigl[ p_{l_{\gamma}\mu}^{M} - i\mu\: p_{l_{\gamma}\mu}^{E} \Bigr].
\end{array}
\label{eq.app.integ.1}
\end{equation}
We calculate
\begin{equation}
\begin{array}{llll}
\vspace{1.5mm}
  \varepsilonbf^{(1)} \cdot \vb{I}_{1} & = &
%
  - \sqrt{\displaystyle\frac{\pi}{2}}\:
  \displaystyle\sum\limits_{l_{\gamma}=1}\,
    (-i)^{l_{\gamma}}\, \sqrt{2l_{\gamma}+1}\;
  \displaystyle\frac{1}{\sqrt{2}}\,
  \displaystyle\sum\limits_{\mu = \pm 1}\,
    \Bigl[ p_{l_{\gamma}\mu}^{M} - i\, \mu\, p_{l_{\gamma}\mu}^{E} \Bigr], \\

  \varepsilonbf^{(2)} \cdot \vb{I}_{1} & = &
%
  -\, \sqrt{\displaystyle\frac{\pi}{2}}\:
  \displaystyle\sum\limits_{l_{\gamma}=1}\,
    (-i)^{l_{\gamma}}\, \sqrt{2l_{\gamma}+1}\;
  \displaystyle\frac{i}{\sqrt{2}}\,
  \displaystyle\sum\limits_{\mu = \pm 1}\,
    \Bigl[ \mu\, p_{l_{\gamma}\mu}^{M} - i\, p_{l_{\gamma}\mu}^{E} \Bigr],
\end{array}
\label{eq.app.integ.2}
\end{equation}
and summation over vectors of polarization is
\begin{equation}
\begin{array}{llllll}
  \displaystyle\sum\limits_{\alpha=1,2}
    \varepsilonbf^{(\alpha)} \cdot \vb{I}_{1} & = &
%
%
  -\, \sqrt{\displaystyle\frac{\pi}{2}}\:
  \displaystyle\sum\limits_{l_{\gamma}=1}\,
    (-i)^{l_{\gamma}}\, \sqrt{2l_{{\gamma}+1}}\;
  \displaystyle\sum\limits_{\mu = \pm 1}\,
    \Bigl[
      \displaystyle\frac{1 + i\, \mu}{\sqrt{2}}\, p_{l_{\gamma}\mu}^{M} +
      \displaystyle\frac{1 - i\,\mu}{\sqrt{2}}\, p_{l_{\gamma}\mu}^{E}
    \Bigr].
\end{array}
\label{eq.app.integ.3}
\end{equation}
There is property
\begin{equation}
\begin{array}{llllll}
\vspace{1.0mm}
  \displaystyle\sum\limits_{\mu = \pm 1}\,
    \Bigl[
      \displaystyle\frac{1 - i\,\mu}{\sqrt{2}}\, p_{l_{\gamma}\mu}^{E}
    \Bigr] =

  \displaystyle\sum\limits_{\mu = \pm 1}\,
    \Bigl[
      \displaystyle\frac{1 + i\,\mu}{\sqrt{2}}\, p_{l_{\gamma},\, -\mu}^{E}
    \Bigr], \\

  \displaystyle\sum\limits_{\mu = \pm 1}\,
    \Bigl[
      \displaystyle\frac{1 + i\, \mu}{\sqrt{2}}\, p_{l_{\gamma}\mu}^{M} +
      \displaystyle\frac{1 - i\,\mu}{\sqrt{2}}\, p_{l_{\gamma}\mu}^{E}
    \Bigr] =

  \displaystyle\sum\limits_{\mu = \pm 1}\,
    \Bigl[
      \displaystyle\frac{1 + i\, \mu}{\sqrt{2}}\, p_{l_{\gamma}\mu}^{M} +
      \displaystyle\frac{1 + i\,\mu}{\sqrt{2}}\, p_{l_{\gamma},\, -\mu}^{E}
    \Bigr] =

  \displaystyle\sum\limits_{\mu = \pm 1}\,
    \displaystyle\frac{1 + i\, \mu}{\sqrt{2}}\,
    \Bigl[ p_{l_{\gamma}\mu}^{M} + p_{l_{\gamma},\, -\mu}^{E} \Bigr].
\end{array}
\label{eq.app.integ.4}
\end{equation}
Then one can write Eq.~(\ref{eq.app.integ.3}) as
\begin{equation}
\begin{array}{llll}
  \displaystyle\sum\limits_{\alpha=1,2} \varepsilonbf^{(\alpha)} \cdot \vb{I}_{1} =
  \sqrt{\displaystyle\frac{\pi}{2}}\:
  \displaystyle\sum\limits_{l_{\gamma}=1}\, (-i)^{l_{\gamma}}\, \sqrt{2l_{\gamma}+1}\;
  \displaystyle\sum\limits_{\mu=\pm 1} \mu\,h_{\mu}\, \bigl(p_{l_{\gamma}, \mu}^{M} + p_{l_{\gamma}, -\mu}^{E} \bigr).
\end{array}
\label{eq.app.integ.5}
\end{equation}
%
%
%
We calculate properties
\begin{equation}
\begin{array}{llll}
  (\varepsilonbf_{\rm x} + \varepsilonbf_{\rm z})\,  \displaystyle\sum\limits_{\alpha=1,2} \Bigl[ \vb{I}_{1} \times \varepsilonbf^{(\alpha)} \Bigr] =
  \sqrt{\displaystyle\frac{\pi}{2}}\:
  \displaystyle\sum\limits_{l_{\gamma}=1}\, (-i)^{l_{\gamma}}\, \sqrt{2l_{\gamma}+1}\;
  \displaystyle\sum\limits_{\mu=\pm 1} \mu\,h_{\mu}\, \bigl(p_{l_{\gamma}, \mu}^{M} - p_{l_{\gamma}, -\mu}^{E} \bigr).
\end{array}
\label{eq.app.integ.6}
\end{equation}

\subsection{Case of $l_{i}=0$, $l_{f}=1$, $l_{\gamma}=1$
\label{sec.app.simplestcase}}

In a case of $l_{i}=0$, $l_{f}=1$, $l_{\gamma}=1$ integrals (\ref{eq.app.integrals.4}), (\ref{eq.app.integrals.7}) are simplified to
\begin{equation}
\begin{array}{llllll}
\vspace{0.5mm}
  \vb{I}_{1} =
  -i\, \sqrt{\displaystyle\frac{3\pi}{2}}\:
  \displaystyle\sum\limits_{\mu = \pm 1}
    \xibf_{\mu}\, \mu\, \times
    \Bigl[ p_{l_{\gamma}\mu}^{M} - i\mu\: p_{l_{\gamma}\mu}^{E} \Bigr], &

  I_{3} =
  -i\, \sqrt{\displaystyle\frac{3\pi}{2}}\:
  \displaystyle\sum\limits_{\mu = \pm 1}
    \Bigl[ \mu\,\tilde{p}_{l_{\gamma}\mu}^{M} (c_{A}) - i\, \tilde{p}_{l_{\gamma}\mu}^{E} (c_{A}) \Bigr], \\

  I_{2} =
  -i\, \sqrt{\displaystyle\frac{3\pi}{2}}\:
  \displaystyle\sum\limits_{\mu = \pm 1}
    \Bigl[ \mu\,\tilde{p}_{l_{\gamma}\mu}^{M} (-c_{p}) - i\, \tilde{p}_{l_{\gamma}\mu}^{E} (-c_{p}) \Bigr], &

  I_{4} =
  -i\, \sqrt{\displaystyle\frac{3\pi}{2}}\:
  \displaystyle\sum\limits_{\mu = \pm 1}
    \Bigl[ \mu\, \breve{p}_{l_{\gamma}\mu}^{M} (c_{A}) - i\, \breve{p}_{l_{\gamma}\mu}^{E} (c_{A}) \Bigr],
\end{array}
\label{eq.app.simplestcase.1}
\end{equation}
where [see from Eqs.~(\ref{eq.app.integrals.4}), (\ref{eq.app.integrals.7})]
\begin{equation}
\begin{array}{lll}
\vspace{3mm}
  p_{l_{\gamma}\mu}^{M} = - I_{M}(0, 1, 1, 1, \mu) \cdot J_{1}(0, 1, 1), \\

\vspace{3mm}
  p_{l_{\gamma}\mu}^{E} =
    \sqrt{\displaystyle\frac{2}{3}}\, I_{E} (0,1,1,1,0, \mu) \cdot J_{1}(0,1,0) -
    \sqrt{\displaystyle\frac{1}{3}}\, I_{E} (0,1,1,1,2, \mu) \cdot J_{1}(0,1,2), \\

\vspace{3mm}
  \tilde{p}_{l_{\gamma}\mu}^{M} (c) = \tilde{I}\,(0,1,1,1, \mu) \cdot \tilde{J}\, (c, 0,1,1), \\

\vspace{3mm}
  \tilde{p}_{l_{\gamma}\mu}^{E} (c) =
    \sqrt{\displaystyle\frac{2}{3}} \tilde{I}\,(0,1,1,0,\mu) \cdot \tilde{J}\,(c, 0,1,0) -
    \sqrt{\displaystyle\frac{1}{3}} \tilde{I}\,(0,1,1,2,\mu) \cdot \tilde{J}\,(c, 0,1,2), \\

\vspace{3mm}
  \breve{p}_{l_{\gamma}\mu}^{M} (c_{A}) = \tilde{I}\,(0,1,1,1, \mu) \cdot \breve{J}\, (c_{A}, 0,1,1), \\

  \breve{p}_{l_{\gamma}\mu}^{E} (c_{A}) =
    \sqrt{\displaystyle\frac{2}{3}} \tilde{I}\,(0,1,1,0,\mu) \cdot \breve{J}\,(c_{A}, 0,1,0) -
    \sqrt{\displaystyle\frac{1}{3}} \tilde{I}\,(0,1,1,2,\mu) \cdot \breve{J}\,(c_{A}, 0,1,2).
\end{array}
\label{eq.app.simplestcase.2}
\end{equation}

The angular integrals are calculated in Appendix B in Ref.~\cite{Maydanyuk.2012.PRC} [see Eqs.~(B1)--(B10) in that paper].
Results of calculation of angular integrals are
%
%
\begin{equation}
\begin{array}{llllll}
\vspace{0.8mm}
  I_{E}\, (0, 1, 1, 1, 0, \mu) = \sqrt{\displaystyle\frac{1}{24\pi}}, &
  I_{M}\, (0, 1, 1, 1, \mu) = 0, &
  I_{E}\, (0, 1, 1, 1, 2, \mu) = \displaystyle\frac{47}{240} \sqrt{\displaystyle\frac{3}{2\pi}}, \\

  \tilde{I}\, (0, 1, 1, 0, \mu) = 0, &
  \tilde{I}\, (0, 1, 1, 1, \mu) = \displaystyle\frac{\mu}{2\sqrt{2\pi}}, &
  \tilde{I}\, (0, 1, 1, 2, \mu) = 0,
\end{array}
\label{eq.app.simplestcase.3}
\end{equation}
and matrix elements (\ref{eq.app.simplestcase.2}) are simplified to
%
\begin{equation}
\begin{array}{lllllllll}
\vspace{2mm}
  p_{l_{\gamma}\mu}^{M} = 0, &
  p_{l_{\gamma}\mu}^{E} =
    \displaystyle\frac{1}{6} \sqrt{\displaystyle\frac{1}{\pi}} \cdot J_{1}(0,1,0) -
    \displaystyle\frac{47}{240} \sqrt{\displaystyle\frac{1}{2\pi}} \cdot J_{1}(0,1,2), \\

\vspace{2mm}
  \tilde{p}_{1 \mu}^{M} (c) = \displaystyle\frac{\mu}{2\sqrt{2\pi}} \cdot \tilde{J}\, (c, 0,1,1), &
  \tilde{p}_{1 \mu}^{E} (c) = 0, \\

  \breve{p}_{1\mu}^{M} (c_{A}) = \displaystyle\frac{\mu}{2\sqrt{2\pi}} \cdot \breve{J}\, (c_{A}, 0,1,1), &
  \breve{p}_{1\mu}^{E} (c_{A}) = 0.
\end{array}
\label{eq.app.simplestcase.4}
\end{equation}

Now we calculate integrals in Eqs.~(\ref{eq.app.simplestcase.4}).
For $p_{l_{\gamma}=1, \mu}^{M}$ and $p_{l_{\gamma}=1, \mu}^{E}$ we obtain:
\begin{equation}
\begin{array}{llllll}
  \vb{I}_{1} =
  -i\, \sqrt{\displaystyle\frac{3\pi}{2}}\:
  \displaystyle\sum\limits_{\mu = \pm 1}
    \xibf_{\mu}\, \mu\, \times
    \Bigl[- i\mu\: p_{l_{\gamma}=1,\, \mu}^{E} \Bigr] =


  -\, \sqrt{\displaystyle\frac{3\pi}{2}}\:
  p_{l_{\gamma}=1,\, \mu}^{E}\,
  (\xibf_{-1} + \xibf_{+1}).
\end{array}
\label{eq.app.simplestcase.5}
\end{equation}
Taking into account summation of vectors of polarizations (\ref{eq.app.polarization.1.1})
\[
\begin{array}{ccc}
  \mathbf{\xi}_{-1} = \displaystyle\frac{1}{\sqrt{2}}\,
                      \bigl(\varepsilonbf^{(1)} - i\varepsilonbf^{(2)}\bigr), &
  \mathbf{\xi}_{+1} = -\displaystyle\frac{1}{\sqrt{2}}\,
                      \bigl(\varepsilonbf^{(1)} + i\varepsilonbf^{(2)}\bigr), &
  \mathbf{\xi}_{0} = \varepsilonbf^{(3)},
\end{array}
\]
we simplify Eq.~(\ref{eq.app.simplestcase.5}) as
\begin{equation}
\begin{array}{llllll}
  \vb{I}_{1} =

  -\, \sqrt{\displaystyle\frac{3\pi}{2}}\:
  p_{l_{\gamma}=1,\, \mu}^{E} \cdot
  \bigl( - i\, \sqrt{2}\, \varepsilonbf^{(2)} \bigr) =

  i\, \sqrt{3\pi}\:
  p_{l_{\gamma}=1,\, \mu}^{E} \cdot \varepsilonbf^{(2)}.
\end{array}
\label{eq.app.simplestcase.6}
\end{equation}

From Eqs.~(\ref{eq.app.integ.2}) we calculate products of integrals on vectors of polarizations:
\begin{equation}
\begin{array}{llll}
\vspace{0.5mm}
  \varepsilonbf^{(1)} \cdot \vb{I}_{1} & = &
  \Bigl\{
  - \sqrt{\displaystyle\frac{\pi}{2}}\:
  \displaystyle\sum\limits_{l_{\gamma}=1}\,
    (-i)^{l_{\gamma}}\, \sqrt{2l_{\gamma}+1}\;
  \displaystyle\frac{1}{\sqrt{2}}\,
  \displaystyle\sum\limits_{\mu = \pm 1}\,
    \Bigl[ p_{l_{\gamma}\mu}^{M} - i\, \mu\, p_{l_{\gamma}\mu}^{E} \Bigr] \Bigr\}_{l_{\gamma}=1}\; = \\


\vspace{0.5mm}
  & = &
  i\, \sqrt{\displaystyle\frac{\pi}{2}}\: \sqrt{\displaystyle\frac{3}{2}}\,
  \displaystyle\sum\limits_{\mu = \pm 1}\,
    \Bigl[ p_{l_{\gamma}=1, \mu}^{M} - i\, \mu\, p_{l_{\gamma}=1, \mu}^{E} \Bigr] =


  i\, \displaystyle\frac{\sqrt{3 \pi}}{2}\,
  p_{l_{\gamma}=1, \mu}^{E}\,
  \displaystyle\sum\limits_{\mu = \pm 1}\,
    (- i\, \mu\, ) = 0,
\end{array}
\label{eq.app.simplestcase.7}
\end{equation}
\begin{equation}
\begin{array}{llll}
\vspace{0.5mm}
  \varepsilonbf^{(2)} \cdot \vb{I}_{1} & = &
  \Bigl\{
    -\, \sqrt{\displaystyle\frac{\pi}{2}}\:
  \displaystyle\sum\limits_{l_{\gamma}=1}\,
    (-i)^{l_{\gamma}}\, \sqrt{2l_{\gamma}+1}\;
  \displaystyle\frac{i}{\sqrt{2}}\,
  \displaystyle\sum\limits_{\mu = \pm 1}\,
    \Bigl[ \mu\, p_{l_{\gamma}\mu}^{M} - i\, p_{l_{\gamma}\mu}^{E} \Bigr]
  \Bigr\}_{l_{\gamma}=1}\; = \\


\vspace{0.5mm}
  & = &
  -\, \sqrt{\displaystyle\frac{\pi}{2}}\, \sqrt{\displaystyle\frac{3}{2}}\,
  \displaystyle\sum\limits_{\mu = \pm 1}\,
    \Bigl[ \mu\, p_{l_{\gamma}=1, \mu}^{M} - i\, p_{l_{\gamma}=1, \mu}^{E} \Bigr] =

  -\, \displaystyle\frac{\sqrt{3\pi}}{2}\,
  \displaystyle\sum\limits_{\mu = \pm 1}\, \Bigl[ - i\, p_{l_{\gamma}=1, \mu}^{E} \Bigr]\; = 


  i\, \sqrt{3\pi}\, p_{l_{\gamma}=1, \mu}^{E}.
\end{array}
\label{eq.app.simplestcase.8}
\end{equation}
Therefore, we write down the final solutions:
\begin{equation}
\begin{array}{llllll}
  \varepsilonbf^{(1)} \cdot \vb{I}_{1} = 0, &
  \varepsilonbf^{(2)} \cdot \vb{I}_{1} =
  \displaystyle\sum\limits_{\alpha=1,2} \varepsilonbf^{(\alpha)} \cdot \vb{I}_{1} =
  i\, \sqrt{3\pi}\, p_{l_{\gamma}=1, \mu}^{E}.
\end{array}
\label{eq.app.simplestcase.9}
\end{equation}
We obtain property
\begin{equation}
\begin{array}{llllll}
  \vb{I}_{1} =
  i\, \sqrt{3\pi}\:
  p_{l_{\gamma}=1,\, \mu}^{E} \cdot \varepsilonbf^{(2)} =

  \Bigl[ \displaystyle\sum\limits_{\alpha=1,2} \varepsilonbf^{(\alpha)} \cdot \vb{I}_{1} \Bigr] \cdot \varepsilonbf^{(2)}.
\end{array}
\label{eq.app.simplestcase.10}
\end{equation}
Taking solution (\ref{eq.app.simplestcase.4}) into account, rewritten via raadial integrals:
%
\[
\begin{array}{lllll}
  p_{l_{\gamma}\mu}^{E} =
    \displaystyle\frac{1}{6} \sqrt{\displaystyle\frac{1}{\pi}} \cdot J_{1}(0,1,0) -
    \displaystyle\frac{47}{240} \sqrt{\displaystyle\frac{1}{2\pi}} \cdot J_{1}(0,1,2),
\end{array}
\]
we obtain
\begin{equation}
\begin{array}{llllll}
\vspace{0.5mm}
  \vb{I}_{1} =
  \varepsilonbf^{(2)} \cdot i\, \displaystyle\frac{\sqrt{3}}{6}\,
  \Bigl\{
    J_{1}(0,1,0) -
    \displaystyle\frac{47}{40} \sqrt{\displaystyle\frac{1}{2}} \cdot J_{1}(0,1,2)
  \Bigr\}, \\

  \varepsilonbf^{(1)} \cdot \vb{I}_{1} = 0, \\

  \varepsilonbf^{(2)} \cdot \vb{I}_{1} =
  \displaystyle\sum\limits_{\alpha=1,2} \varepsilonbf^{(\alpha)} \cdot \vb{I}_{1} =
  i\, \displaystyle\frac{\sqrt{3}}{6}\,
  \Bigl\{
    J_{1}(0,1,0) -
    \displaystyle\frac{47}{40} \sqrt{\displaystyle\frac{1}{2}} \cdot J_{1}(0,1,2)
  \Bigr\}.
\end{array}
\label{eq.app.simplestcase.11}
\end{equation}

\section{Form factor of deuteron in cluster approach
\label{sec.app.clusterformfactor}} 

\subsection{Normalization of the deuteron wave function} 

The deuteron wave function

    \begin{equation}
    \vspace{0.5mm}
        \phi(\pmb{r}) = A \dfrac{e^{-\kappa r}}{r} Y_{lm}(\theta, \varphi).
    \end{equation}
    where $\kappa = \sqrt{\dfrac{2 m | E_d |}{\hbar ^ 2}}$ and $\pmb{r}$ - Jacobi vector of the relative position of the nucleons inside deuteron.
    
    Normalization condition:
    \begin{equation}
    \begin{array}{lllll}
    \vspace{0.5mm}
        \displaystyle \int |\phi(\pmb{r})| ^ 2 dV = 1,\;
        A^2 \int \dfrac{e ^ {-2 \kappa r}}{r^2} Y_{l'm'}^{*}(\theta, \varphi) Y_{lm}(\theta, \varphi)dV = 1,
        \end{array}
    \end{equation}

    \begin{equation}
    \begin{array}{lllll}
    \vspace{0.5mm}
        \displaystyle A^2 \int_{0}^{\infty} \dfrac{e ^ {-2 \kappa r}}{r^2} r^2 dr \int Y_{l'm'}^{*}(\theta, \varphi) Y_{lm}(\theta, \varphi) d\Omega = 1, \;
        \delta_{l, l'} \delta_{m, m'} A^2 \int_{0}^{\infty} e ^ {-2 \kappa r} dr = 1,
        \end{array}
    \end{equation}

    \begin{equation}
    \begin{array}{lllll}
    \vspace{0.5mm}
        -\dfrac{A^2}{2\kappa} e ^ {-2 \kappa r} \bigg|_{r = 0}^{r \rightarrow \infty} = 1, \;
        \frac{A^2}{2\kappa} = 1.
    \end{array}
    \end{equation}

    Thus we have the normalization multiplier:
    \begin{equation}
    \vspace{5mm}
        A = \sqrt{2\kappa},
    \end{equation}
    and the final form of the deuteron wave function:
    \begin{equation}
    \vspace{0.5mm}
        \phi(\pmb{r}) = \sqrt{2\kappa} \dfrac{e ^ {- \kappa r}}{r} Y_{l,m} (\theta, \varphi) = \sqrt{\dfrac{\kappa} {2\pi}} \dfrac{e ^ {- \kappa r}}{r},
    \end{equation}
%
where we have chosen $l = 0$.

\subsection{Calculation of the form factors} 

We have to calculate the following form factors :
\begin{equation}
\begin{array}{lll}
\vspace{0.5mm}
        F_{1} (\pmb{k}_\gamma) & = & \biggl\langle \phi(\pmb{r}) \biggl|
        \exp \biggl\{ -\dfrac{i}{\sqrt{2}} (\pmb{k}_\gamma \pmb{r})\biggr\} \biggr| \phi(\pmb{r}) \biggr\rangle, \\
        F_{2} (\pmb{k}_\gamma) & = & \biggl\langle \phi(\pmb{r}) \biggl|
        \exp \biggl\{ -\dfrac{i}{\sqrt{2}} (\pmb{k}_\gamma \pmb{r})\biggr\} (\varepsilon^{(\alpha)}, \pi_{\pmb{r}}^{*}) \biggr| \phi(\pmb{r}) \biggr\rangle.
    \end{array}
\end{equation}
We calculate the first form factor as
%

    \begin{equation}
    \begin{array}{lll}
    \vspace{0.5mm}
        F_{1} (\pmb{k}_\gamma)  = \displaystyle \int \phi^{*}(\pmb{r}) \exp \biggl\{ -\dfrac{i}{\sqrt{2}} (\pmb{k}_\gamma \pmb{r})\biggr\} \phi(\pmb{r}) d\pmb{r} = \dfrac{\kappa}{2 \pi} \int \dfrac{e^{-2\kappa r}}{r^2} e^{-i/\sqrt{2} \, \pmb{k}_\gamma \pmb{r}}
        r^2 dr d\Omega.
    \end{array}
    \end{equation}

    Now we use the following expansion
    \begin{equation}
    \begin{array}{llll}
    \vspace{0.5mm}
        & \displaystyle e^{i \pmb{k}\pmb{r}} = \sum_{l=0}^{\infty} i^l j_l(kr) \sum_{m=-l}^{m=l} Y_{lm}^*(\theta_{k}, \varphi_{k}) Y_{lm} (\theta_{r}, \varphi_{r}), \\
        & \displaystyle e^{-i/\sqrt{2} \, \pmb{k}_\gamma \pmb{r}} =
        \sum_{l=0}^{\infty} (i^l)^* j_l \biggl(\dfrac{k_\gamma r}{\sqrt{2}}\biggr) \sum_{m=-l}^{m=l} Y_{lm}^* (\theta_{k}, \varphi_{k}) Y_{lm} (\theta_{r}, \varphi_{r}).
    \end{array}
    \end{equation}

    We will have for our integral
    \begin{equation}
        \begin{array}{ll}
            \vspace{0.5mm}
            \displaystyle F_{1} (\pmb{k}_\gamma) & = \dfrac{\kappa}{2 \pi} \displaystyle \sum_{l'=0}^\infty \bigl(i^{l'}\bigr)^*
            \displaystyle \int_0^\infty j_{l'} \bigl(k_\gamma r /\sqrt{2}\bigr) \dfrac{e^{-2\kappa r}}{r^2} r^2 dr \; \int d\Omega \; \sum_{m'=-l'}^{m=l'} Y_{l'm'} (\theta, \varphi) Y_{l'm'}^* (\theta, \varphi) = \\
            & = \dfrac{\kappa}{2 \pi}
        \displaystyle \int_0^\infty j_{0} \bigl(k_\gamma r /\sqrt{2}\bigr) e^{-2\kappa r} dr.
        \end{array}
    \end{equation}
    %
    and for the Bessel spherical functions we have $j_0(x) = \sin (x) / x$:
    \begin{equation}
    \begin{array}{ll}
    \vspace{0.5mm}
        F_{1} (\pmb{k}_\gamma) & =  \dfrac{\sqrt{2}\kappa}{2 \pi k_\gamma}
        \displaystyle \int_0^\infty \dfrac{\sin \bigl(k_\gamma r / \sqrt{2}\bigr)}{r} e^{-2\kappa r} dr,
    \end{array}
    \end{equation}

    Let us focus on the integral itself
    \begin{equation}
    \begin{array}{ll} 
    \vspace{0.5mm}
        I & = \displaystyle \int_0^\infty \dfrac{\sin \bigl(k_\gamma r / \sqrt{2}\bigr)}{r} e^{-2\kappa r} dr =
        \displaystyle \int_0^\infty \dfrac{\sin at}{t} e^{-t} dt,
    \end{array}
    \end{equation}
    where $t = 2 \kappa r$, $a = k_{\gamma} / 2 \sqrt{2} \kappa$.

    \begin{equation}
    \begin{array}{lll}
    \vspace{0.5mm}
        I (a) &= \displaystyle \int_0^\infty \dfrac{\sin at}{t} e^{-t} dt, \;
        \dfrac{dI}{da} = \dfrac{d}{da} \biggl[\displaystyle \int_0^\infty \dfrac{\sin at}{t} e^{-t} dt\biggr], \;
        \dfrac{dI}{da} = \displaystyle \int_0^\infty \cos at \; e^{-t} dt = I_1.
    \end{array}
    \end{equation}

    For the integral $I_1$ we have
    \begin{equation}
    \begin{array}{ll}
    \vspace{0.5mm}
        I_1 = \displaystyle \int_0^\infty \cos at \; e^{-t} dt = - \displaystyle \int_0^\infty \cos at \; d(e^{-t}), \\
        I_1 = - \cos at \; e^{-t} \bigg|_{t = 0}^{t \rightarrow \infty} +
        \displaystyle \int_0^\infty e^{-t} d(\cos at) = 1 - a \displaystyle \int_0^\infty \sin at \; e^{-t} dt, \\
        I_1 = 1 + a \displaystyle \int_0^\infty \sin at \; d(e^{-t}), \;
        I_1 = 1 + a \biggl[ \sin at \; e^{-t} \bigg|_{t = 0}^{t \rightarrow \infty} - \displaystyle \int_0^\infty e^{-t} d(\sin at) \biggr], \\
        I_1 = 1 - a^2 \displaystyle \int_0^\infty \cos at \; e^{-t} dt = 1 - a^2 I_1, \\
        I_1 + a^2 I_1 = 1.
    \end{array}
    \end{equation}
    Finally, we have the following equation for the integral $I_1$
    \begin{equation}
    \vspace{0.5mm}
        I_1 = \dfrac{1}{1+a^2},
    \end{equation}
    so for the $I$ we have the following equation
    %
    \begin{equation}
    \begin{array}{l}
    \vspace{3mm}
        \dfrac{dI}{da} = \dfrac{1}{1 + a^2},
        \displaystyle \int dI = \displaystyle \int \dfrac{da}{1 + a^2}, \\
        I = \arctan a + C.
    \end{array}
    \end{equation}

    
    Let us come back to the form factor and write down the following
    %
    \begin{equation}
    \vspace{0.5mm}
        F_{1} (\pmb{k}_\gamma) =  \dfrac{2\sqrt{2}\kappa}{ k_\gamma}
        \arctan \biggl(\dfrac{k_\gamma}{2\sqrt{2}\kappa}\biggr) + C_1,
    \end{equation}
    to find out $C$ we will use
    %
    \begin{equation}
    \begin{array}{l}
    \vspace{3mm}
        F_{1}(0) = 1, \;
        \lim_{k_\gamma\rightarrow 0} \dfrac{\sqrt{2}\kappa}{2 \pi k_\gamma}
        \arctan \biggl(\dfrac{k_\gamma}{2\sqrt{2}\kappa}\biggr) + C_1 = 1, \\
        \lim_{k_\gamma\rightarrow 0} \dfrac{2\sqrt{2}\kappa}{k_\gamma}
        \arctan \biggl(\dfrac{k_\gamma} {2\sqrt{2}\kappa}\biggr) + C_1 = 1,  \
        1 + C_1 = 1, \\
        C_1 = 0. 
    \end{array}
    \end{equation}

The final result
%
\begin{equation}
\vspace{0.5mm}
  F_{1} (\pmb{k}_\gamma) =  
  \dfrac{2\sqrt{2}\kappa}{ k_\gamma}
  \arctan \biggl(\dfrac{k_\gamma}{2\sqrt{2}\kappa}\biggr).
\end{equation}
%
%



\end{document}